\documentclass[%
reprint,
superscriptaddress,
nofootinbib,
 amsmath,amssymb,
 aps,
 prd,
 twocolumn
]{revtex4-2}

\usepackage{graphicx}

\usepackage{tikz}
\usetikzlibrary{positioning}
\usepackage[percent]{overpic}
\usepackage{dcolumn}
\usepackage{bm}
\usepackage{xcolor}

\usepackage{siunitx}

\usepackage{amsmath}

\usepackage{subcaption}
\usepackage{mathtools}

\newcommand{\dif}{\mathrm{d}}

\newcommand{\til}[1]{\tilde{#1}}

\begin{document}

\title{Spherically Symmetric Fluid Simulations of Black Hole Accretion in Self-Interacting Dark Matter Halos}

\author{Zhe Meng}
\affiliation{School of Physics and Astronomy, Beijing Normal University, Beijing 100875, China}

\author{Tan Chen}
\email[Corresponding author:~]{chentan@bnu.edu.cn}
\affiliation{School of Physics and Astronomy, Beijing Normal University, Beijing 100875, China}
\affiliation{Institute for Frontier in Astronomy and Astrophysics, Beijing Normal University, Beijing 102206, China}

\author{Bocheng Zhu}
\affiliation{Institute of Astrophysics, School of Physics, Zhengzhou University, China
}

\author{Fan Zhou}
\affiliation{School of Physics and Astronomy, Beijing Normal University, Beijing 100875, China}

\author{Bin Hu}
\affiliation{School of Physics and Astronomy, Beijing Normal University, Beijing 100875, China}
\affiliation{Institute for Frontier in Astronomy and Astrophysics, Beijing Normal University, Beijing 102206, China}

\author{Liang Gao}
\affiliation{School of Physics and Astronomy, Beijing Normal University, Beijing 100875, China}
\affiliation{Institute for Frontier in Astronomy and Astrophysics, Beijing Normal University, Beijing 102206, China}
\affiliation{Institute of Astrophysics, School of Physics, Zhengzhou University, China
}

\author{Rong-Gen Cai}
\affiliation{Institute of Fundamental Physics and Quantum Technology \& School of Physical Science and Technology, Ningbo University, Ningbo 315211, China}

\date{\today}

\begin{abstract}

We investigate black hole accretion in self-interacting dark matter (SIDM) halos using a self-gravitating fluid model with thermal conduction. We develop a robust one-dimensional spherically symmetric hydrodynamic code based on an operator-splitting finite-volume method. Simulating both Singular Isothermal Sphere (SIS) and Navarro--Frenk--White (NFW) profiles, we find that black hole growth is regulated by the competition between gravity-driven inflow and SIDM heat transport. Our results demonstrate that an SIS-like environment facilitates rapid accretion, allowing a $100\,\mathrm{M_{\odot}}$ seed to grow to $10^4\,\mathrm{M_{\odot}}$ within $2\,\mathrm{Myr}$. Furthermore, we show that larger initial black hole masses, steeper density profiles, and higher scattering cross sections significantly enhance the accretion rate. This study provides a comprehensive fluid-dynamical picture of black hole growth in SIDM halos.

\end{abstract}

\maketitle

\section{\label{sec:intro} Introduction}
The nature of dark matter (DM) remains one of the most fundamental unresolved problems in modern physics and cosmology. 
Although the cold dark matter (CDM) paradigm—particularly scenarios involving weakly interacting massive particles (WIMPs)—has been remarkably successful in explaining the large-scale structure of the Universe, increasingly stringent limits from experiments have substantially constrained the viable parameter space of many WIMP models~\cite{Arcadi:2017kky,LZ:2024zvo,XENON:2024hup}.
This situation has motivated the exploration of alternative dark matter scenarios, among which self-interacting dark matter (SIDM) has emerged as a particularly compelling framework~\cite{Spergel:1999mh}.
SIDM preserves the large-scale successes of CDM while introducing non-gravitational scattering between dark matter particles. 
Such self-interactions enable energy and momentum transport within the inner regions of dark matter halos, thereby modifying their density and thermal structure~\cite{Vogelsberger:2012ku,Tulin:2017ara,Jiang:2022aqw,Yang:2023jwn,Zhong:2023yzk,Hu:2023oiu,Jia:2026ocr,Li:2026duw}. 
As a result, SIDM has been widely discussed as a possible explanation~\cite{Spergel:1999mh, Kaplinghat:2015aga,Bullock:2017xww} for several small-scale tensions associated with CDM, including the cusp–core problem~\cite{2010AdAst2010E...5D,Yang:2025xsp}, the too-big-to-fail problem~\cite{Boylan-Kolchin:2011qkt,Zavala:2012us}, and the diversity of galactic rotation curves~\cite{Oman:2015xda,Kamada:2016euw,Kaplinghat:2019dhn,Roberts:2024uyw}.
In addition, recent studies of dark matter substructure in strong gravitational lensing systems have provided further motivation for considering SIDM models~\cite{Yang:2021kdf,Yang:2022hkm, Gilman:2022ida, Hou:2025gmv, Li:2025kpb, Lei:2025pky, Kong:2025sqx, Yu:2025tmp}.

Concurrently, the formation and growth of black holes (BHs) represent another major frontier in astrophysics. 
Understanding these processes is essential for explaining the origin of massive black hole seeds at high redshift, as well as the co-evolution of black holes and their host galaxies. 
This question has become even more timely in light of recent JWST discoveries of little red dots (LRDs) at high redshift~\cite{Matthee:2023utn}, some of which may trace rapid early black hole growth, although their physical nature remains under active debate~\cite{Li:2025nvu, Feng:2025rzf,Inayoshi:2025isg,Jiang:2025jtr}.
In the traditional CDM framework, dark matter is effectively collisionless and lacks an efficient dissipation mechanism. 
Consequently, it is difficult for a central black hole to accrete CDM efficiently, since dark matter particles generally remain on collisionless orbits unless additional processes transport them inward~\cite{Read:2002wb,Shapiro:2023gpe}.
SIDM, however, modifies this picture in an important way. 
Frequent particle scattering introduces effective heat transport and dynamical redistribution within the halo~\cite{Balberg:2002ue}.
On longer timescales, this continuous energy exchange can trigger gravothermal core collapse~\cite{Balberg:2002ue}, driving a central density runaway that potentially seeds supermassive black holes~\cite{Feng:2021rst, Feng:2020kxv, Gu:2026zzq}.
Furthermore, this collisional transport can alter the inner halo structure, redistribute energy, and potentially enhance the inflow of dark matter toward the central region, thereby affecting the growth of a black hole embedded in the halo. 
For this reason, SIDM provides an interesting theoretical framework in which black hole growth may differ substantially from that in collisionless CDM halos, especially in the early Universe.

Despite this motivation, the dynamical details of black hole accretion in SIDM halos remain incompletely understood. 
Existing representative studies of SIDM--BH systems include semi-analytic estimates based on Bondi-like accretion arguments~\cite{Ostriker:1999ee, Hennawi:2001be, Hu:2005cd,Feng:2025rzf}, cosmological simulations comparing black hole growth in CDM and SIDM environments~\cite{VMSabarish:2025tya}, or analytic descriptions of the evolution of SIDM density spikes~\cite{Shapiro:2014oha}.
Although these approaches have provided valuable insights, detailed hydrodynamic studies that resolve the interplay between black hole gravity, halo restructuring, and gravothermal heat transport are still lacking. 
As a result, a fully dynamical picture of SIDM accretion onto a central black hole has yet to be established.

To address this problem, we investigate the accretion of SIDM onto a central black hole by modeling the halo as a self-gravitating fluid with thermal conduction. 
We develop a one-dimensional spherically symmetric hydrodynamic code based on the finite volume method. 
To handle the mixed hyperbolic–parabolic character of the system, we adopt an operator-splitting scheme in which the hyperbolic part is solved with a Roe approximate Riemann solver and the conduction term is advanced with an implicit Euler method.
By following the evolution of both Singular Isothermal Sphere (SIS) and Navarro–Frenk–White (NFW)~\cite{Navarro:1995iw} halos, we study the competition between gravitationally driven inflow and conductive heat transport.
We also examine how the accretion process depends on the initial black hole mass, the scattering cross section, and the slope of the density profile, with the aim of clarifying the conditions under which SIDM may facilitate the formation of massive black hole seeds.

This paper is organized as follows.
In Section~\ref{sec:model}, we introduce the hydrodynamic model for the SIDM halo. 
Section~\ref{sec:Numer} describes the numerical method and validation tests. 
In Section~\ref{sec:result}, we present the simulation results, including a comparison between SIS and NFW halos and a study of the dependence on key physical parameters. 
Finally, we summarize and discuss our findings in Section~\ref{sec:discuss}.

\section{\label{sec:model} Physical Model}  
We model SIDM as a conducting fluid. 
Taking velocity moments of the Boltzmann equation gives an infinite hierarchy of moment equations. 
If the pressure tensor is assumed to be approximately isotropic and the hierarchy is closed at second order, the non-conductive part of the system reduces to the Euler equations for
an ideal gas~\cite{Ahn:2004xt,Koda:2011yb}. 
Elastic self-scattering, however, also transports energy. 
We therefore supplement the ideal-fluid equations with a phenomenological conductive heat flux.

For a scattering cross section per unit mass $\sigma_m=\sigma/m$, the collisional mean free path is defined as $\lambda = 1/(\rho \sigma_m)$, with $\rho$ being the density. The gravitational scale height is defined as $H=\sqrt{v^2/(4\pi G \rho)}$, where $v$ is the one-dimensional velocity dispersion~\cite{Balberg:2002ue,Nishikawa:2019lsc}.
The corresponding Knudsen number is $\mathrm{Kn}\equiv \lambda/H$. 
The regimes $\mathrm{Kn}\ll 1$ and $\mathrm{Kn}\gg 1$ are referred to as the short-mean-free-path (SMFP) and long-mean-free-path (LMFP) limits, respectively.

In the SMFP regime, frequent self-interactions efficiently isotropize the velocity distribution locally. The hydrodynamic description, together with a local conductive closure, is therefore motivated by kinetic theory. 
In this regime, SIDM can be treated as a genuine collisional-fluid accretion problem, provided that the flow is well resolved and remains close to local thermodynamic equilibrium.
In the LMFP regime, the situation is more subtle. 
The system is no longer in a strict local hydrodynamic limit, because particles transport energy over distances comparable to, or larger than, the local scale height. 
The LMFP conductivity commonly used in SIDM gravothermal calculations should instead be understood as a phenomenological closure for nonlocal, orbit-averaged heat transport, usually combined with the SMFP expression through an interpolation formula and calibrated for quasi-static halo evolution~\cite{Balberg:2002ue}. 
Thus, in LMFP regions, the conducting-fluid equations should be understood as an effective gravothermal description of orbit-averaged heat transport rather than as a strictly local hydrodynamic closure.
This approximation is appropriate in the weak-inflow regime considered here, where the bulk radial velocity remains small, and the halo remains close to quasi-static equilibrium. 
In this sense, our model describes black-hole accretion in the SMFP region as a collisional fluid flow, while in the LMFP region it applies to the weak-inflow,
near-quasi-static part of the evolution.

In the present work, we specialize this framework to a spherically symmetric SIDM halo surrounding a central black hole. Let $M_{\text{H}}(t,r)$ denote the dark-matter mass enclosed within radius $r$ at time $t$, and let $M_{\text{BH}}(t)$ denote the mass of the central black hole. The SIDM fluid is characterized by the density $\rho(t,r)$, radial velocity $u(t,r)$, and one-dimensional velocity dispersion $v(t,r)$. Under the assumptions stated above, its evolution is governed by the spherically symmetric equations of mass, momentum, and energy conservation, supplemented by a conductive heat flux:

\begin{subequations}\label{eq:spherical_hydro}
    \begin{align}
        \frac{\partial \rho}{\partial t} + u \frac{\partial\rho}{\partial r} =& - \frac{\rho}{r^2} \frac{\partial}{\partial r}(r^2 u) \label{eq:spherical_continuity}, \\
        \frac{\partial u}{\partial t} + u \frac{\partial u}{\partial r} =& - \frac{1}{\rho} \frac{\partial (\rho v^2)}{ \partial r} \notag \\
        &- \frac{G(M_{\text{H}} + M_{\text{BH}})}{r^2}, \label{eq:spherical_momentum}\\
        \frac{\partial (\frac32 v^2)}{\partial t} + u \frac{\partial (\frac32 v^2)}{\partial r} =&  \frac{1}{\rho r^2} \frac{\partial}{\partial r} \left( r^2 \kappa \frac{\partial (v^2)}{\partial r} \right) \notag \\ 
        & - \frac{ v^2}{r^2} \frac{\partial}{\partial r} (r^2 u). \label{eq:spherical_energy}
    \end{align}
\end{subequations} 

Here the local pressure is $p = \rho v^2$, and the temperature is defined by $k_{\text{B}}T = mv^2$, where $m$ is the dark matter particle mass.
The first term on the right-hand side of Eq.~\eqref{eq:spherical_energy} describes heat transport due to SIDM self-interactions. 
In this equation, $\kappa$ denotes the effective conductivity defined by the heat-flux relation $q=-\kappa\frac{\partial v^2}{\partial r}$.
Under the hydrostatic limit, $u=0$, Eqs.~\eqref{eq:spherical_hydro} reduce to the standard gravothermal fluid model used for SIDM halo evolution~\cite{Balberg:2002ue,Nishikawa:2019lsc}.

During the evolution of the SIDM fluid, the enclosed dark matter mass $M_{\text{H}}$ satisfies
\begin{equation}
    \frac{\partial M_{\text{H}}}{\partial r} = 4 \pi r^2 \rho.
\end{equation}
The central black hole mass $M_{\text{BH}}$ grows over time via accretion
\begin{equation}
    \frac{\dif M_{\text{BH}}}{\dif t} = - 4 \pi r_{\text{min}}^2 \rho_{\text{min}} u_{\text{min}},
\end{equation}
where the subscript ``min'' denotes quantities evaluated at the inner boundary of the computational domain.

The thermal conductivity $\kappa$ in Eq.~\eqref{eq:spherical_energy} characterizes the heat transport due to dark matter self-interactions. Following the standard gravothermal fluid model for SIDM~\cite{Balberg:2002ue,Nishikawa:2019lsc}, we adopt an interpolation formula that smoothly connects the SMFP and LMFP regimes $\kappa^{-1} = \kappa_{\text{SMFP}}^{-1} + \kappa_{\text{LMFP}}^{-1}$. In the SMFP regime ($\lambda \ll H$), collisions are frequent and the conductivity satisfies $\kappa_{\text{SMFP}} = 3 b\rho\lambda^2/(2at_r)$, where $a = \sqrt{16/\pi}$, $b = 25\sqrt{\pi}/32$ and $t_r = \lambda / a v $~\cite{Nishikawa:2019lsc}. In the LMFP regime ($\lambda \gg H$), the conductivity satisfies $\kappa_{\text{LMFP}} = 3C\rho H^2/(2t_r)$, where $C = 0.75$ is a calibration parameter~\cite{Nishikawa:2019lsc}. Substituting these into the interpolation formula, the total thermal conductivity is given by
\begin{equation}
    \kappa = \frac{3}{2} \left( \frac{\sigma_m}{b v} + \frac{4\pi G}{a C \rho v^3\sigma_m} \right)^{-1}. \label{eq:kappa}
\end{equation}

\section{\label{sec:Numer} Numerical methods}
In this section, we describe the numerical methodology adopted in this work. The governing equations are discretized using the finite volume method (FVM)~\cite{blazek2015computational,leveque2002finite}. To efficiently handle the different mathematical characteristics of the system, we employ an operator-splitting strategy~\cite{leveque2002finite} that separates the thermal conduction terms from the hyperbolic hydrodynamic terms. The hyperbolic subsystem is solved using a Roe~\cite{roe1981approximate,toro2013riemann} approximate Riemann solver together with explicit Euler time stepping, while the thermal conduction term is treated with an implicit Euler scheme to avoid the severe numerical stiffness associated with diffusive heat transport.

\subsection{Dimensionless form}
For numerical convenience, we rewrite the governing equations in dimensionless form. 
We introduce characteristic scales $r_{\text{0}} $ and $\rho_{\text{0}} $ as the characteristic scales for radius and density, and define the scales for other physical quantities as 
\begin{equation}\label{eq:dimless_scales1}
    M_0 = 4\pi r_0^3 \rho_0, \quad v_0 = (G M_0/r_0)^{1/2},\quad t_0 = (4\pi G \rho_0)^{-1/2},
\end{equation}
together with
\begin{equation}\label{eq:dimless_scales2}
    p_0 = \rho_0 v_0^2,~ \sigma_{m0} = 1/(\rho_0 r_0).
\end{equation}
The specific choice of $r_{\text{0}} $ and $\rho_{\text{0}} $ depends on the halo profile. Under these definitions, the dimensionless hydrodynamic equations are expressed as 
\begin{subequations}\label{eq:dimless_eqs}
    \begin{align}
        \frac{\partial \til\rho}{\partial \til t} + \til u \frac{\partial\til\rho}{\partial \til r} =& - \frac{\til\rho}{\til r^2} \frac{\partial}{\partial \til r}(\til r^2 \til u), \label{eq:dimless_continuity} \\
        \frac{\partial \til u}{\partial \til t} + \til u \frac{\partial \til u}{\partial \til r} =& - \frac{1}{\til\rho}\frac{\partial (\til \rho \til v^2)}{\partial \til r}\notag\\ 
        &- \frac{\til M_{\text{H}} + \til M_{\text{BH}}}{\til r^2}, \label{eq:dimless_momentum}\\
        \frac{\partial (\frac32 \til v^2)}{\partial \til t} + \til u \frac{\partial (\frac32 \til v^2)}{\partial \til r} =& \frac{1}{\til \rho \til r^2} \frac{\partial}{\partial \til r} \left( \til r^2 \til \kappa \frac{\partial (\til v^2)}{\partial \til r} \right) \notag\\
        &- \frac{ \til v^2}{\til r^2} \frac{\partial}{\partial \til r} (\til r^2 \til u).\label{eq:dimless_energy}
    \end{align}
\end{subequations}
The dimensionless enclosed halo mass and black hole mass satisfy
\begin{subequations}\label{eq:dimless_eqs1}
    \begin{align}
        \frac{\partial \til M_{\text{H}}}{\partial \til r} &= \til r^2 \til \rho, \\
        \frac{\dif \til M_{\text{BH}}}{\dif \til t} &= - \til r_{\text{min}}^2 \til \rho_{\text{min}} \til u_{\text{min}},
    \end{align}
\end{subequations}
while the dimensionless thermal conductivity is expressed as
\begin{equation}
    \til\kappa = \frac{3}{2} \left( \frac{\til \sigma_m}{b \til v} + \frac{1}{a C \til \rho \til v^3\til \sigma_m} \right)^{-1}.
\end{equation}
Here and throughout, quantities marked with a tilde denote dimensionless variables.

\begin{figure*}[t]
    \centering
    \begin{subfigure}[b]{0.4\textwidth}
        \includegraphics[width=1\textwidth]{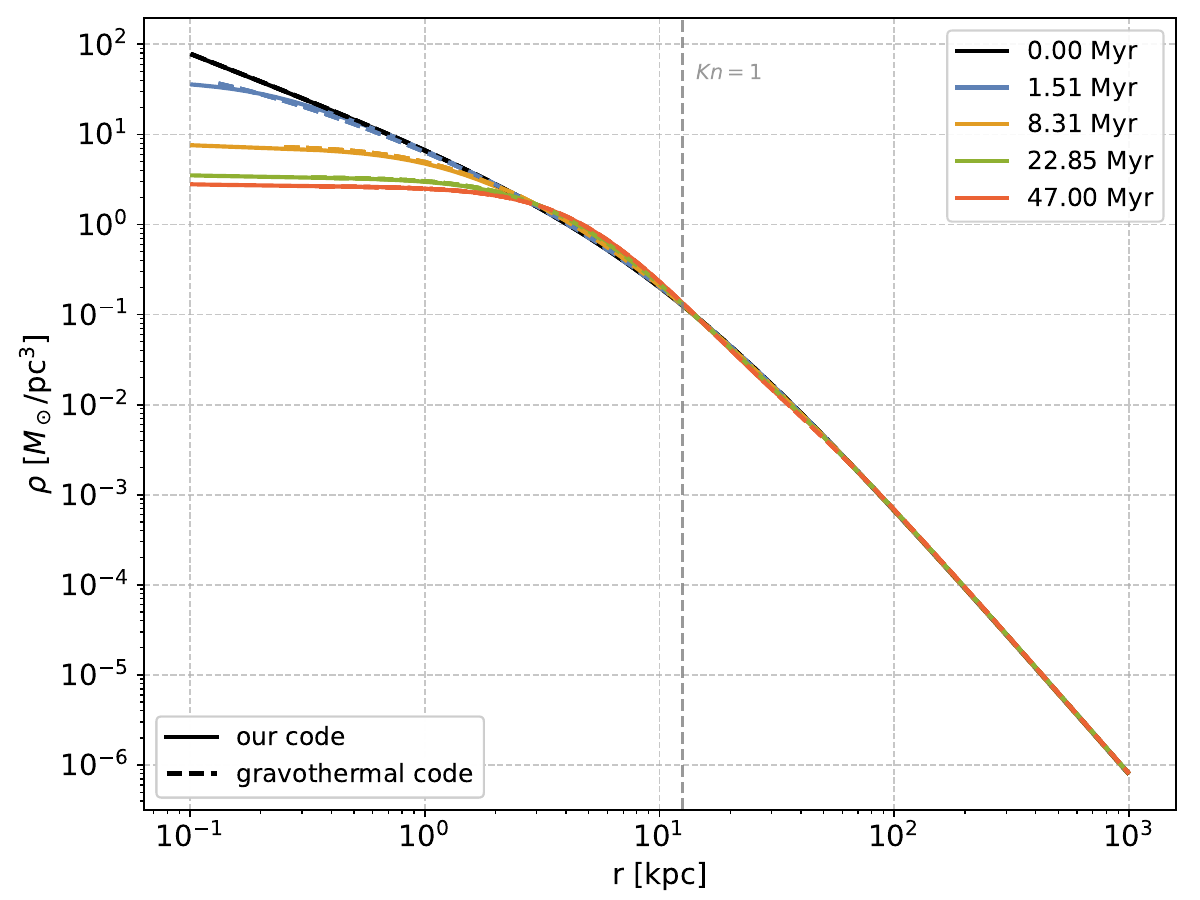}
    \end{subfigure}
    \begin{subfigure}[b]{0.4\textwidth}
        \includegraphics[width=1\textwidth]{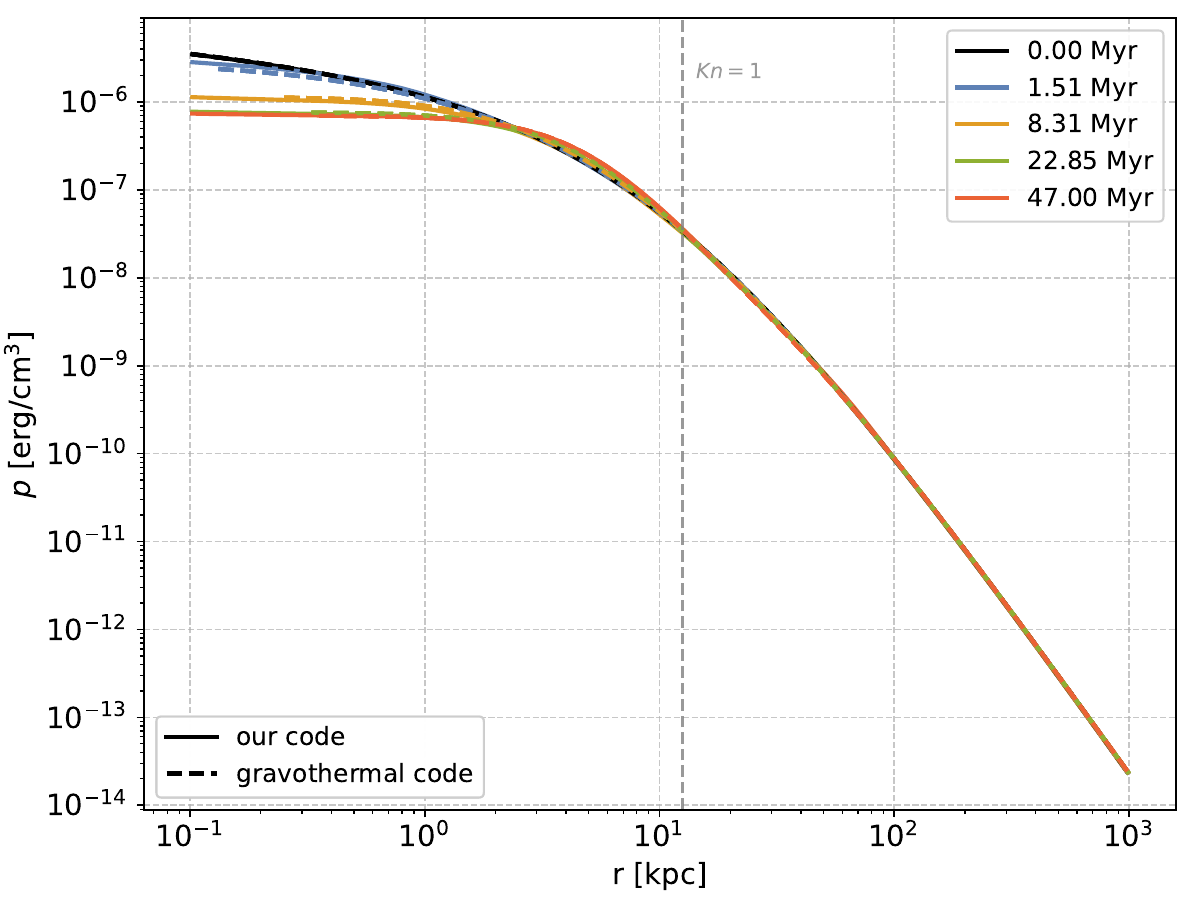}
    \end{subfigure}

    \begin{subfigure}[b]{0.4\textwidth}
        \includegraphics[width=1\textwidth]{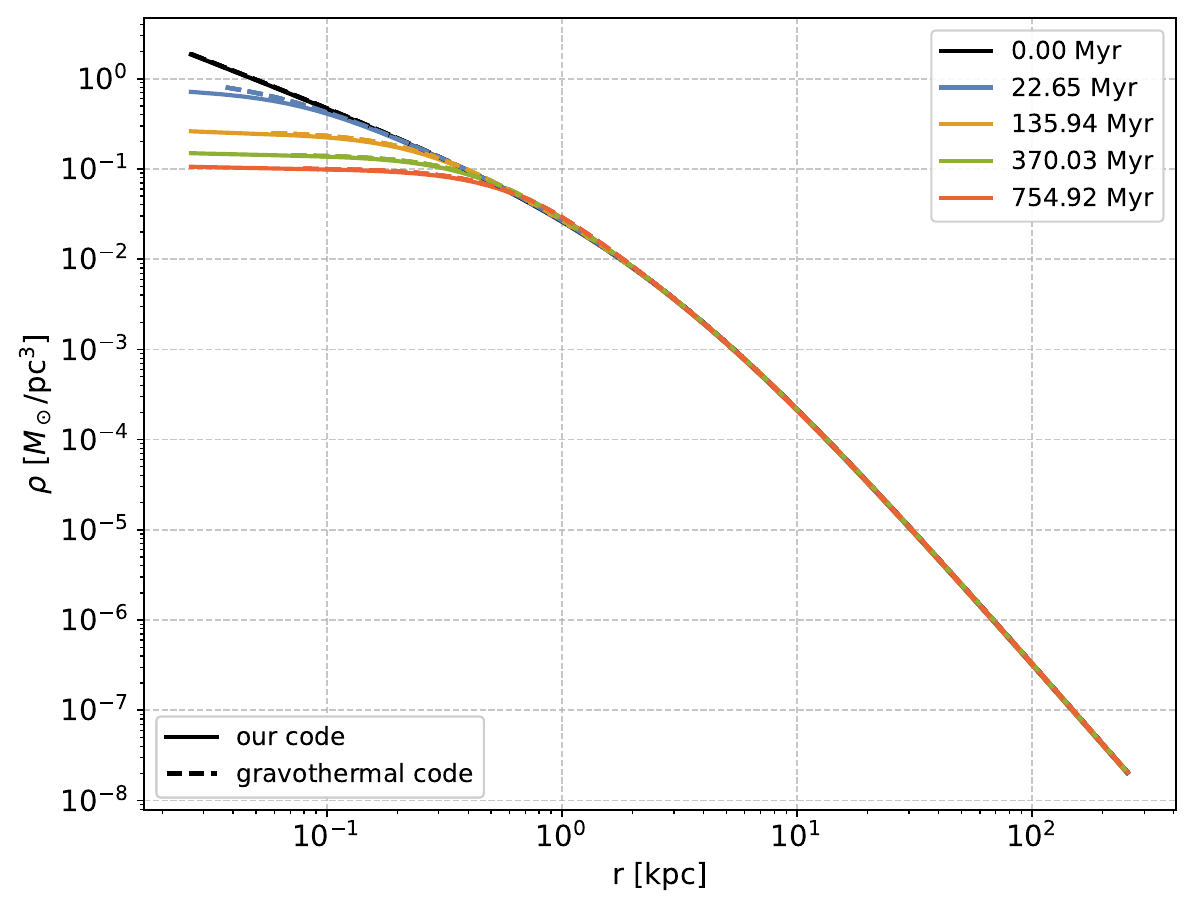}
    \end{subfigure}
    \begin{subfigure}[b]{0.4\textwidth}
        \includegraphics[width=1\textwidth]{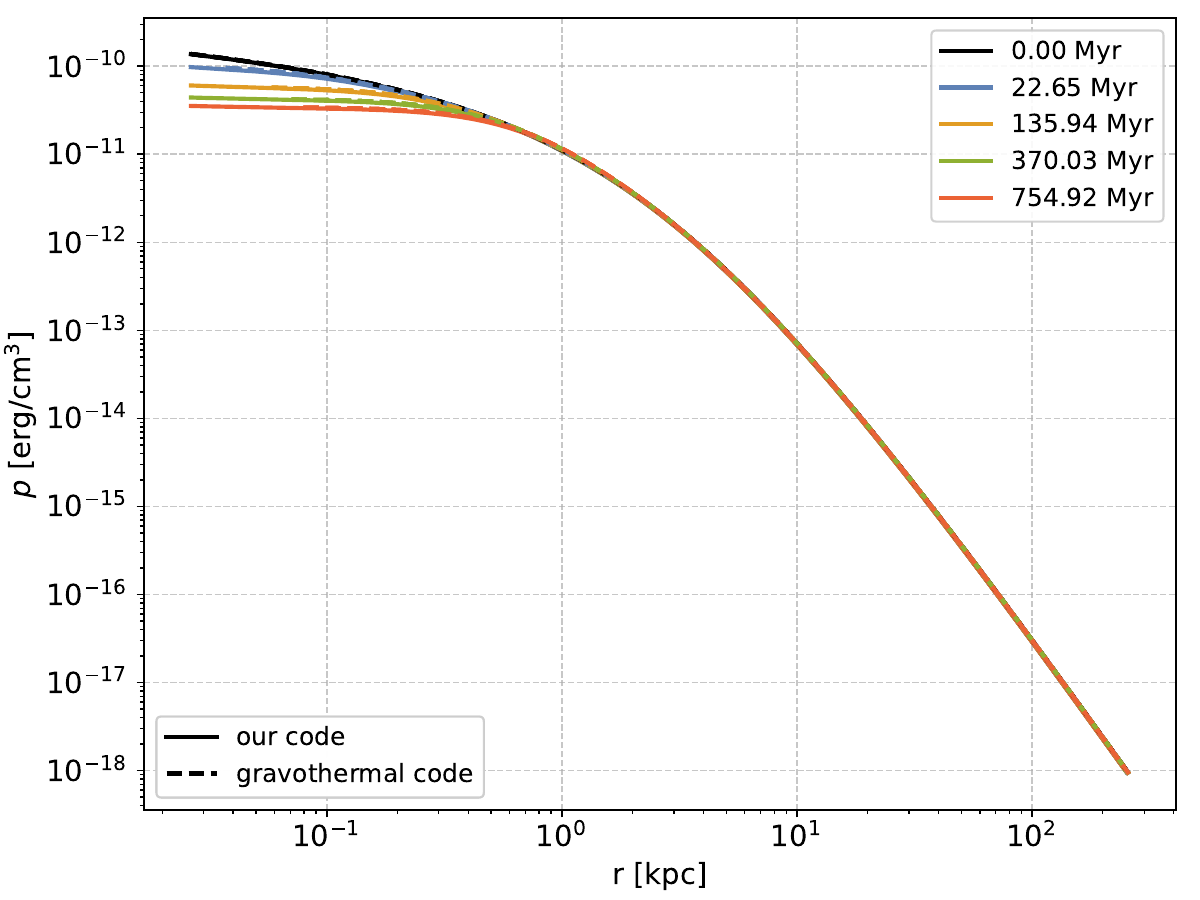}
    \end{subfigure}
    \caption{Comparison of our numerical results with those obtained from the gravothermal fluid code of Nishikawa et al.~\cite{Nishikawa:2019lsc}. The top row corresponds to an NFW halo with $\rho_s = 0.8\,\mathrm{M_{\odot}}/\mathrm{pc}^3$, $r_s = 10\,\mathrm{kpc}$ and $\sigma_m = 5\,\mathrm{cm}^2/\mathrm{g}$. The bottom row corresponds to an NFW halo with $\rho_s = 0.0194\,\mathrm{M_{\odot}}/\mathrm{pc}^3$, $r_s = 2.586\,\mathrm{kpc}$ and $\sigma_m = 5\,\mathrm{cm}^2/\mathrm{g}$. The left and right panels show the density and pressure evolution, respectively. Solid curves represent our results, whereas dashed curves show the results of Nishikawa et al.~\cite{Nishikawa:2019lsc}. Different colors indicate different time snapshots. The vertical gray dashed lines indicate the location of the $Kn=1$ transition, separating the SMFP and LMFP regimes on the left and right, respectively. }
    \label{fig:validation}
\end{figure*}

\subsection{Finite volume method}
To employ the finite volume method, we first rewrite the dimensionless equations in the spherically symmetric conservation form
\begin{equation}\label{eq:conservation_equation}
    \partial_{\til t}\mathbf{U} + \frac{1}{\til r ^2} \partial_{\til r} [\til r^2 \mathbf{F}(\mathbf{U})] = \mathbf{S}(\mathbf{U}),
\end{equation}
where the state vector $\mathbf{U}$ is defined as
\begin{equation}
    \mathbf{U} =
    \begin{bmatrix}
         \til \rho \\
         \til \rho \til u \\
         \til \rho \til E
    \end{bmatrix}.
\end{equation}
The source vector $\mathbf{S}$ is given by
\begin{equation}
    \mathbf{S} = 
    \begin{bmatrix}
        0 \\
        - \til \rho \frac{\til M_{\text{H}} + \til M_{\text{BH}}}{\til r ^2} + \frac{2 \til p} {\til r} \\
        - \til \rho \frac{\til M_{\text{H}} + \til M_{\text{BH}}}{\til r^2} \til u
    \end{bmatrix},
\end{equation}
and the flux vector $\mathbf{F}$ is given by
\begin{equation}
    \mathbf{F} = \mathbf{F}_{\text{c}} + \mathbf{F}_{\text{diff}} = 
    \begin{bmatrix}
         \til \rho \til u \\
         \til \rho \til u^2 + \til p \\
         \til u (\til \rho \til E  + \til p )
    \end{bmatrix}
    +
    \begin{bmatrix}
        0\\
        0\\
        - \til \kappa \partial_{\til r} \til v^2
    \end{bmatrix},
\end{equation}
where $\til E = \frac32 \til v^2 + \til u^2/2$ is the total specific energy. Here, $\mathbf{F}_{\text{c}}$ and $\mathbf{F}_{\text{diff}}$ denote the convective and diffusive fluxes, respectively. The convective part is hyperbolic in nature and describes the propagation of physical quantities at finite characteristic speeds. By contrast, the diffusive part is parabolic and represents thermal conduction. From a numerical perspective, this parabolic term introduces significant stiffness by imposing a stringent stability constraint on explicit time integration, for which the allowable time step scales quadratically with the spatial resolution.

\begin{figure*}[t]
    \centering
    \begin{subfigure}[b]{0.4\textwidth}
        \includegraphics[width=\textwidth]{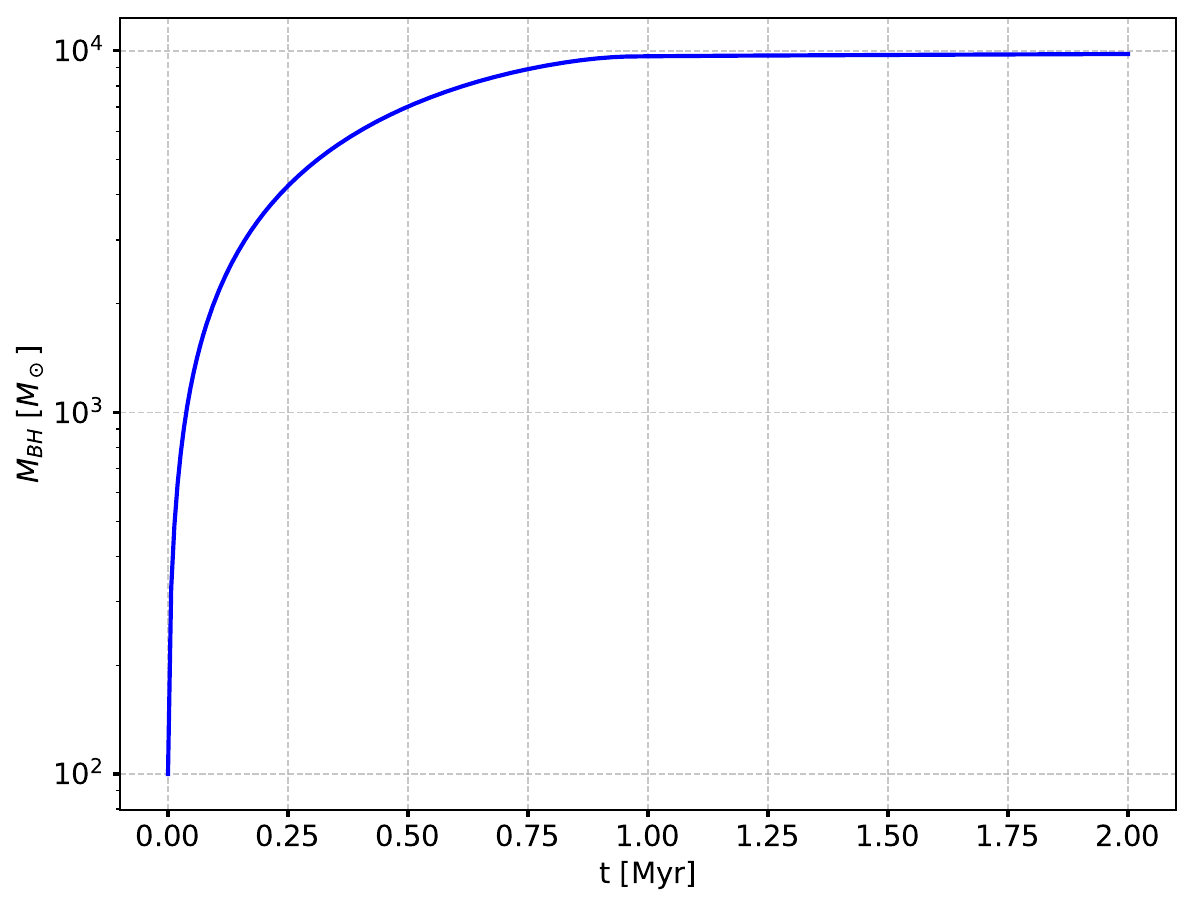}
    \end{subfigure}%
    \begin{subfigure}[b]{0.4\textwidth}
        \includegraphics[width=\textwidth]{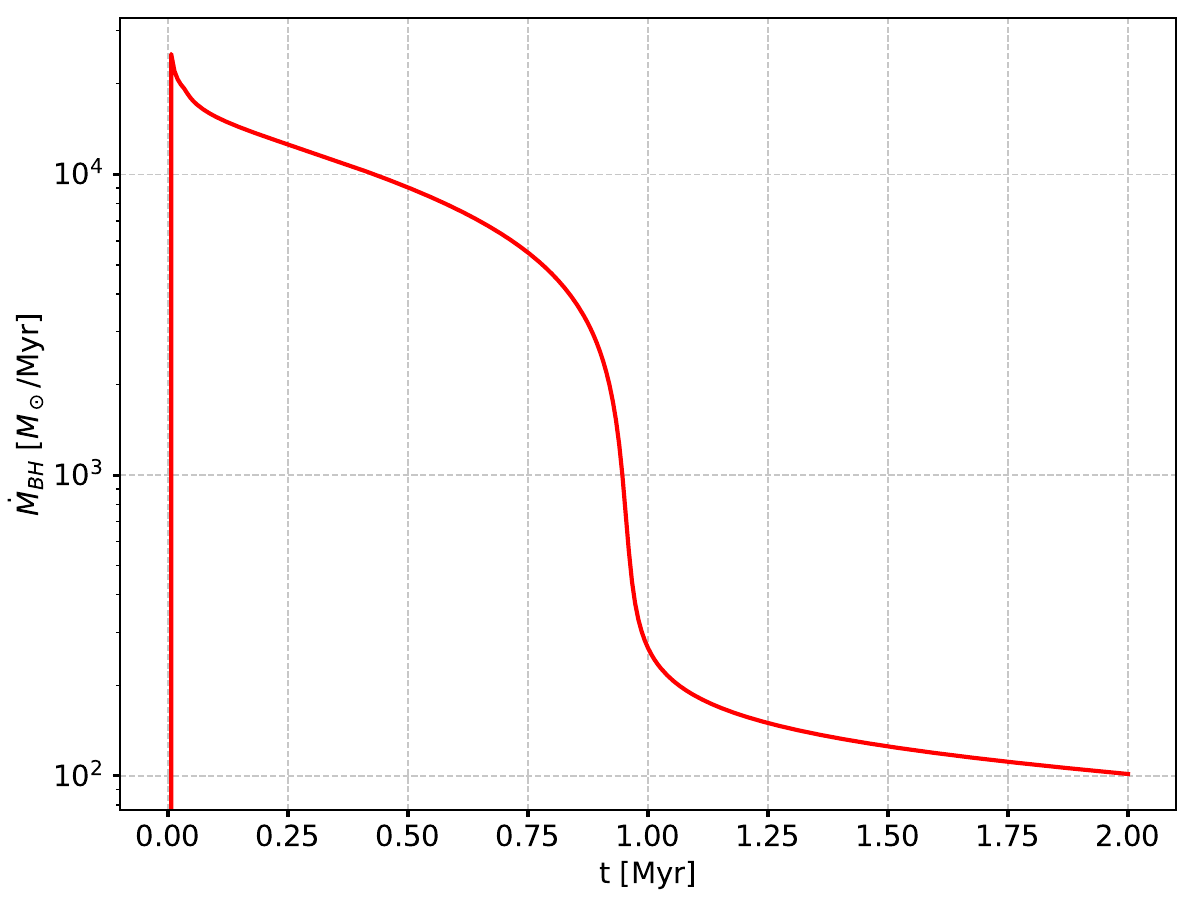}
    \end{subfigure}
    \begin{subfigure}[b]{0.33333\textwidth}
        \includegraphics[width=\textwidth]{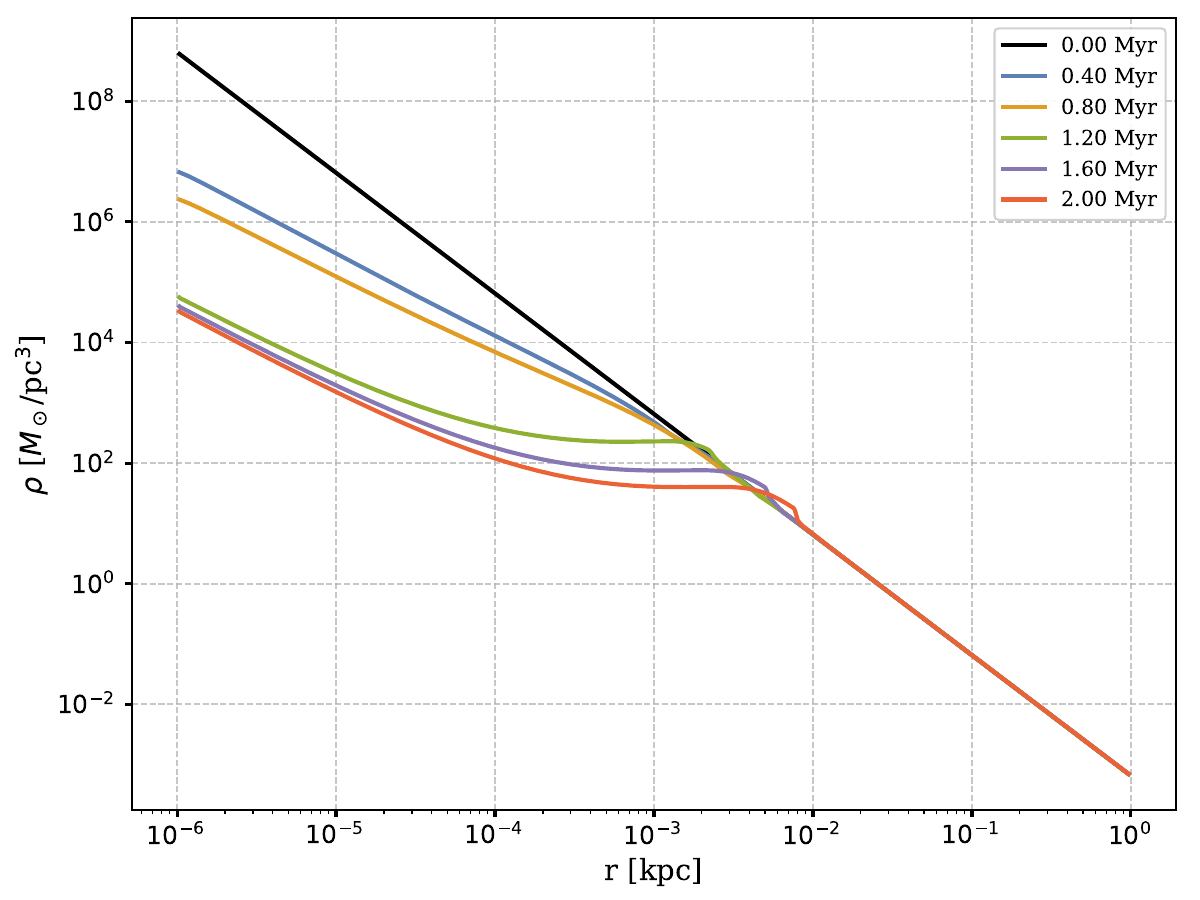}
    \end{subfigure}%
    \begin{subfigure}[b]{0.33333\textwidth}
        \includegraphics[width=\textwidth]{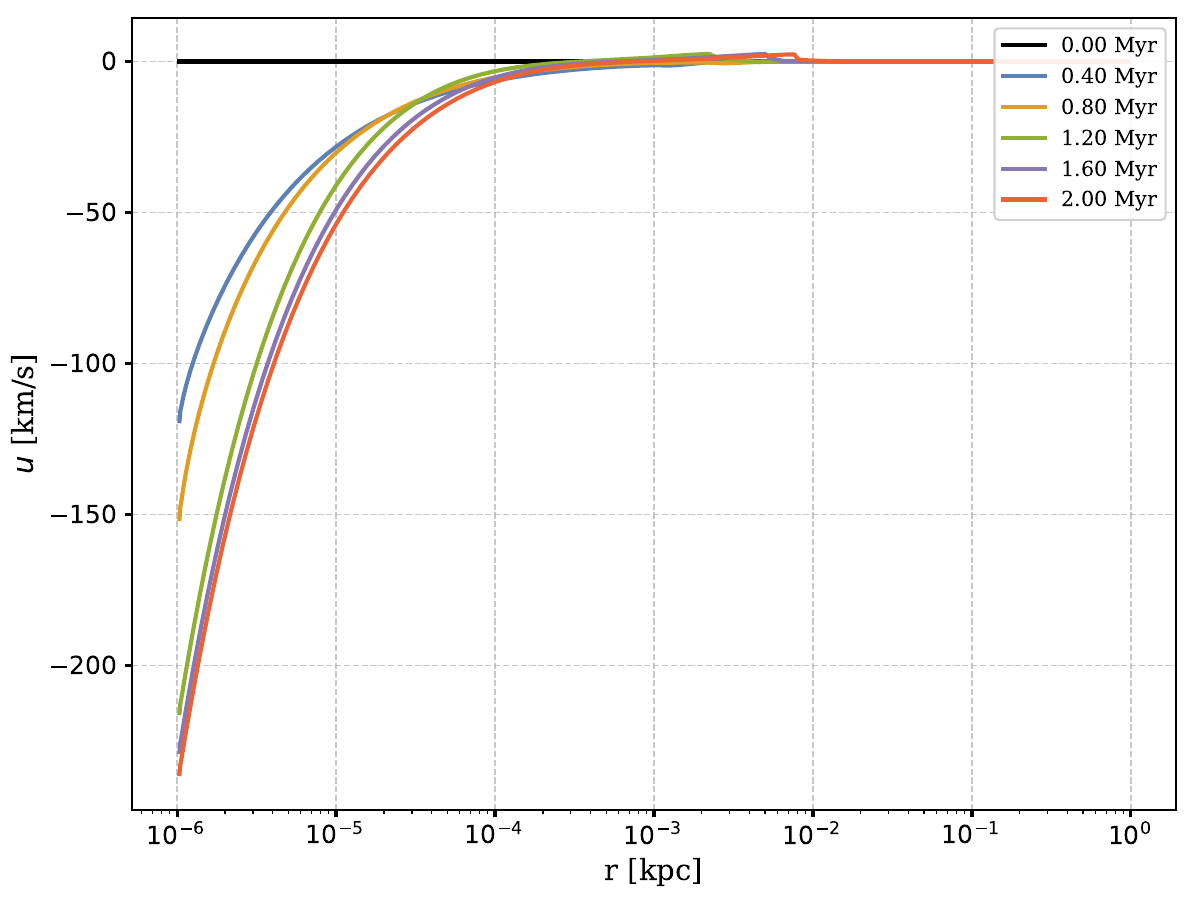}
    \end{subfigure}%
    \begin{subfigure}[b]{0.33333\textwidth}
        \includegraphics[width=\textwidth]{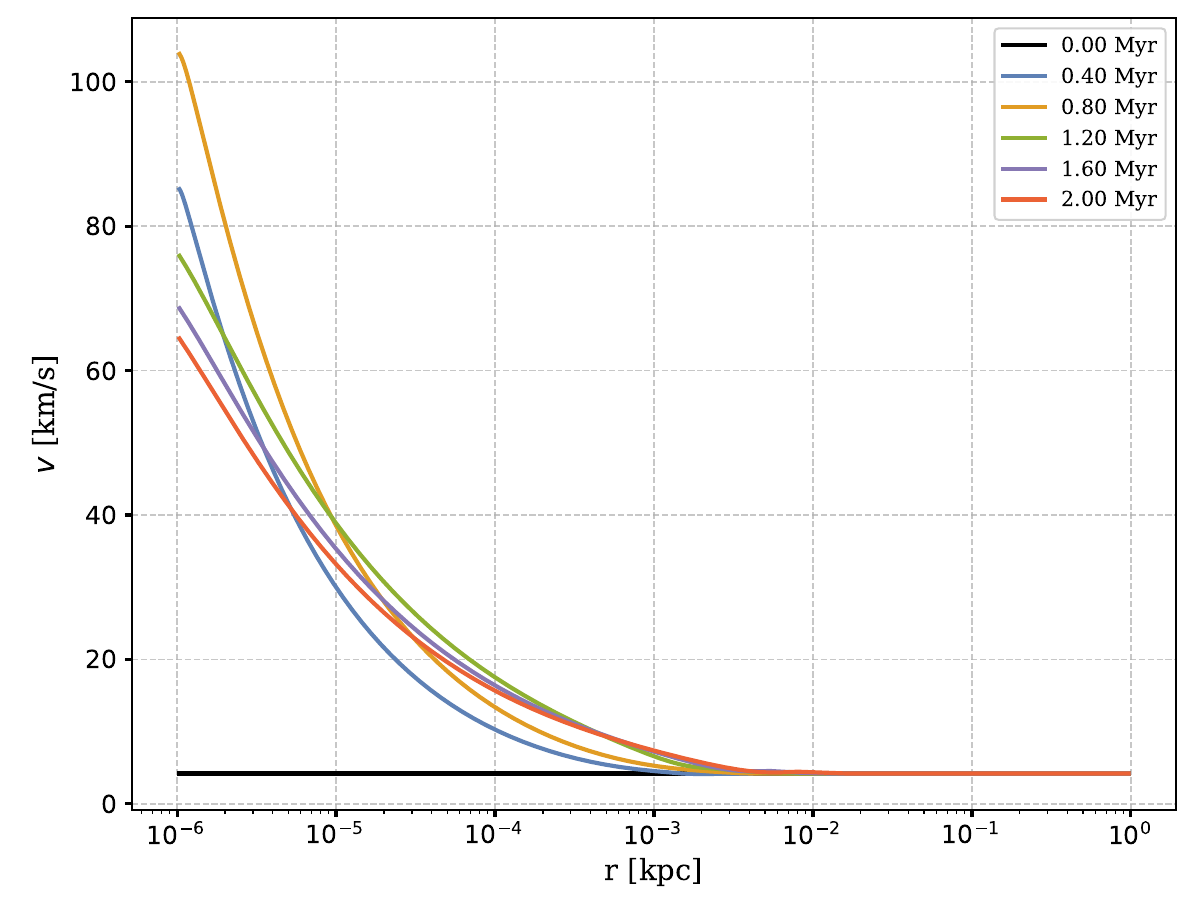}
    \end{subfigure}
    \caption{Simulation results for an SIS halo ($c_s = 4.2 \, \mathrm{km/s}$) with a total mass of $10^6\,\mathrm{M_{\odot}}$ at redshift $z = 25$. The top row shows the evolution of the black hole mass (left) and accretion rate (right). The bottom row shows the radial profiles of the density $\rho$, radial velocity $u$ and velocity dispersion $v$. Different colors indicate different time snapshots.}
    \label{fig:SIS_results}
\end{figure*}

The spatial domain $[\til r_{\text{min}},\til r_{\text{max}}]$ is partitioned into $N$ computational cells. The state vector $\mathbf{U}^n_i$ and the source term $\mathbf{S}^n_i$ represent cell-averaged quantities in the $i$-th cell at the $n$-th time step, while the numerical fluxes $\mathbf{F}^n_{i+1/2}$ are defined at the interfaces between adjacent cells. Here, $i$ and $n$ denote the spatial and temporal indices, respectively.
Consistent with the spherically symmetric conservation law in Eq.~\eqref{eq:conservation_equation}, the dimensionless cell volume and interface areas are given by $\til V_i = (\til r_{i+1/2}^3 - \til r_{i-1/2}^3)/3$ and $\til A_{i+1/2} = \til r_{i+1/2}^2$, respectively.
To advance the solution in time, we employ a first-order operator-splitting method~\cite{leveque2002finite}. 
For each dimensionless time step $\Delta \til t$ determined by the CFL condition for the hyperbolic part, the evolution proceeds in two stages.
In the first stage, we consider only the hyperbolic part of the system and neglect the diffusive flux $\mathbf{F}_\text{diff}$. The solution is advanced to an intermediate state $\mathbf{U}^{*}_i$ using a first-order explicit Euler scheme
\begin{equation}
    \frac{\mathbf{U}^{*}_i - \mathbf{U}^{n}_i}{\Delta \til t} = 
    \mathbf{S}^n_i-  \frac{1}{\til V_i}\left( \til A_{i+\frac12} \mathcal{F}_{\text{c}}(\mathbf{U}_{i+\frac12}^n) - \til A_{i-\frac12} \mathcal{F}_{\text{c}}(\mathbf{U}_{i-\frac12}^n) \right),
\end{equation}
To evaluate the numerical fluxes $\mathcal{F}_{\text{c}}(\mathbf{U}_{i+1/2}^n)$, the cell-averaged variables are first reconstructed at the interfaces using the Monotonized Central (MC) slope limiter~\cite{van1977towards,leveque2002finite}. The interfacial fluxes are then computed with an approximate Roe Riemann solver~\cite{roe1981approximate,toro2013riemann}. Detailed expressions for the hyperbolic solver are given in Appendix~\ref{app:hyperbolic}.
In the second stage, to circumvent the severe time-step constraints imposed by the diffusive terms, we employ a backward Euler scheme to advance the intermediate state $\mathbf{U}^{*}_i$ to the final state $\mathbf{U}^{n+1}_i$. 
Since the diffusive fluxes account solely for thermal conduction, the density $\til \rho$ and radial velocity $\til u$ remain fixed at their intermediate values during this sub-step. 
Consequently, only the velocity dispersion squared $\til v^2$ needs to be updated:
\begin{equation}
    \frac{(\til v^2)^{n+1}_i - (\til v^2)^{*}_i}{\Delta \til t} = \mathcal{D} \left[ (\til v^2)^{n+1}_i \right],
\end{equation}
where $\mathcal{D}$ represents the discretized thermal conduction operator normalized by the intermediate density $\til \rho^{*}$.
This treatment reduces the implicit update to a scalar tridiagonal system, which can be solved efficiently via the Thomas algorithm. A detailed derivation of the implicit scheme is provided in Appendix \ref{app:implicit}.

To initialize the simulation, we prescribe the initial density profile of the dark matter halo $\til \rho(\til r,0)$ together with the initial mass of the central black hole $\til M_{\rm BH}(0)$. 
We assume that the initial radial velocity satisfies $\til u(\til r,0) = 0$. 
The initial pressure profile $\til p(\til r,0)$  is determined from the hydrostatic equilibrium, such that the pressure gradient balances the gravitational force of the dark matter halo. At both inner and outer boundaries, we impose zero-gradient boundary conditions to minimize spurious numerical reflections and prevent artificial fluxes from entering the domain.

\begin{figure*}[t]
    \centering
    \begin{subfigure}[b]{0.4\textwidth}
        \includegraphics[width=\textwidth]{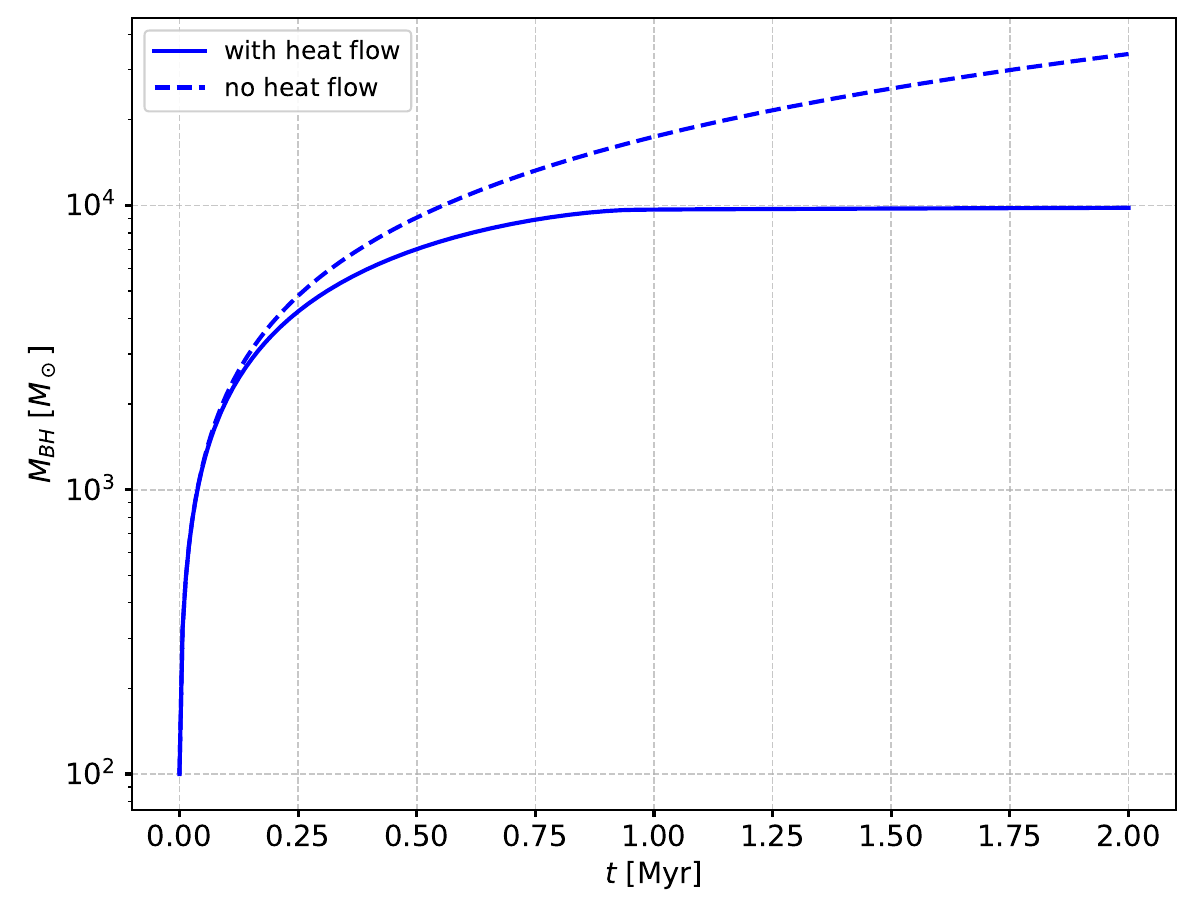}
    \end{subfigure}%
    \begin{subfigure}[b]{0.4\textwidth}
        \includegraphics[width=\textwidth]{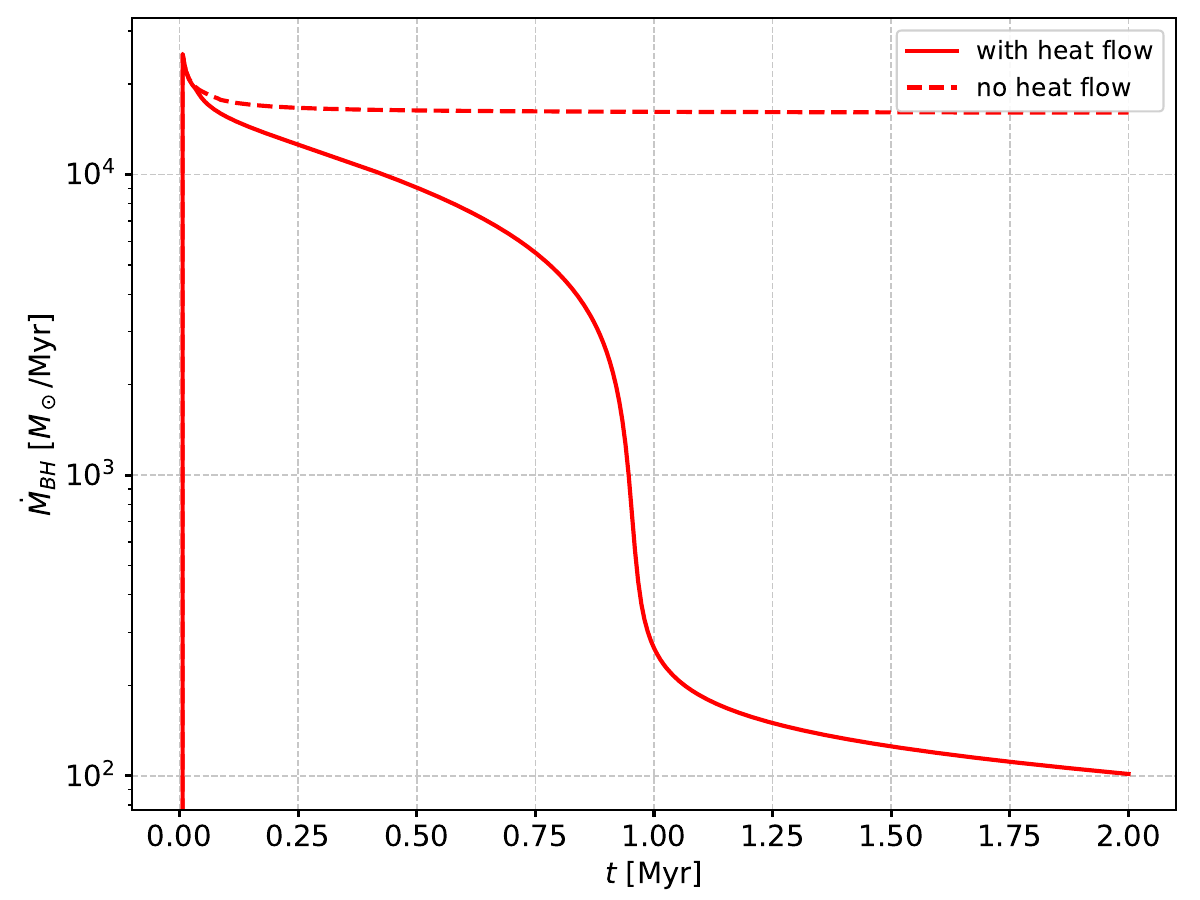}
    \end{subfigure}
    \begin{subfigure}[b]{0.33333\textwidth}
        \includegraphics[width=\textwidth]{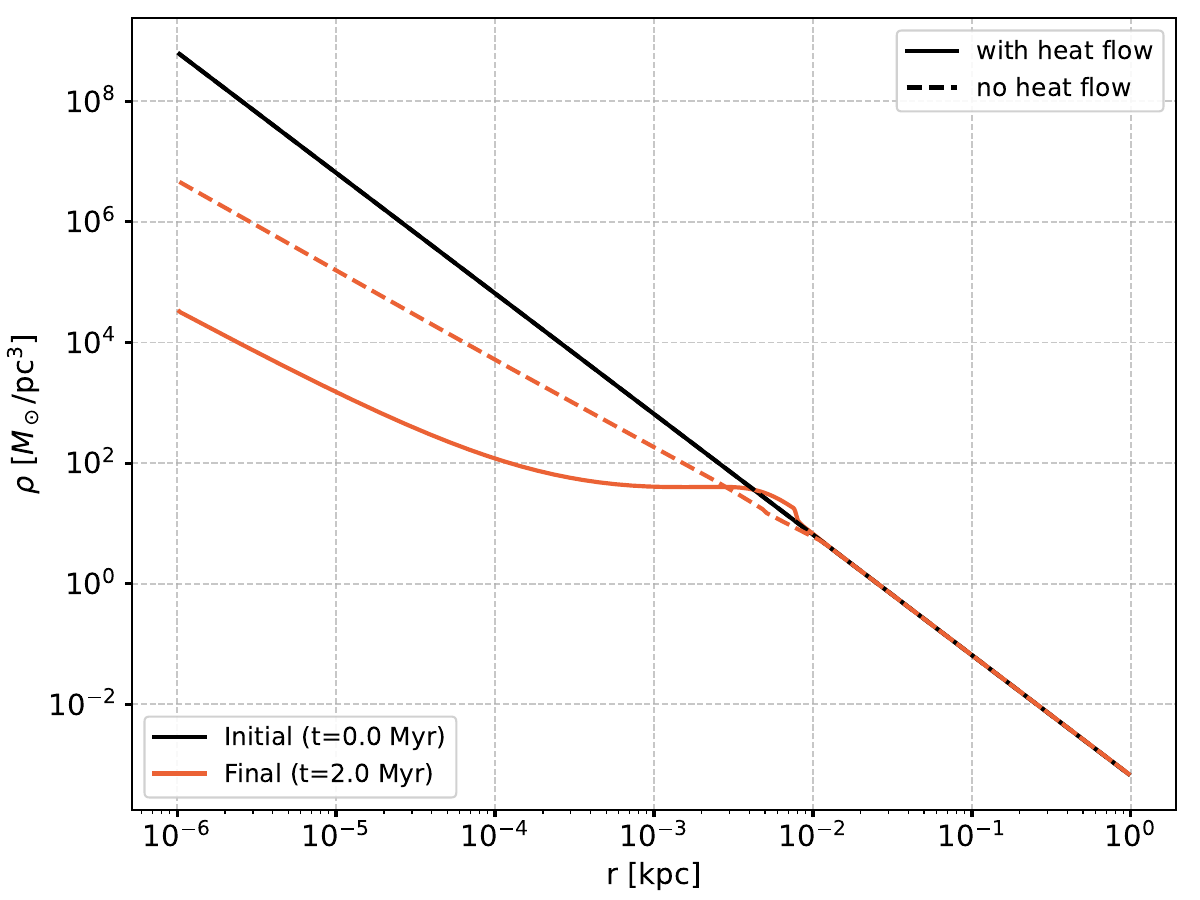}
    \end{subfigure}%
    \begin{subfigure}[b]{0.33333\textwidth}
        \includegraphics[width=\textwidth]{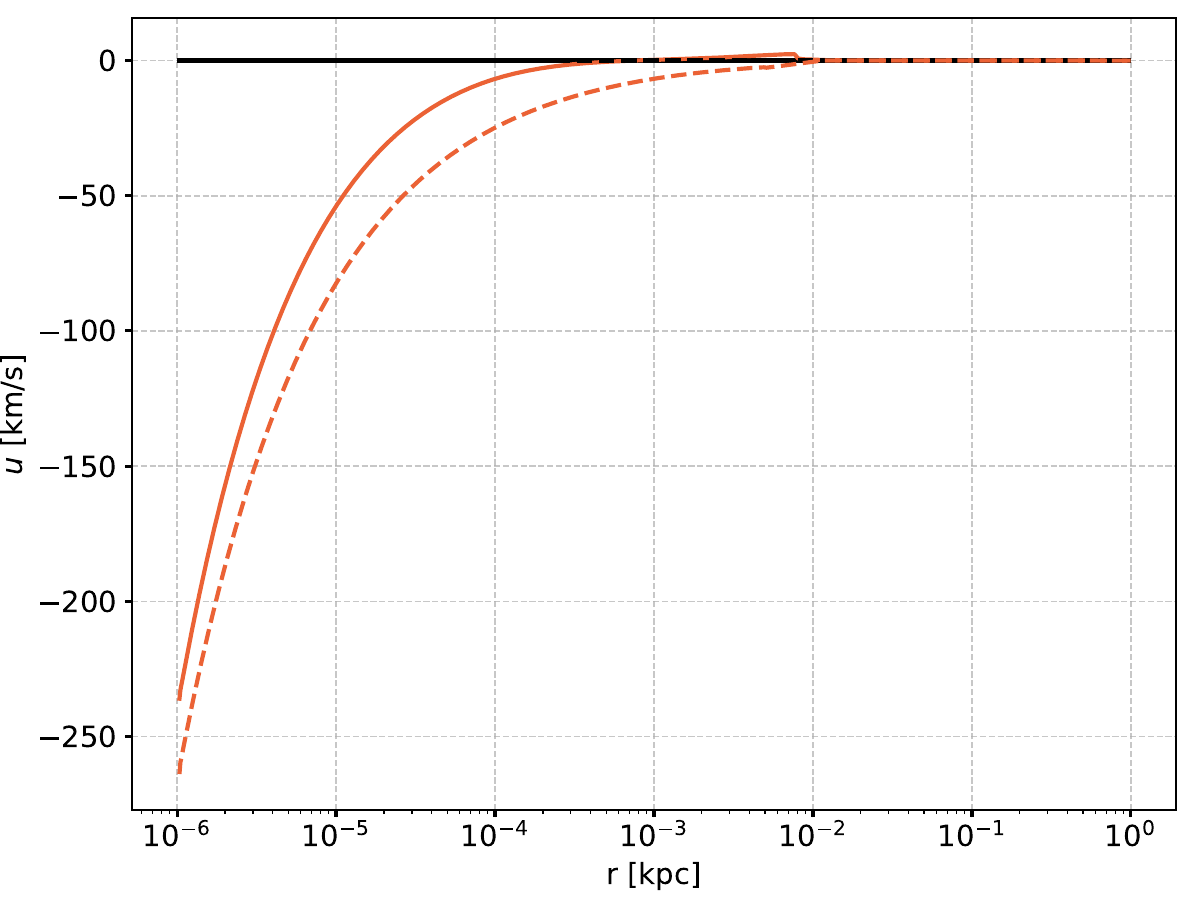}
    \end{subfigure}%
    \begin{subfigure}[b]{0.33333\textwidth}
        \includegraphics[width=\textwidth]{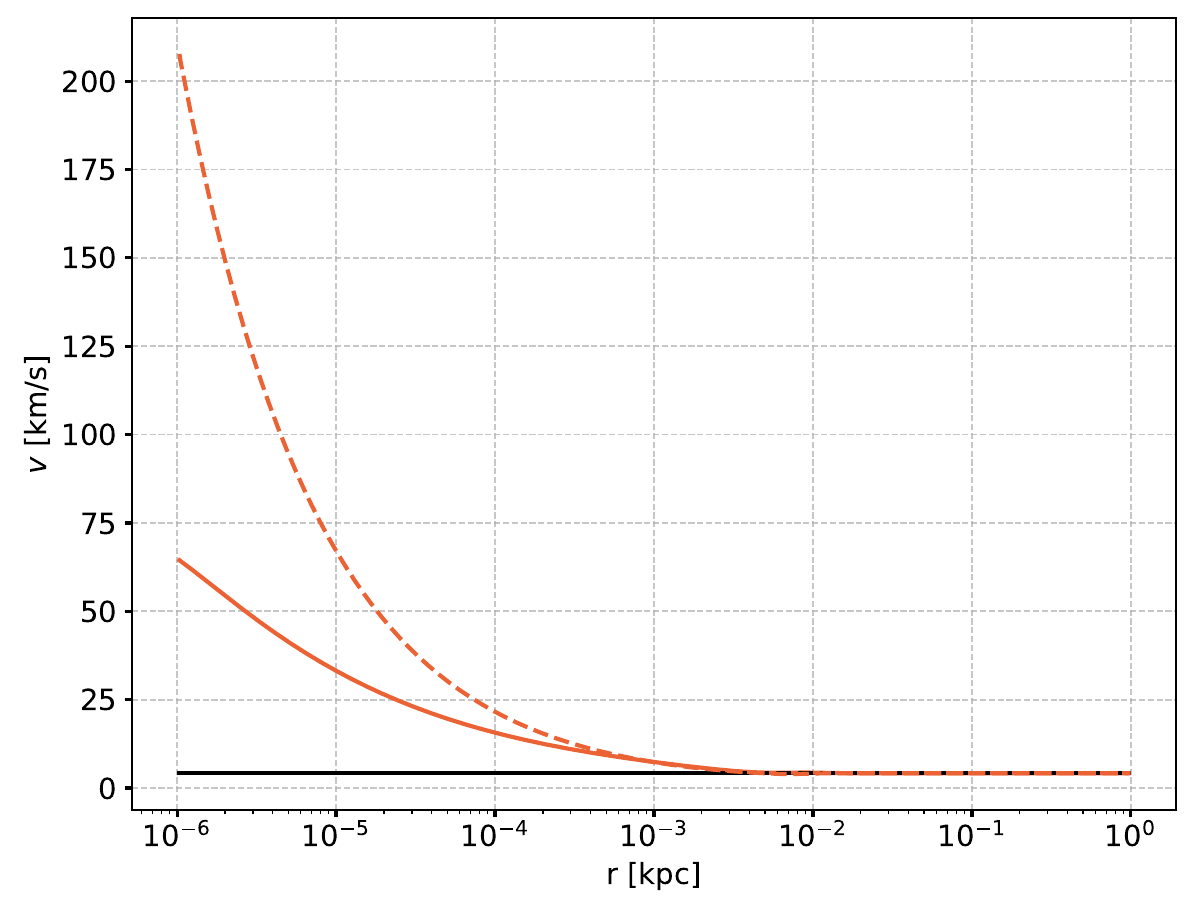}
    \end{subfigure}
    \caption{Comparison of the SIS halo evolution with and without heat flow. Parameters are identical to those in Figure~\ref{fig:SIS_results}. Solid and dashed lines distinguish cases with and without heat flow. In the bottom row, black and red lines represent the initial and final time snapshots, respectively.}
    \label{fig:compare_SIS}
\end{figure*}

\subsection{Validation without black holes}
To validate our numerical implementation, we temporarily switch off the gravitational influence of the central black hole and compare the SIDM core-formation results produced by our code with those obtained from the gravothermal fluid code of Nishikawa et al.~\cite{Nishikawa:2019lsc,Outmezguine:2022bhq,Gad-Nasr:2023gvf}. 
We perform two benchmark tests that probe different transport regimes, characterized by the Knudsen number $Kn$.

In the first test, we initialize an NFW dark matter halo $\rho_{\text{NFW}}(r) = \rho_s/[(r/r_s)(1+(r/r_s))^{2}]$ with $\rho_s = 0.8\,\mathrm{M_{\odot}}/\mathrm{pc}^3$ and $r_s = 10\,\mathrm{kpc}$, and adopt a self-interaction cross section of $\sigma_m = 5\,\mathrm{cm}^2/\mathrm{g}$. 
These parameters are chosen as a controlled validation setup in which the inner halo lies mainly in the SMFP regime, $\mathrm{Kn}<1$.
In this regime, self-interactions are frequent, and SIDM behaves as a collisional fluid.
In the second test, we adopt the default parameters of the gravothermal
fluid code~\cite{Nishikawa:2019lsc} (i.e., an NFW profile with $\rho_s = 0.0194\,\mathrm{M_{\odot}}/\mathrm{pc}^3$, $r_s = 2.586\,\mathrm{kpc}$, and self-interaction cross section $\sigma_m = 5\,\mathrm{cm}^2/\mathrm{g}$), which correspond to the LMFP regime, $Kn>1$.

Figure~\ref{fig:validation} displays the resulting density and pressure profiles for both cases, showing excellent agreement with the reference results. 
The top panels represent the first test, where a vertical gray dashed line at $Kn=1$ separates the SMFP regime (left) from the LMFP regime (right). The bottom panels show the second test, which resides entirely within the LMFP regime ($Kn>1$). 
We note that the benchmark code in Ref.~\cite{Nishikawa:2019lsc} is formulated in a Lagrangian framework with moving mass shells. 
As a result, the radial positions of the innermost data points drift outward over time, as shown in the plotted profiles. Notably, our results remain highly consistent with the benchmark solution even in the LMFP regime ($Kn > 1$). This consistency indicates that our numerical scheme reproduces the gravothermal evolution captured by the reference model in both the SMFP and LMFP limits. 
The reason our fluid-based code remains consistent in the LMFP regime is that, for the parameters adopted here, the SIDM halo undergoes quasi-static evolution and the characteristic flow speeds remain too small to produce substantial departures from local equilibrium. 

\begin{figure*}[t]
    \centering
    \begin{subfigure}[b]{0.4\textwidth}
        \includegraphics[width=\textwidth]{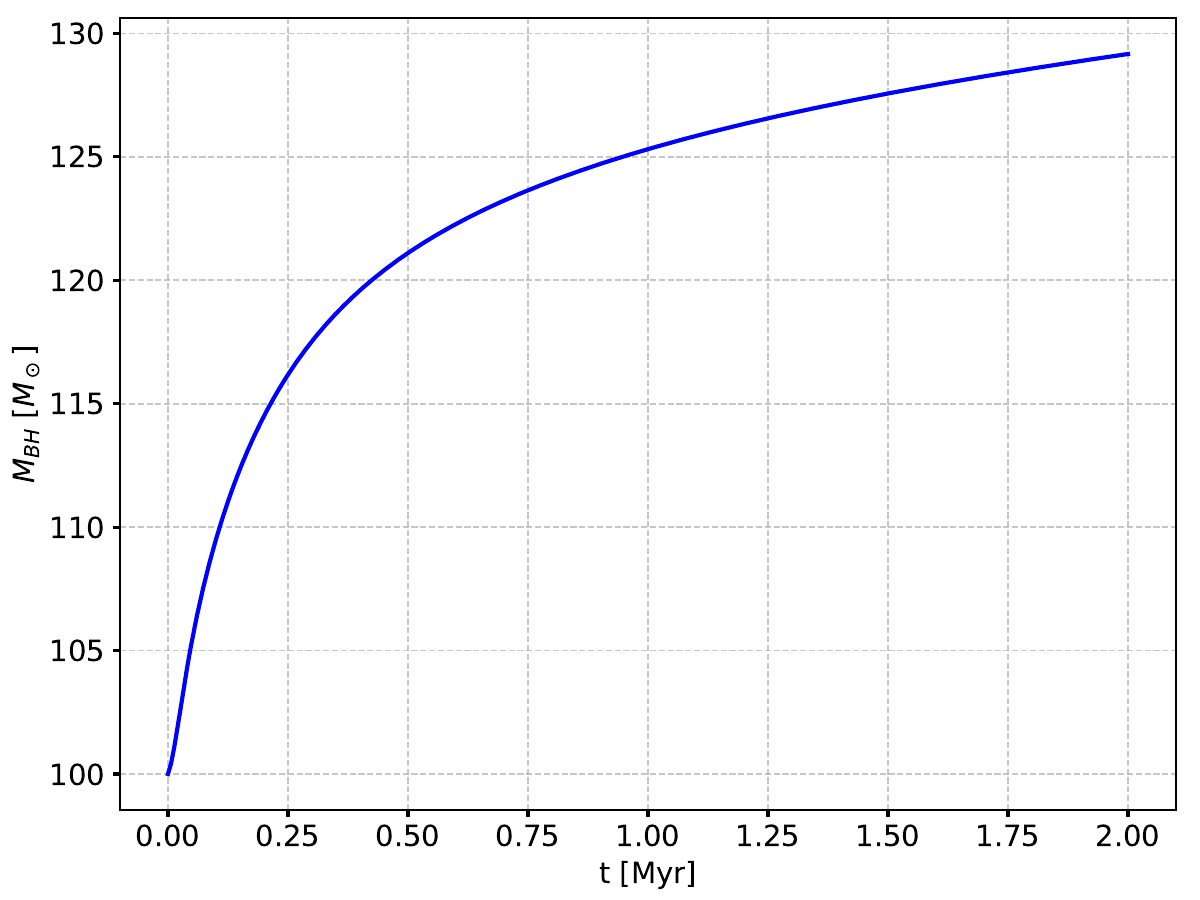}
    \end{subfigure}%
    \begin{subfigure}[b]{0.4\textwidth}
        \includegraphics[width=\textwidth]{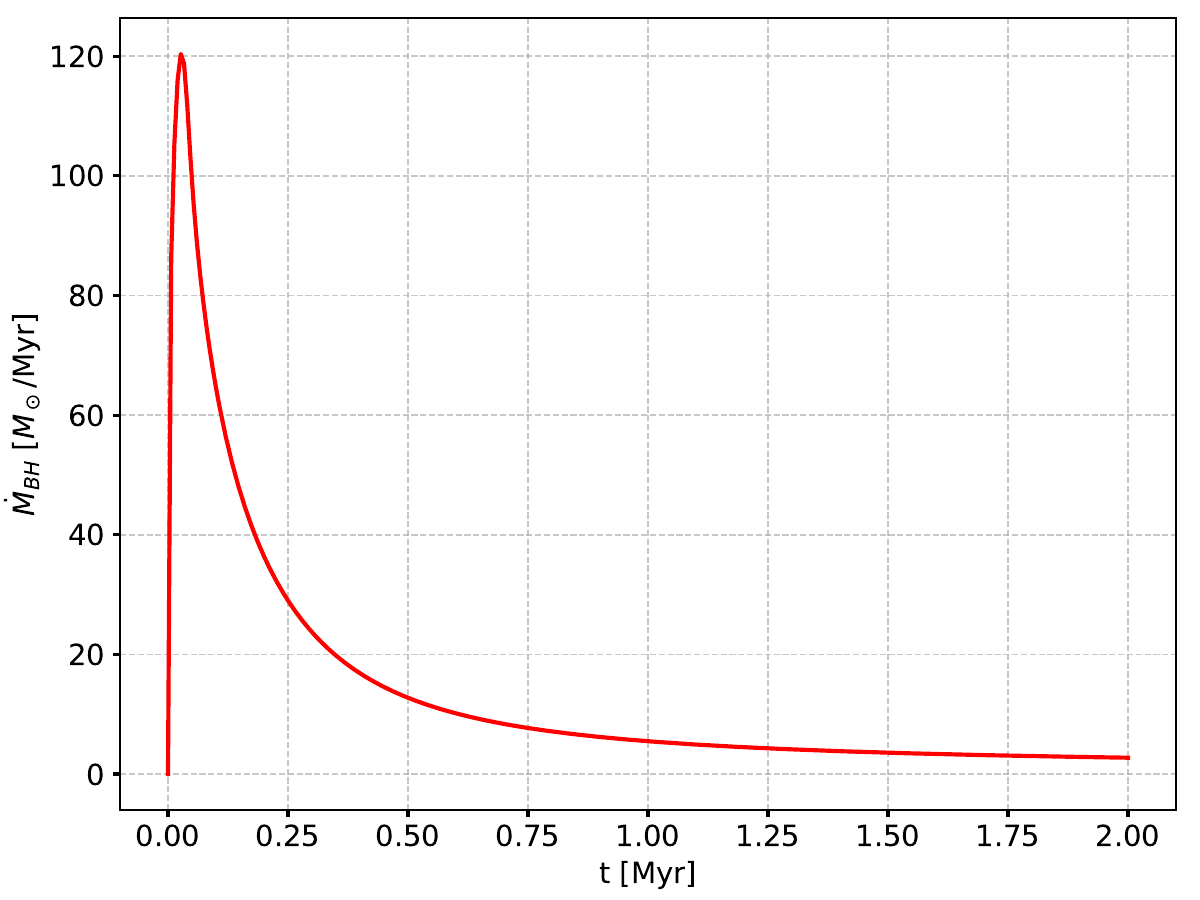}
    \end{subfigure}
    \begin{subfigure}[b]{0.33333\textwidth}
        \includegraphics[width=\textwidth]{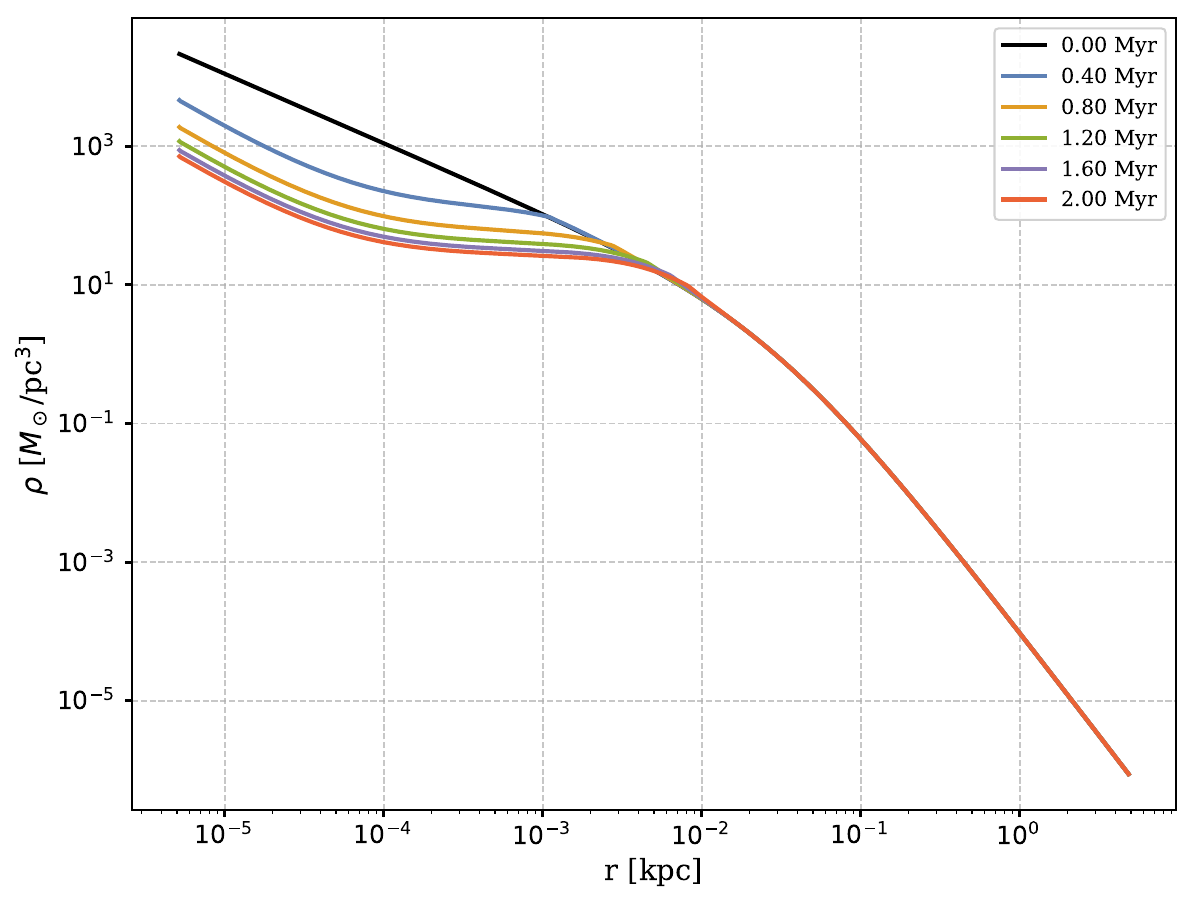}
    \end{subfigure}%
    \begin{subfigure}[b]{0.33333\textwidth}
        \includegraphics[width=\textwidth]{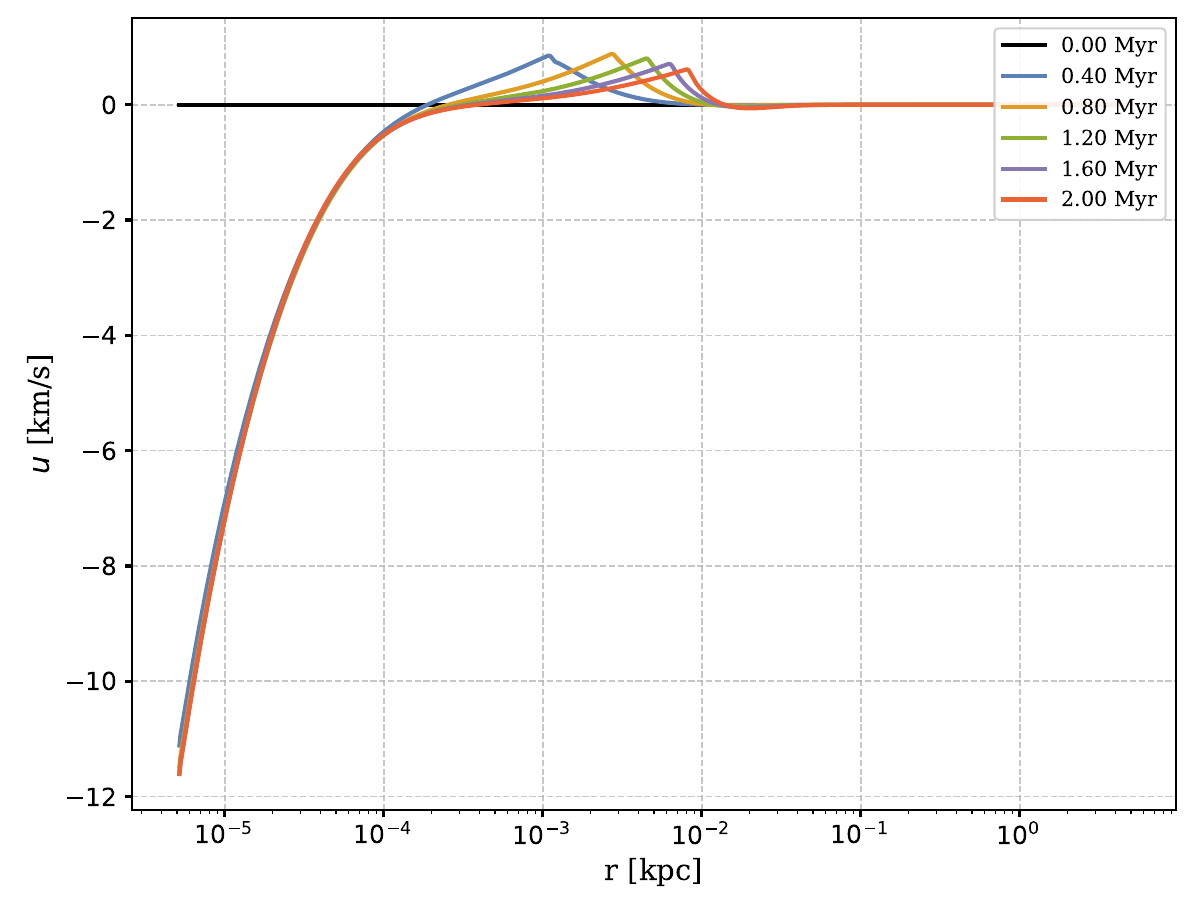}
    \end{subfigure}%
    \begin{subfigure}[b]{0.33333\textwidth}
        \includegraphics[width=\textwidth]{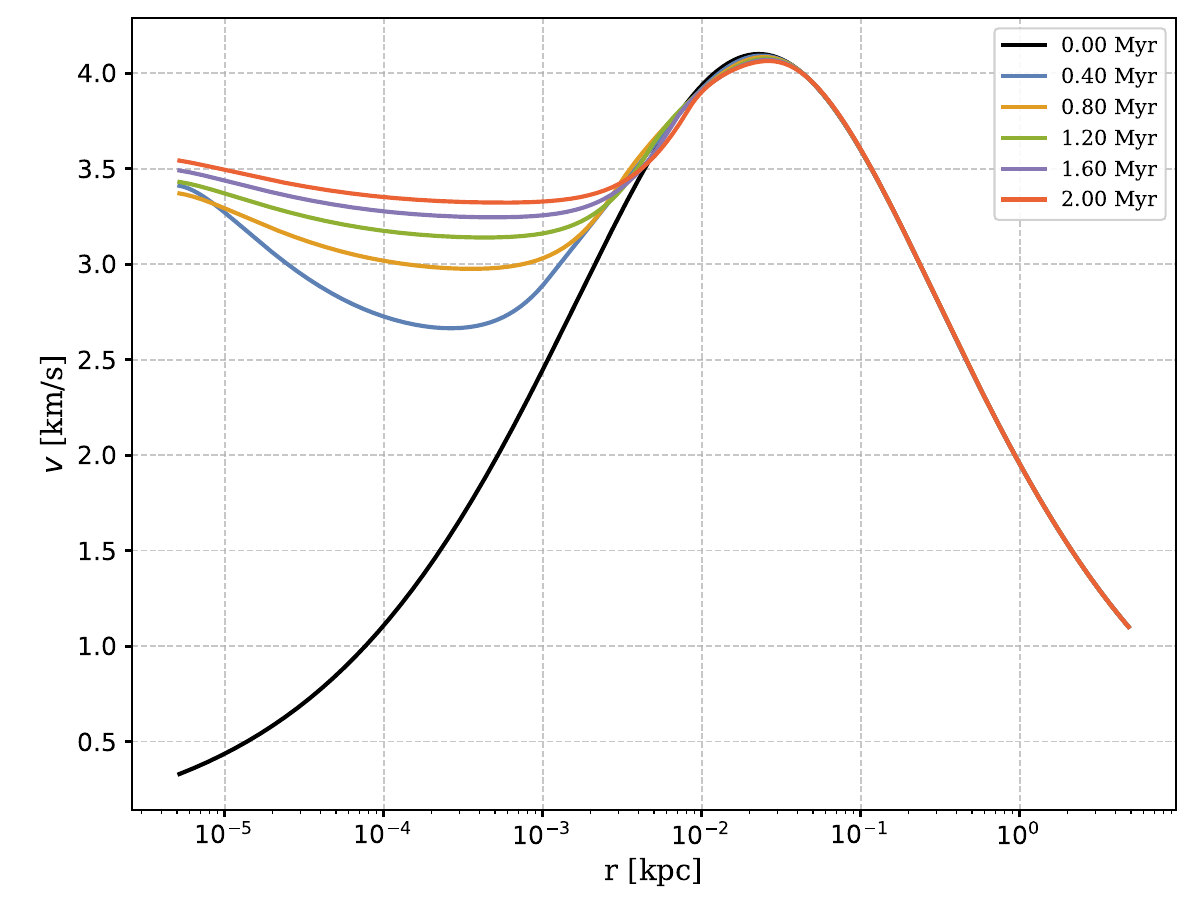}
    \end{subfigure}
    \caption{Simulation results for an NFW halo ($r_s = 0.03\,\mathrm{kpc}$, $\rho_s = 3.7\,\mathrm{M_{\odot}}/\mathrm{pc}^3$) with a total mass of $10^6\,\mathrm{M_{\odot}}$ at redshift $z = 25$. The top row shows the time evolution of the black hole mass (left) and accretion rate (right). The bottom row shows the radial profiles of the density $\rho$, radial velocity $u$ and velocity dispersion $v$. Different colors indicate different time snapshots.}
    \label{fig:NFW_results}
\end{figure*}

\begin{figure*}[htbp]
    \centering
    \begin{subfigure}[b]{0.4\textwidth}
        \includegraphics[width=\textwidth]{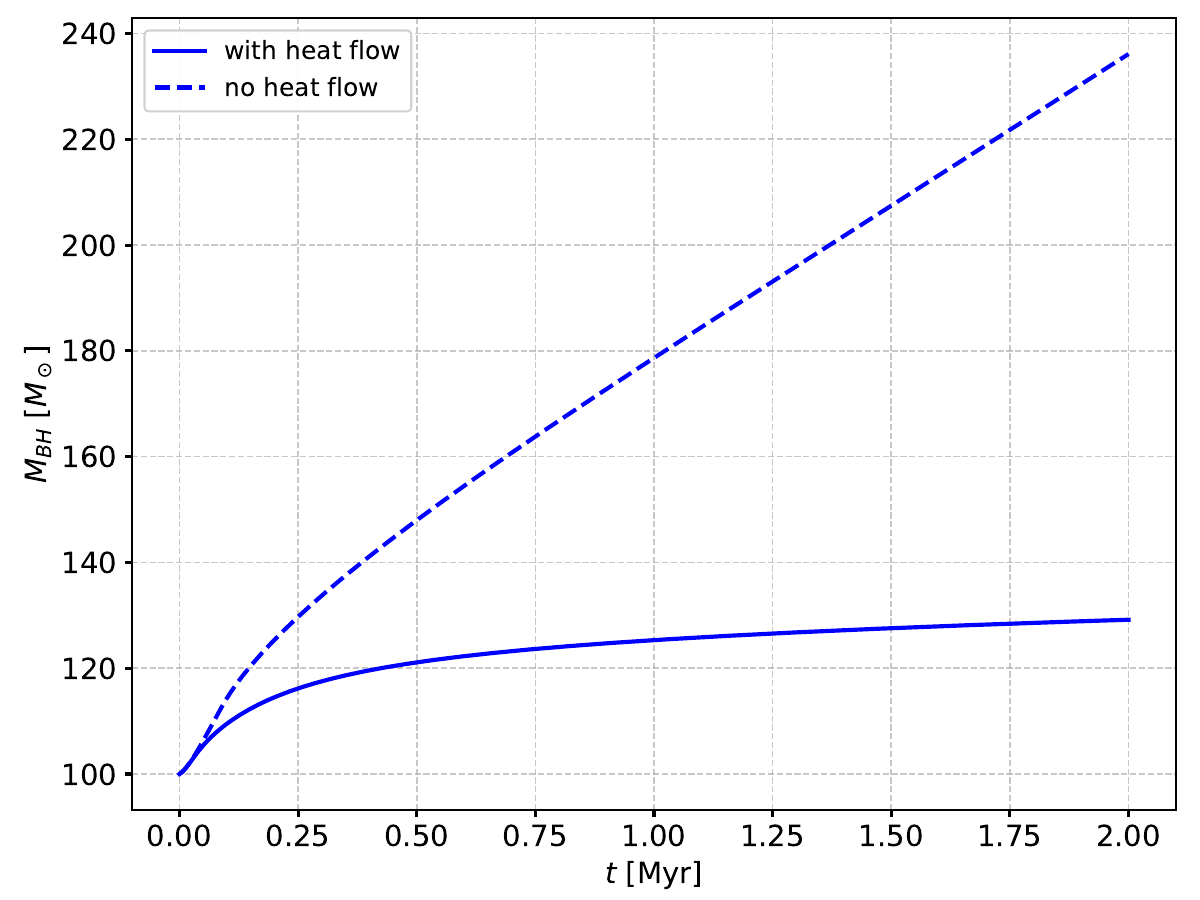}
    \end{subfigure}%
    \begin{subfigure}[b]{0.4\textwidth}
        \includegraphics[width=\textwidth]{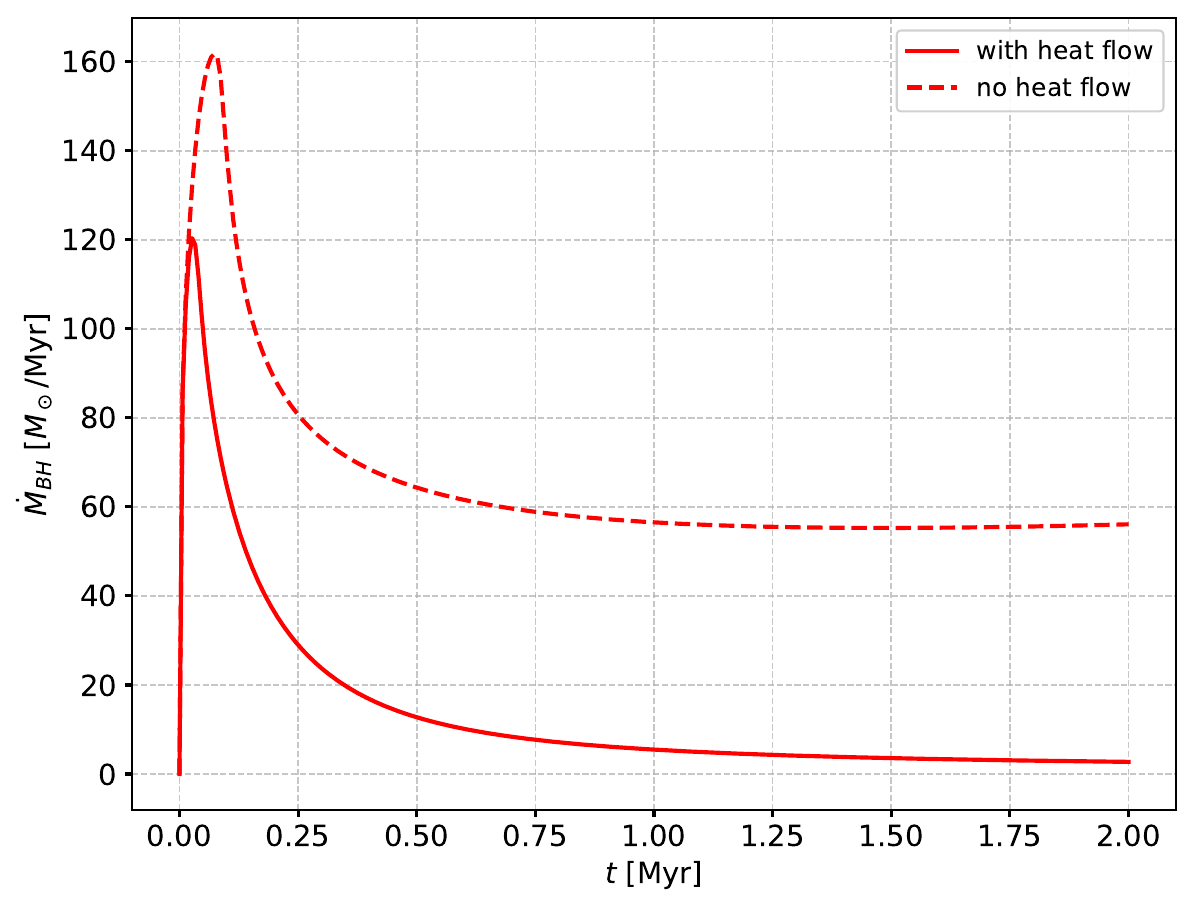}
    \end{subfigure}
    \begin{subfigure}[b]{0.33333\textwidth}
        \includegraphics[width=\textwidth]{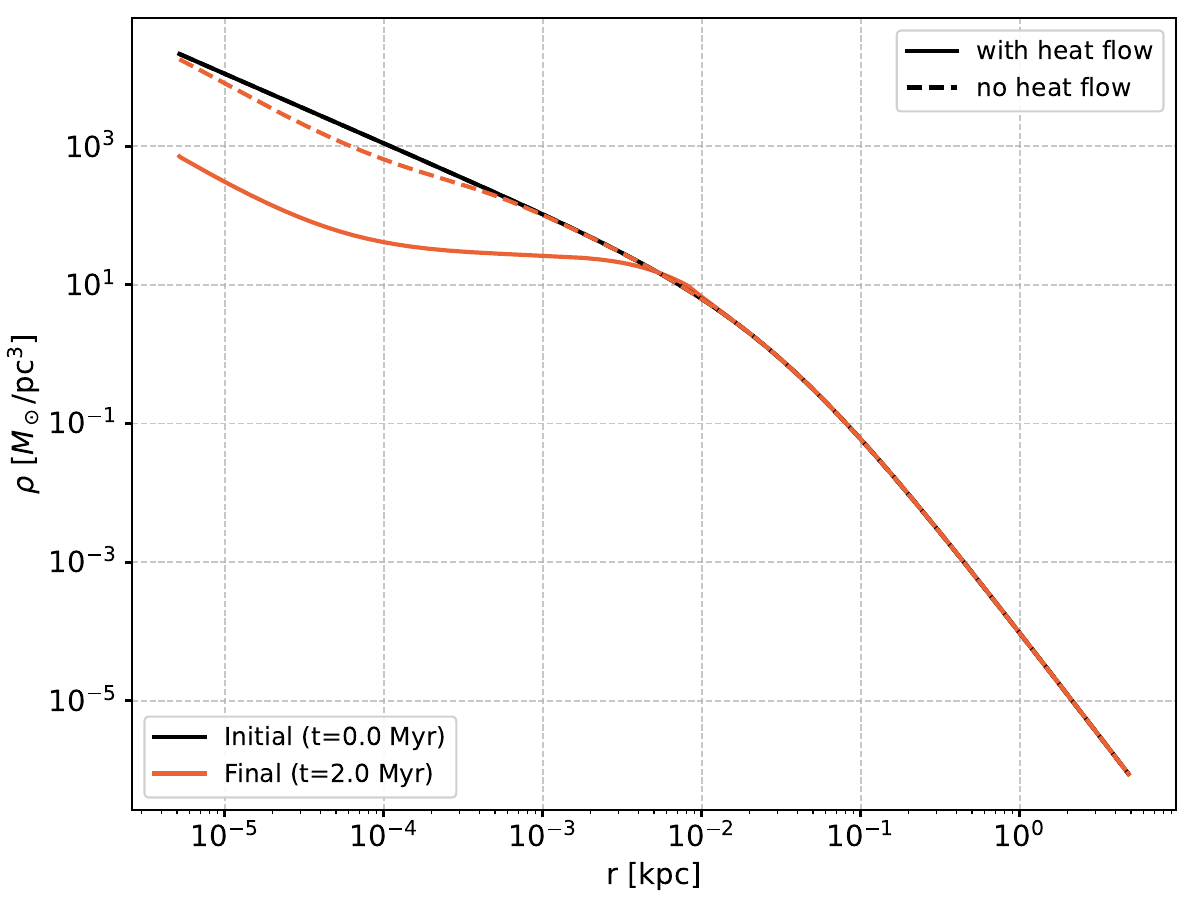}
    \end{subfigure}%
    \begin{subfigure}[b]{0.33333\textwidth}
        \includegraphics[width=\textwidth]{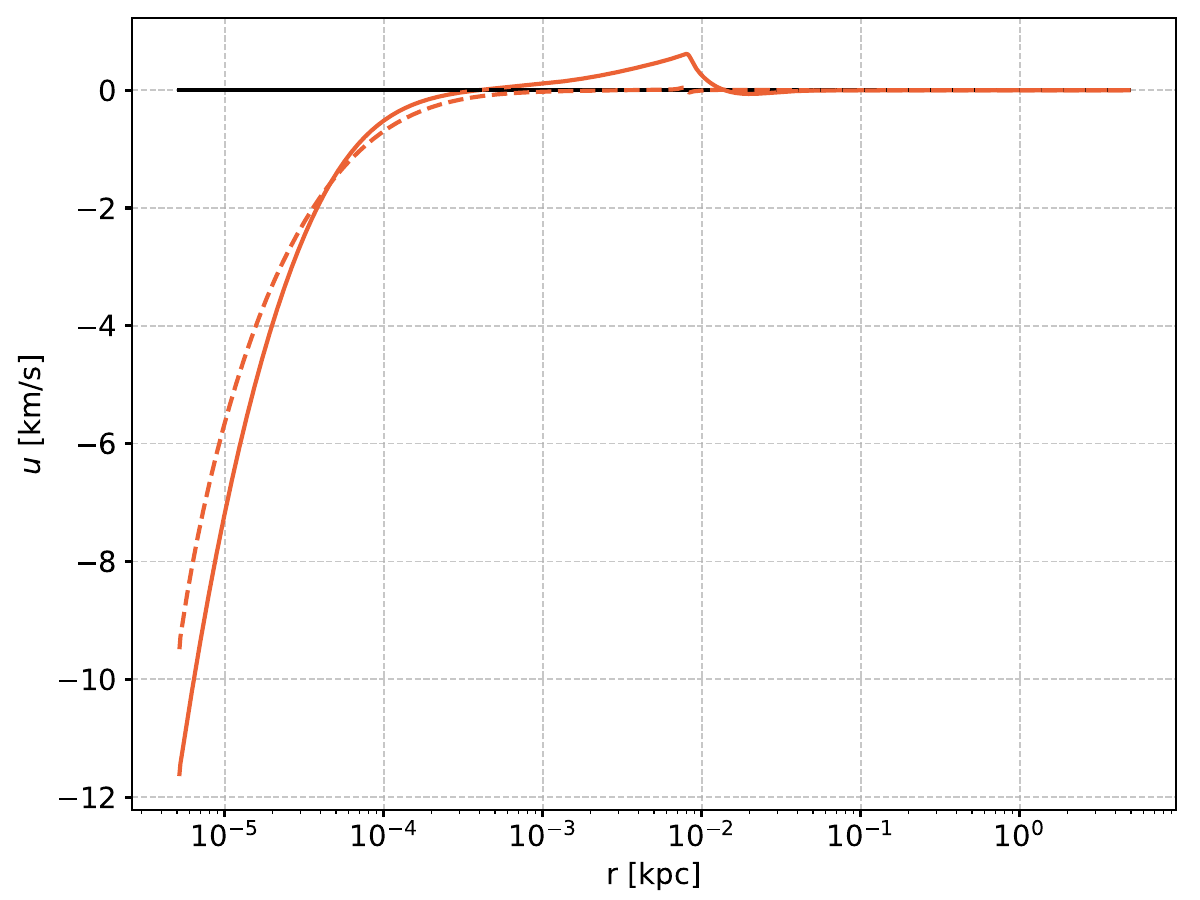}
    \end{subfigure}%
    \begin{subfigure}[b]{0.33333\textwidth}
        \includegraphics[width=\textwidth]{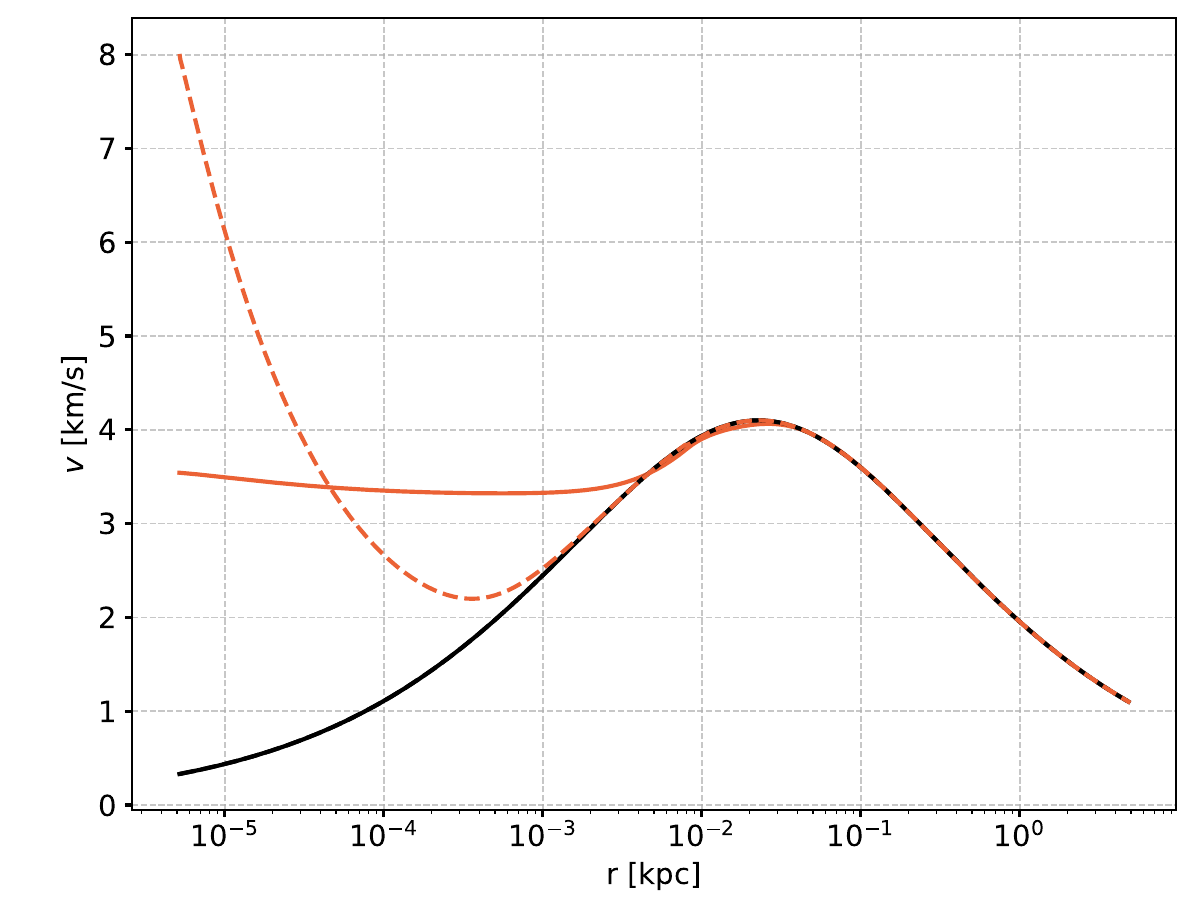}
    \end{subfigure}
    \caption{Comparison of the NFW profile evolution with and without heat flow. Parameters are identical to those in Figure~\ref{fig:NFW_results}. Solid and dashed lines distinguish cases with and without heat flow. In the bottom row, black and red lines represent the initial and final time snapshots, respectively.}
    \label{fig:compare_NFW}
\end{figure*}

\section{\label{sec:result} Results}

In this section, we present and analyze the results of our numerical simulations regarding the accretion of SIDM onto the central black hole. We begin by examining the dynamical evolution within two typical halo environments—the Singular Isothermal Sphere (SIS) and the Navarro--Frenk--White (NFW) profile—to identify how different initial density profiles influence the accretion process. Subsequently, we perform a systematic parametric study to investigate how the initial black hole mass, the self-interaction cross section, and the density power-law index affect the efficiency of mass growth. For all parameter sets, numerical convergence and robustness are confirmed with respect to the locations of both the inner and outer boundaries, as well as grid refinement. Together, these simulations elucidate the competition between gravitational attraction and thermal conduction, providing insights into the viability of SIDM accretion as a mechanism for the formation of supermassive black hole seeds.

\subsection{\label{subsec_NFW_SIS} Accretion in SIS and NFW halos}

We consider two dark matter halos, each with a total mass of $10^6\,\mathrm{M_{\odot}}$ at redshift $z = 25$, hosting a central black hole with an initial mass of $100\,\mathrm{M_{\odot}}$.
These parameters are adopted as a representative configuration reflecting the typical properties of black hole seeds and their host environments in the high-redshift universe. 
The first halo is modeled by a Singular Isothermal Sphere (SIS) density profile $\rho_{\text{SIS}}(r) = c_s^2/(2\pi G r^2)$ with sound speed $c_s = 4.2\,\mathrm{km}/\mathrm{s}$. 
The second halo follows an NFW density profile $\rho_{\text{NFW}}(r) = \rho_s/[(r/r_s)(1+(r/r_s))^{2}]$, with scale radius $r_s = 0.03\,\mathrm{kpc}$ and characteristic density $\rho_s = 3.7\,\mathrm{M_{\odot}}/\mathrm{pc}^3$. 
For both halos, the self-interaction cross section is set to $\sigma_m = 50\,\mathrm{cm}^2/\mathrm{g}$ as an exploratory small-scale SIDM benchmark, motivated by dwarf-scale studies where such large cross sections can remain viable at low velocities, especially in velocity-dependent SIDM models~\cite{Tulin:2017ara,Yang:2022mxl,Jia:2026ocr}.
These simulations yield the time evolution of the black hole mass and accretion rate, as well as the radial profiles of the dark matter density, radial flow velocity and velocity dispersion at different times.

Figure~\ref{fig:SIS_results} shows the simulation results for the SIS profile. 
The computational domain is $[0.001\,\mathrm{pc},1000\,\mathrm{pc}]$, and the system is evolved for $2\,\mathrm{Myr}$. 
The top row of Figure~\ref{fig:SIS_results} displays the black hole mass growth and accretion rate, while the bottom row shows the evolution of the density, radial velocity and velocity dispersion profiles.
The black hole initially undergoes a phase of rapid accretion, followed by a gradual decline toward a quasi-steady state. 
This behavior can be understood as follows: initially, the SIS profile obeys a power law $\rho \propto r^{-2}$, with a high central density and a spatially constant velocity dispersion. 
Under these conditions, the strong gravitational pull of the black hole drives a large initial accretion rate and maintains a spike-like central concentration. 
This inner structure is continuously regulated by SIDM accretion and conductive heat transport.
As the dark matter flows toward the central black hole, the local supply becomes insufficient to maintain the initial rate, leading to a depletion of the inner density. At the same time, the velocity dispersion increases because of compressional heating associated with the compression term in the energy equation. 
Part of this thermal energy is then transported outward by heat conduction, which redistributes energy from the inner halo to larger radii and thereby further reduces the central density.
Figure~\ref{fig:compare_SIS} illustrates the influence of heat conduction by contrasting it with the adiabatic case. Through a comparison of the cases with heat flow (solid lines) and without it (dashed lines), it is evident that heat conduction leads to a significant reduction in the central density, inflow velocity, and velocity dispersion, consistent with the physical mechanism described above.
As a result of the combined effects of mass depletion and conductive heat transport, the accretion rate gradually declines and eventually approaches a lower, quasi-steady value. 
By the end of the simulation, the black hole mass reaches $\sim 10^4\,\mathrm{M_{\odot}}$, suggesting that SIDM accretion in SIS-like halos may provide an efficient channel for the formation of massive black hole seeds. 

Figure~\ref{fig:NFW_results} presents the corresponding results for the NFW profile. 
The computational domain spans $[0.005\,\mathrm{pc},5000\,\mathrm{pc}]$, and the system is likewise evolved for $2\,\mathrm{Myr}$. 
In contrast to the SIS profile, the NFW halo has a shallower inner density cusp, $\rho \propto r^{-1}$, together with a lower central density and a smaller initial velocity dispersion. 
Owing to its non-uniform initial velocity-dispersion profile, thermal conduction efficiently transports heat into the colder inner region, leading to a rapid increase in the central velocity dispersion and altering the inflow.
Under the initial gravitational influence of the central black hole, dark matter flows into the inner region, leading to a rapid increase in central velocity dispersion. 
The resulting inner density profile remains centrally peaked and spike-like rather than developing a flat core, reflecting the continued gravitational influence of the central black hole. 
This central concentration is nevertheless regulated by SIDM heat transport, which raises the central velocity dispersion, increases pressure support, and gradually reduces the innermost density. 
Consequently, the decline in central density leads to a gradual reduction in the accretion rate. 
This effect is further illustrated in Figure~\ref{fig:compare_NFW}, which contrasts the impact of thermal conduction against the adiabatic case.
It is evident that heat conduction flattens the velocity dispersion profile and suppresses the central density. 
As the evolution proceeds, the central velocity dispersion continues to rise under the combined effects of thermal conduction and compressional heating. 
Since both the inner density and the inflow velocity are smaller than those in the SIS halo, the NFW profile exhibits a lower initial accretion rate.
As in the SIS run, the accretion rate gradually declines with time as the halo evolves.

\begin{figure*}[htbp]
    \centering
    \begin{subfigure}[b]{0.3333\textwidth}
        \includegraphics[width=1\textwidth]{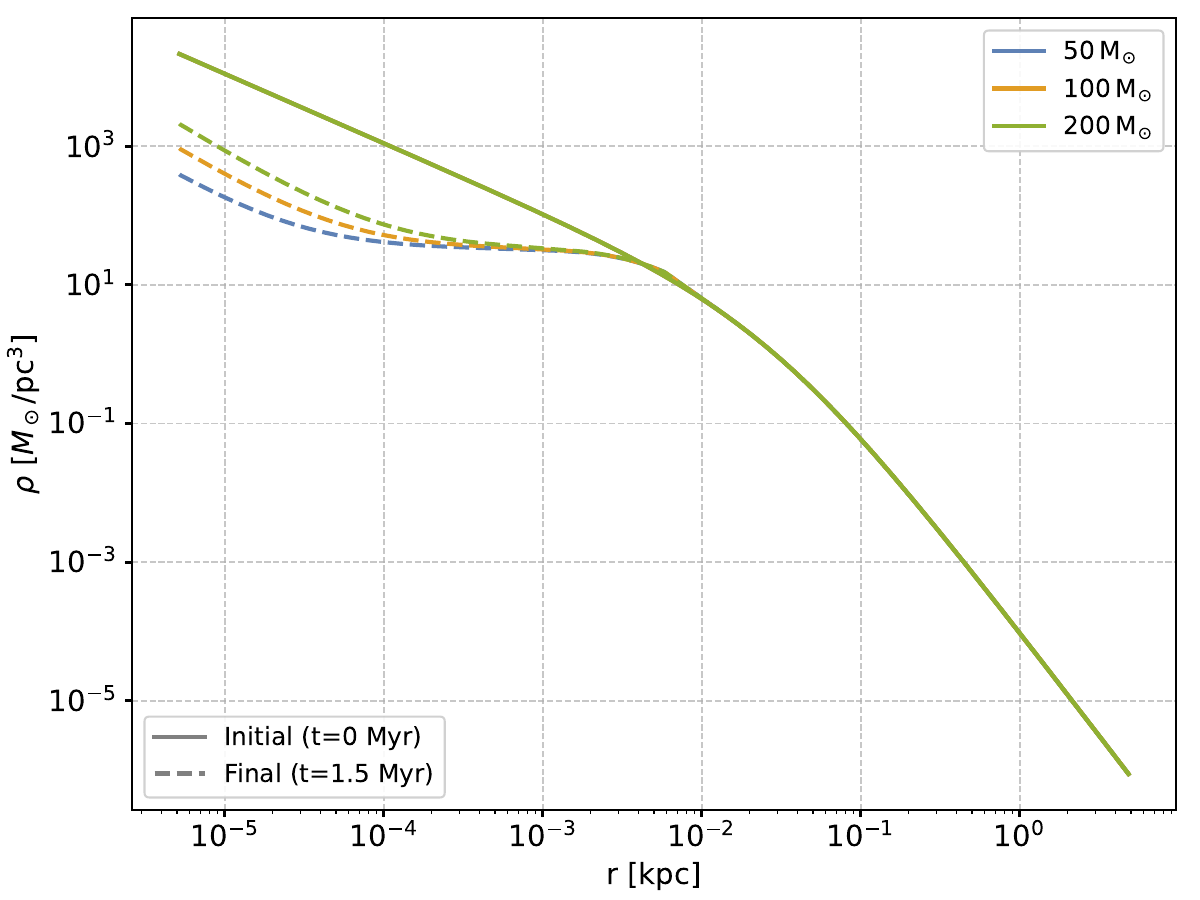}
    \end{subfigure}%
    \begin{subfigure}[b]{0.3333\textwidth}
        \includegraphics[width=1\textwidth]{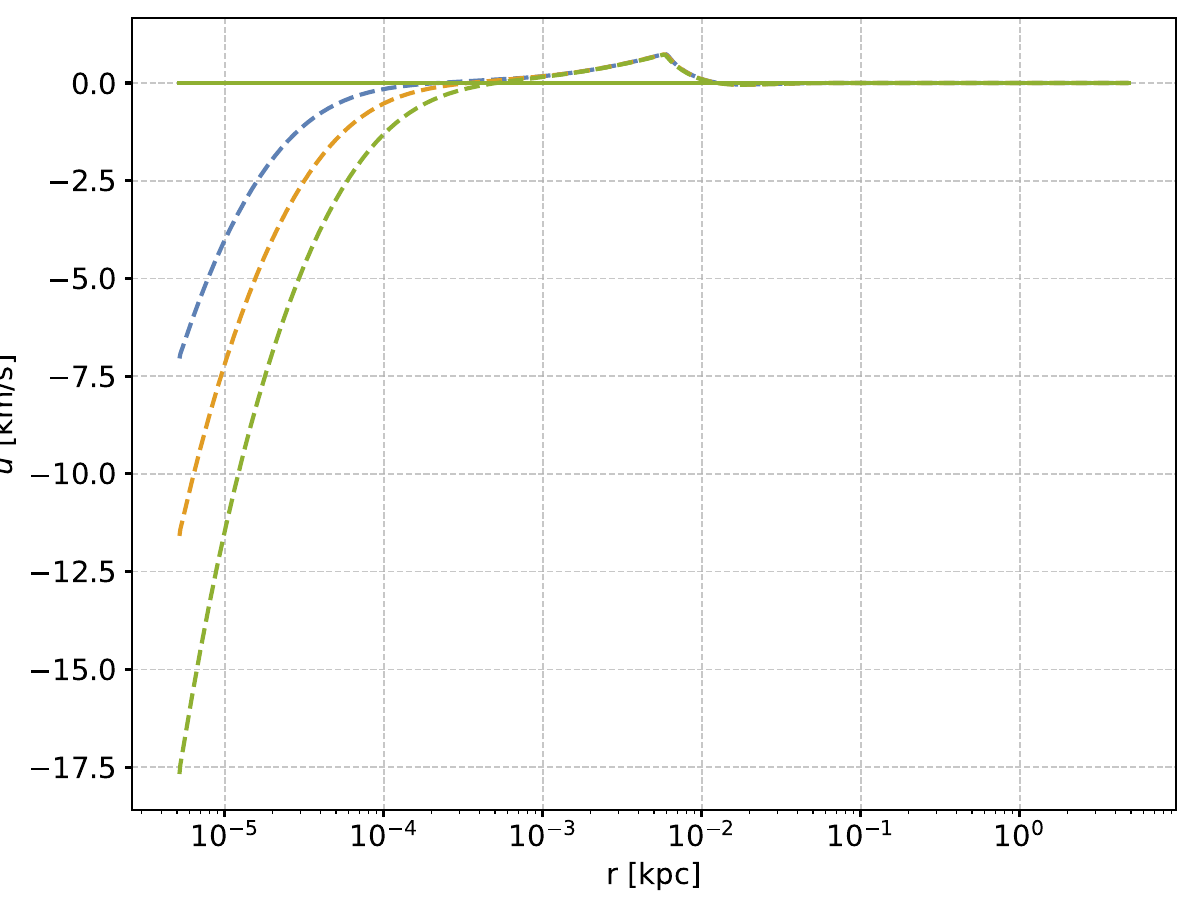}
    \end{subfigure}%
    \begin{subfigure}[b]{0.3333\textwidth}
        \includegraphics[width=1\textwidth]{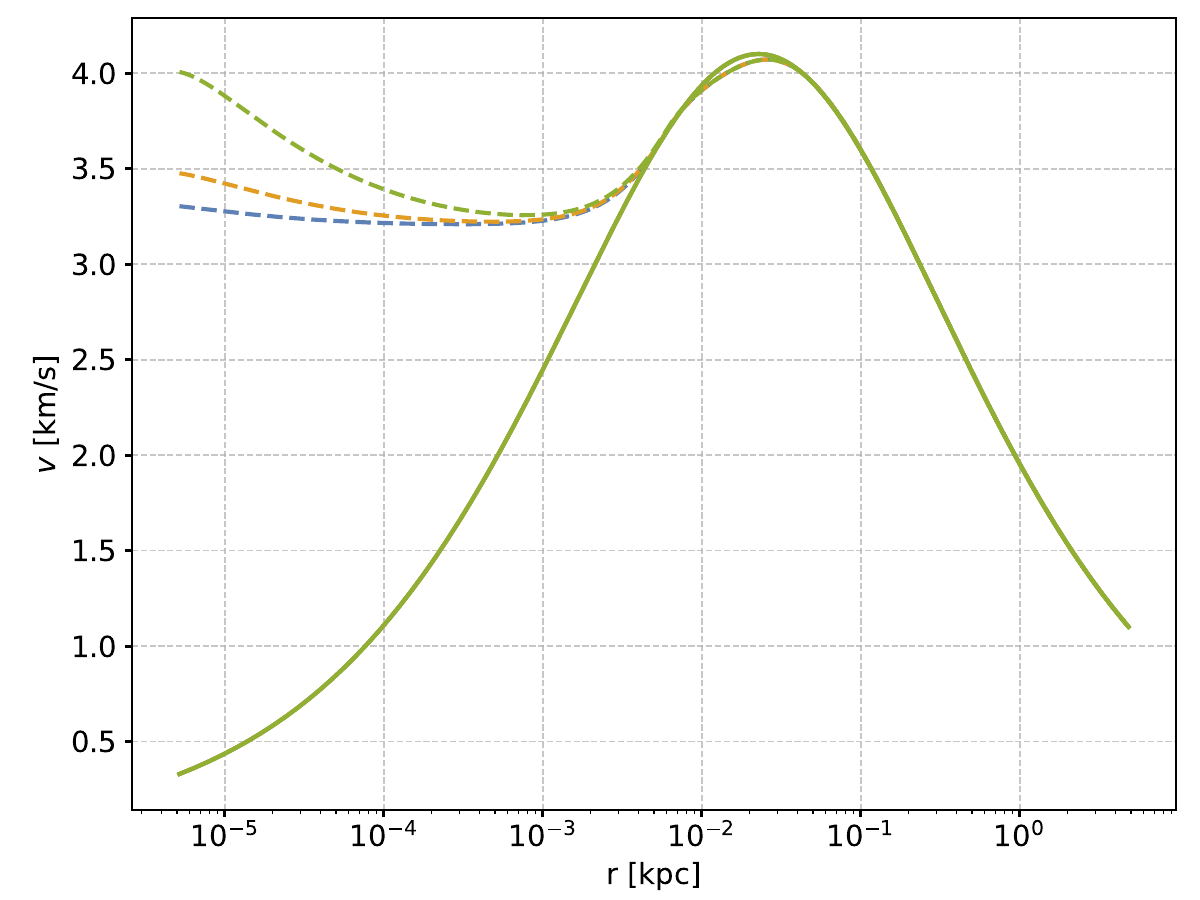}
    \end{subfigure} 
    \begin{subfigure}[b]{0.3333\textwidth}
        \includegraphics[width=1\textwidth]{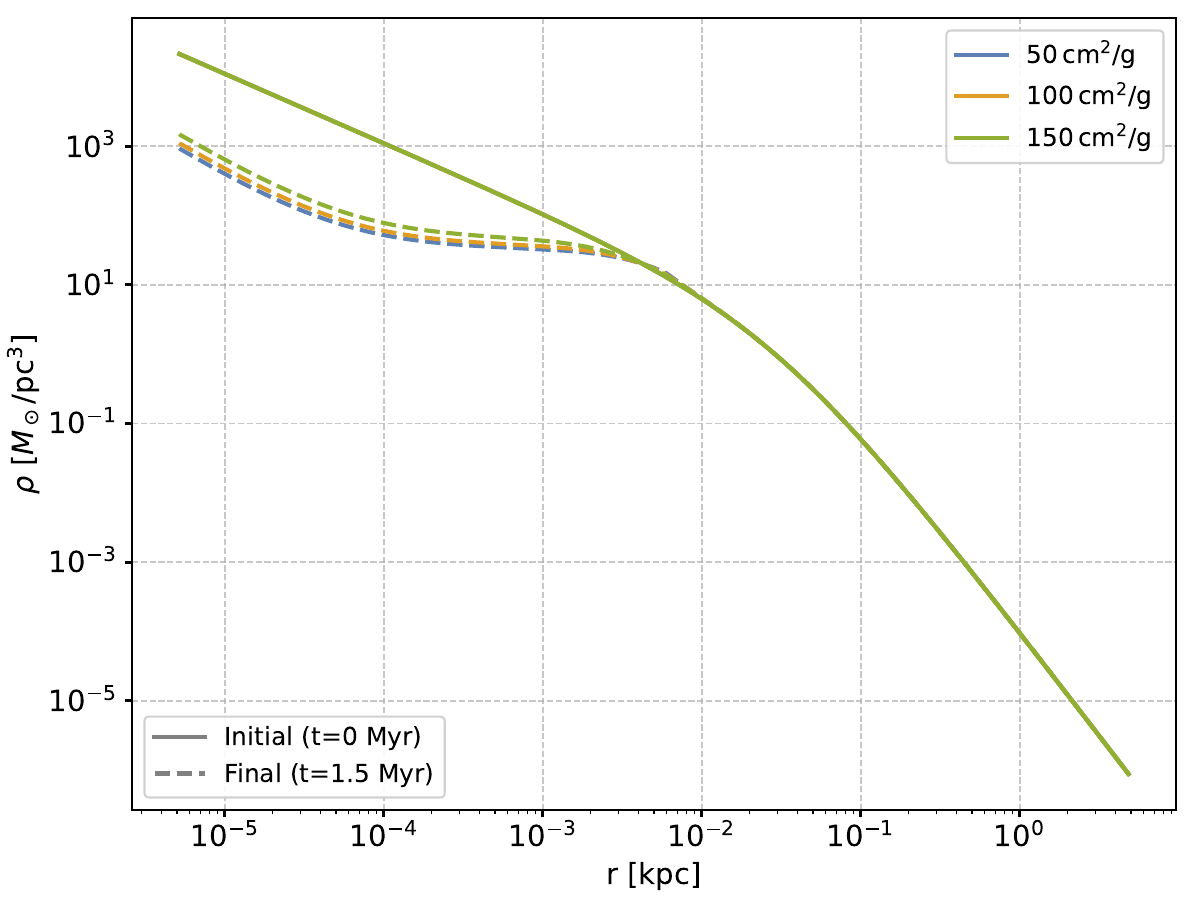}
    \end{subfigure}%
    \begin{subfigure}[b]{0.3333\textwidth}
        \includegraphics[width=1\textwidth]{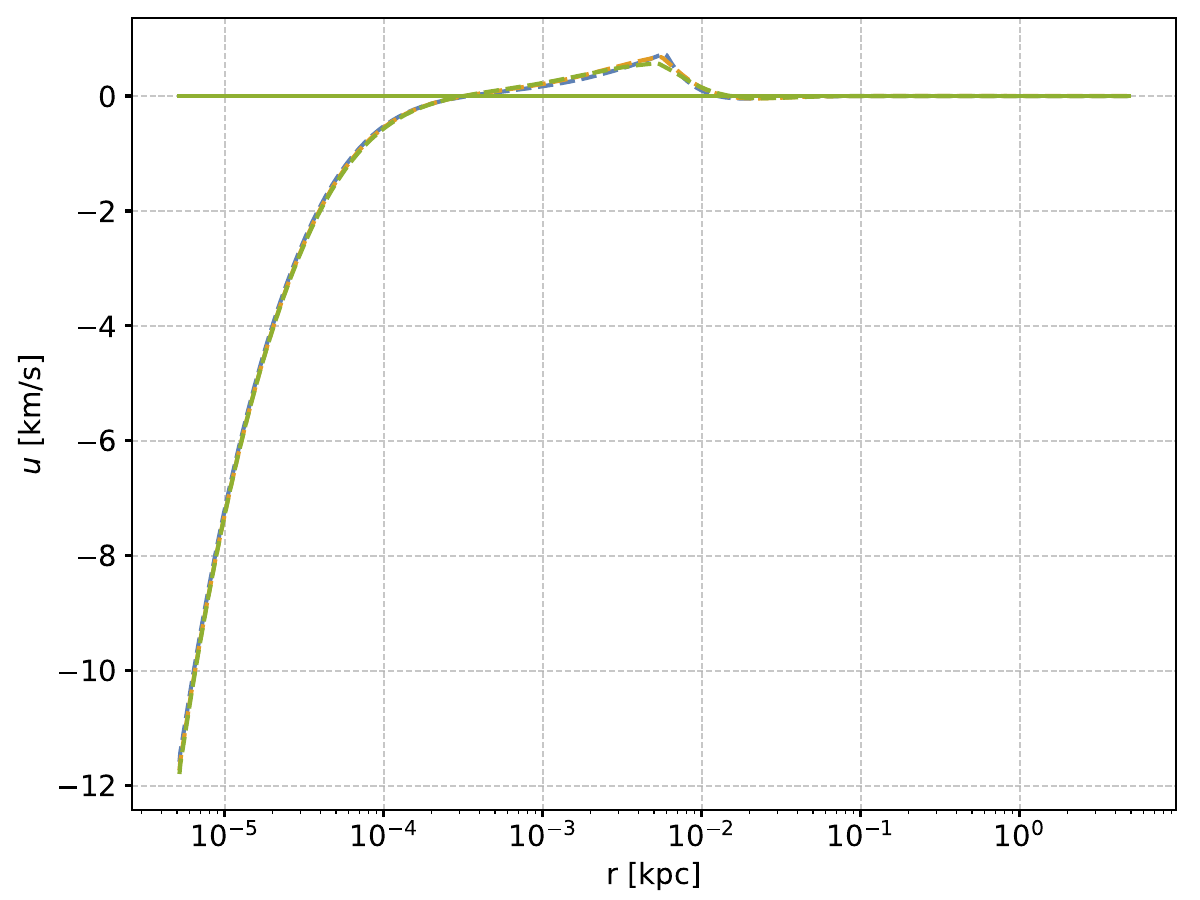}
    \end{subfigure}%
    \begin{subfigure}[b]{0.3333\textwidth}
        \includegraphics[width=1\textwidth]{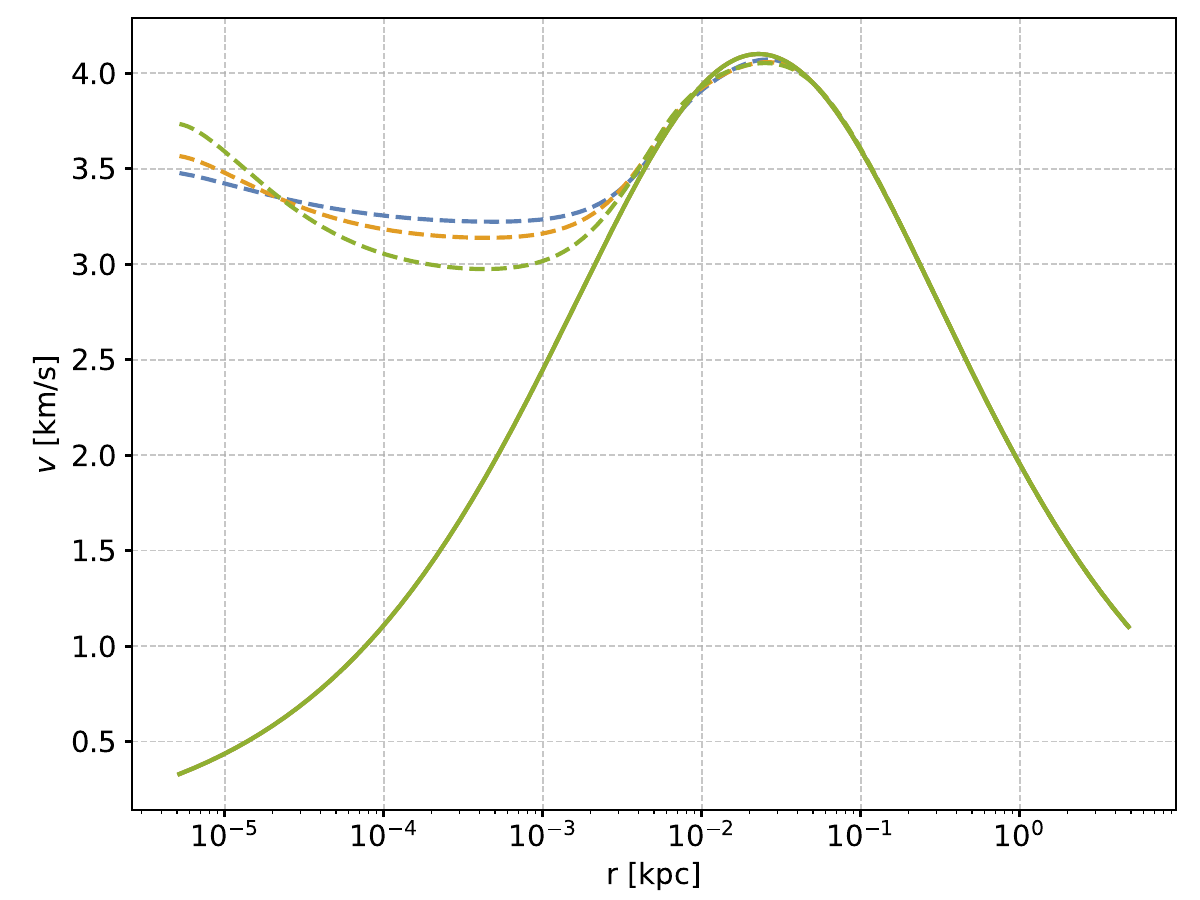}
    \end{subfigure} 
    \begin{subfigure}[b]{0.3333\textwidth}
        \includegraphics[width=1\textwidth]{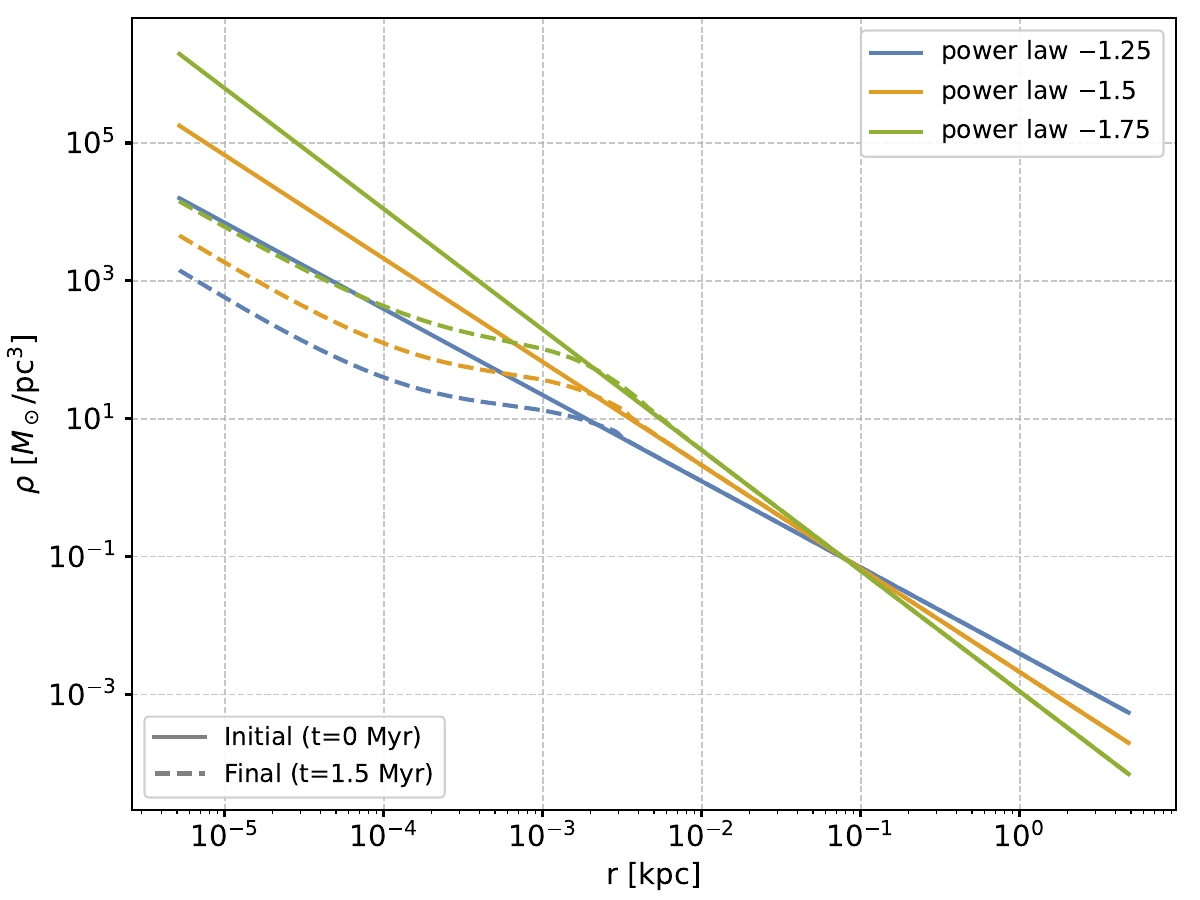}
    \end{subfigure}%
    \begin{subfigure}[b]{0.3333\textwidth}
        \includegraphics[width=1\textwidth]{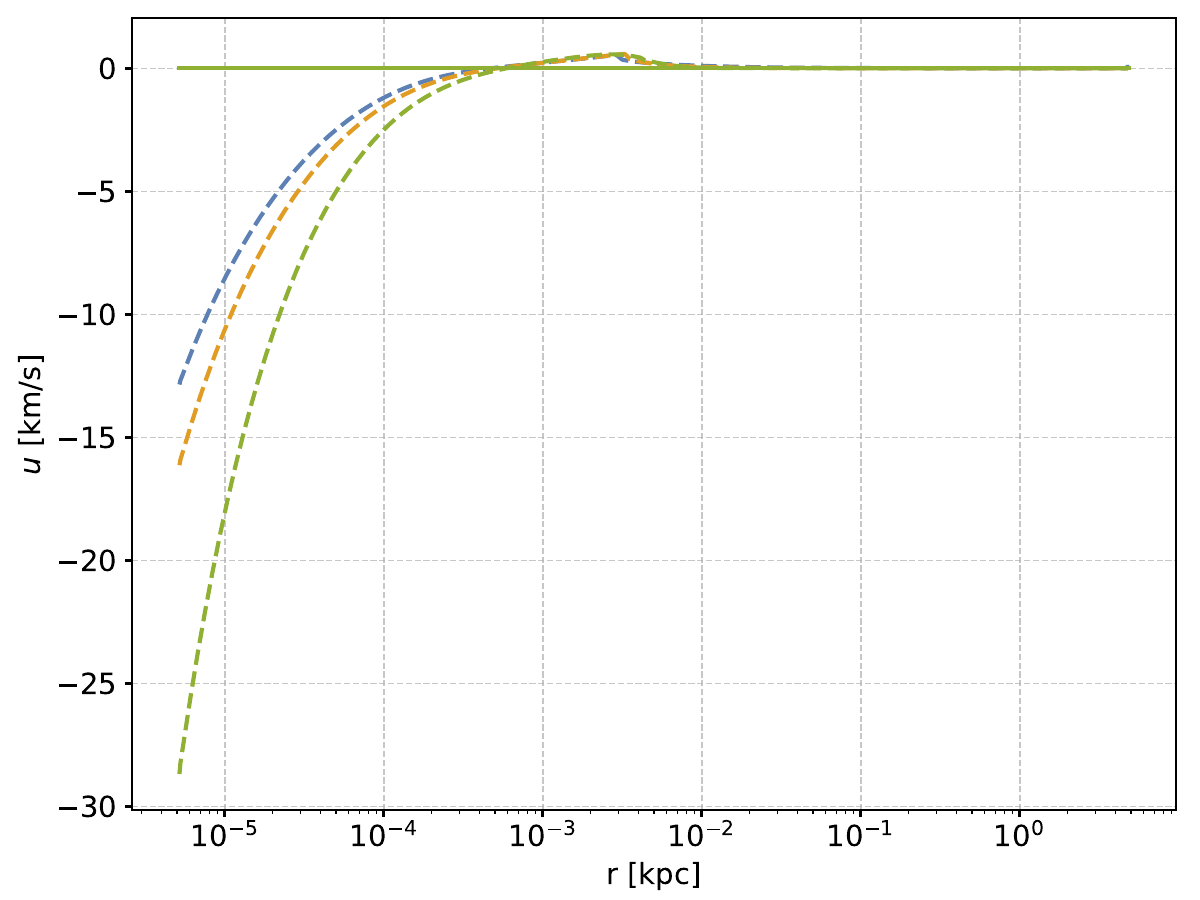}
    \end{subfigure}%
    \begin{subfigure}[b]{0.3333\textwidth}
        \includegraphics[width=1\textwidth]{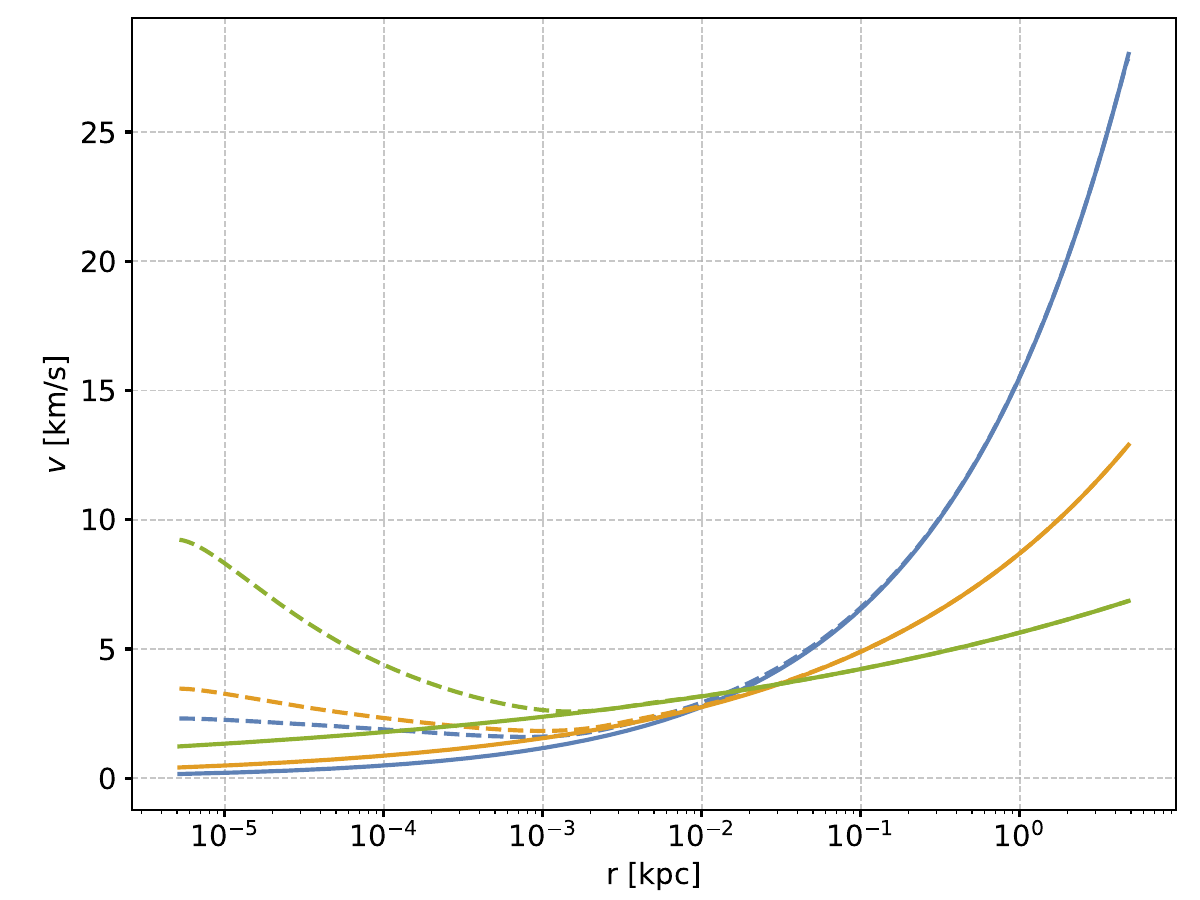}
    \end{subfigure}
    \caption{Evolution of physical quantities under various parameter configurations. The baseline setup is an NFW profile ($r_s = 0.03\,\mathrm{kpc}$, $\rho_s = 3.7\,\mathrm{M_{\odot}}/\mathrm{pc}^3$) with a total mass of $10^6\,\mathrm{M_{\odot}}$ at redshift $z = 25$. Rows (top to bottom) correspond to variations in initial black hole mass, scattering cross section, and density power-law index, respectively. Columns (left to right) display the evolution of density, radial velocity, and velocity dispersion. Line colors distinguish different parameter values within each row, while line styles represent the time evolution: solid lines for the initial profiles and dashed lines for final states.}
    \label{fig:physical_parameters_v_linear}
\end{figure*}

\subsection{\label{subsec_para} Dependence of SIDM accretion on physical parameters}
In this subsection, we examine the effects of several key parameters on SIDM accretion, namely the initial black hole mass, the self-interaction cross section, and the slope of the initial density profile. 
Using the NFW halo model introduced in Section~\ref{subsec_NFW_SIS} as the baseline setup, we perform three sets of simulations in which one parameter is varied at a time. 
Specifically, we consider initial black hole masses of $50$, $100$ and $200\,\mathrm{M_{\odot}}$, self-interaction cross sections of $50$, $100$ and $150\,\mathrm{cm}^2/\mathrm{g}$, and power-law initial density profiles with inner slopes $\rho \propto r^{-1.25}$, $r^{-1.5}$, and $r^{-1.75}$. In the last case, the total halo mass is fixed at $10^6\,\mathrm{M_{\odot}}$. 
All other parameters are the same as in the baseline simulation.

\begin{figure*}[htbp]
    \centering
    \begin{subfigure}[b]{0.4\textwidth}
        \includegraphics[width=1\textwidth]{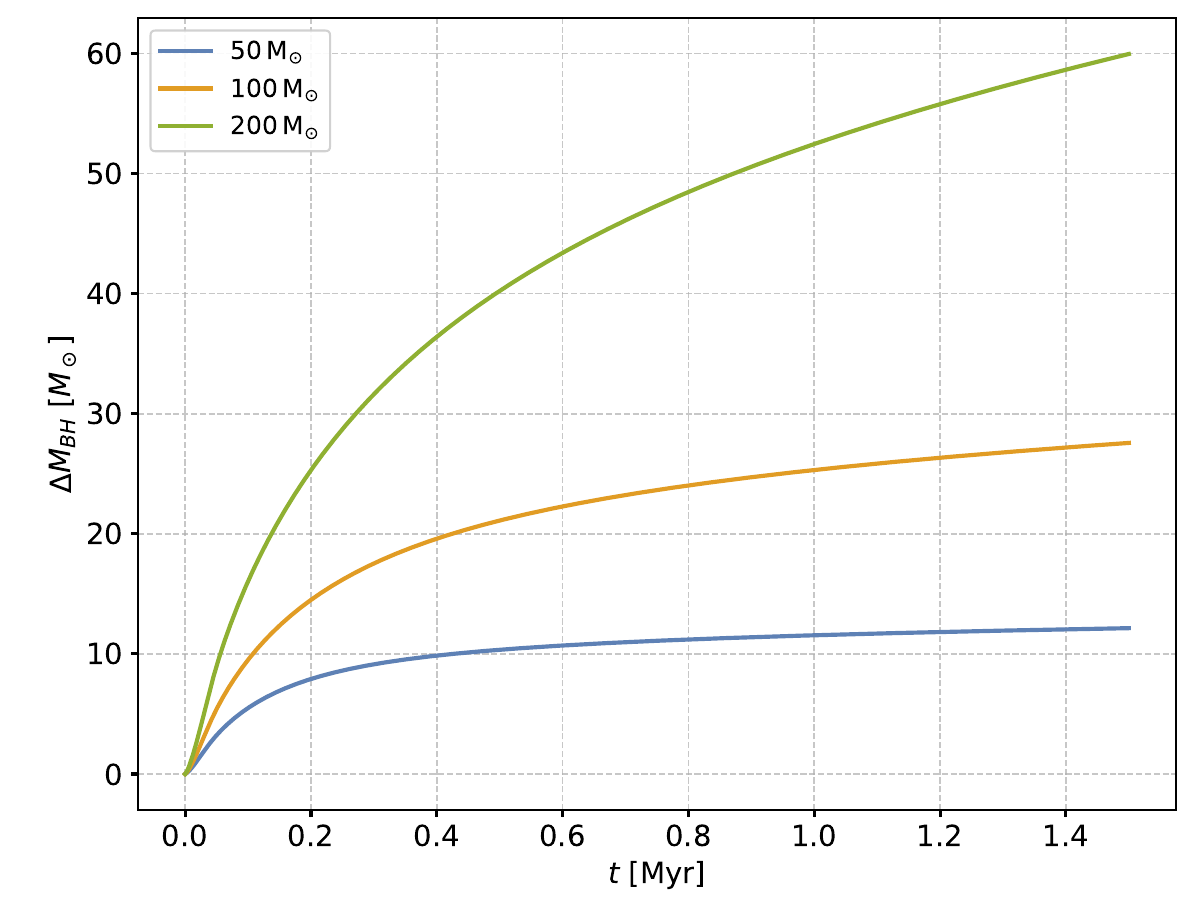}
    \end{subfigure}%
    \begin{subfigure}[b]{0.4\textwidth}
        \includegraphics[width=1\textwidth]{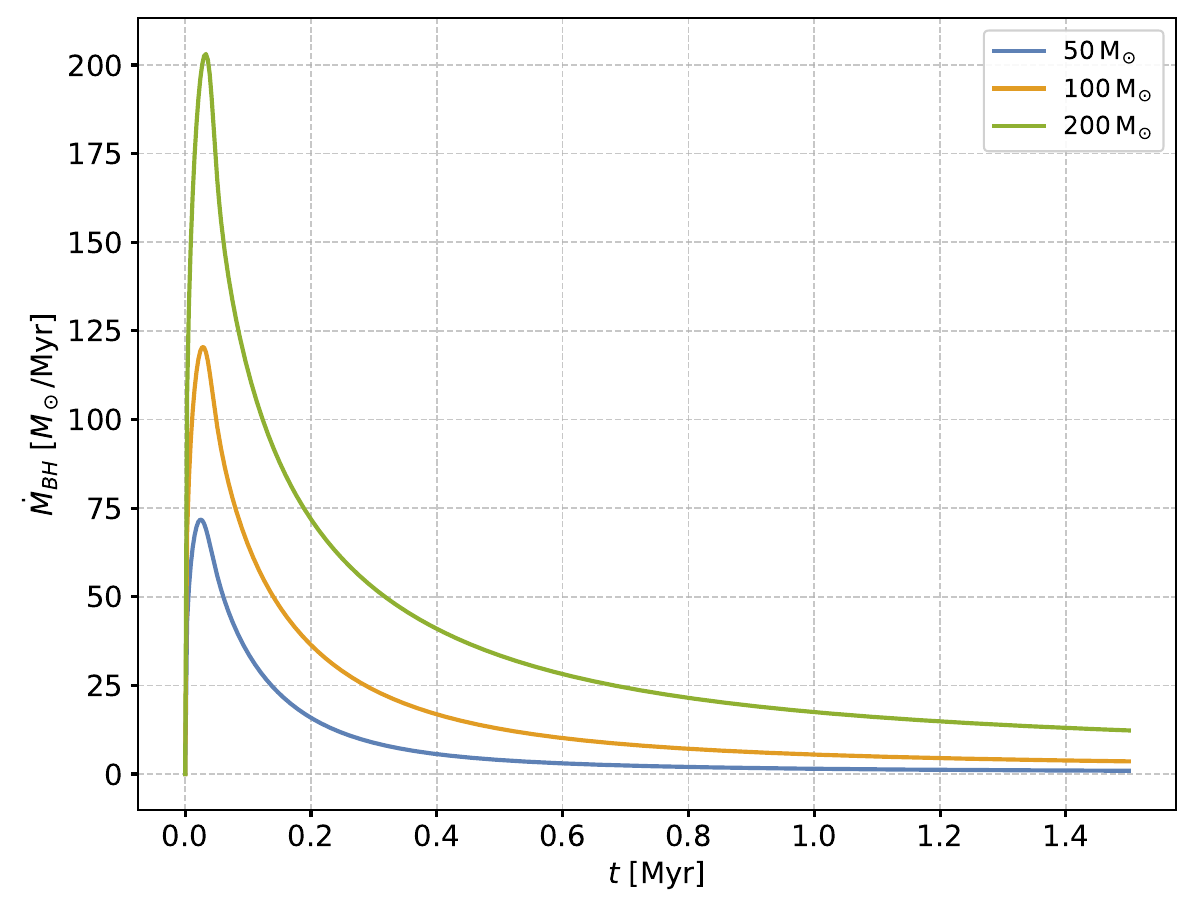}
    \end{subfigure} 
    \begin{subfigure}[b]{0.4\textwidth}
        \includegraphics[width=1\textwidth]{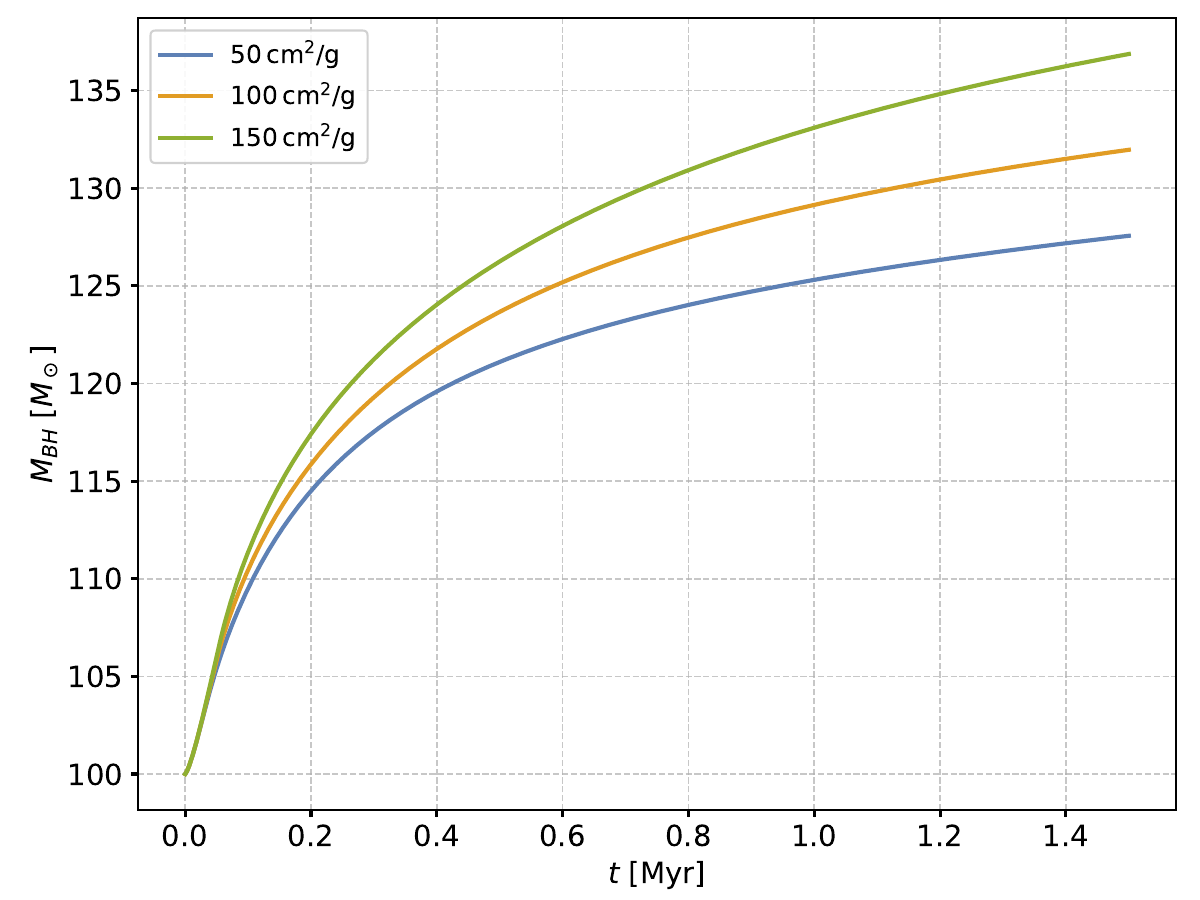}
    \end{subfigure}%
    \begin{subfigure}[b]{0.4\textwidth}
        \includegraphics[width=1\textwidth]{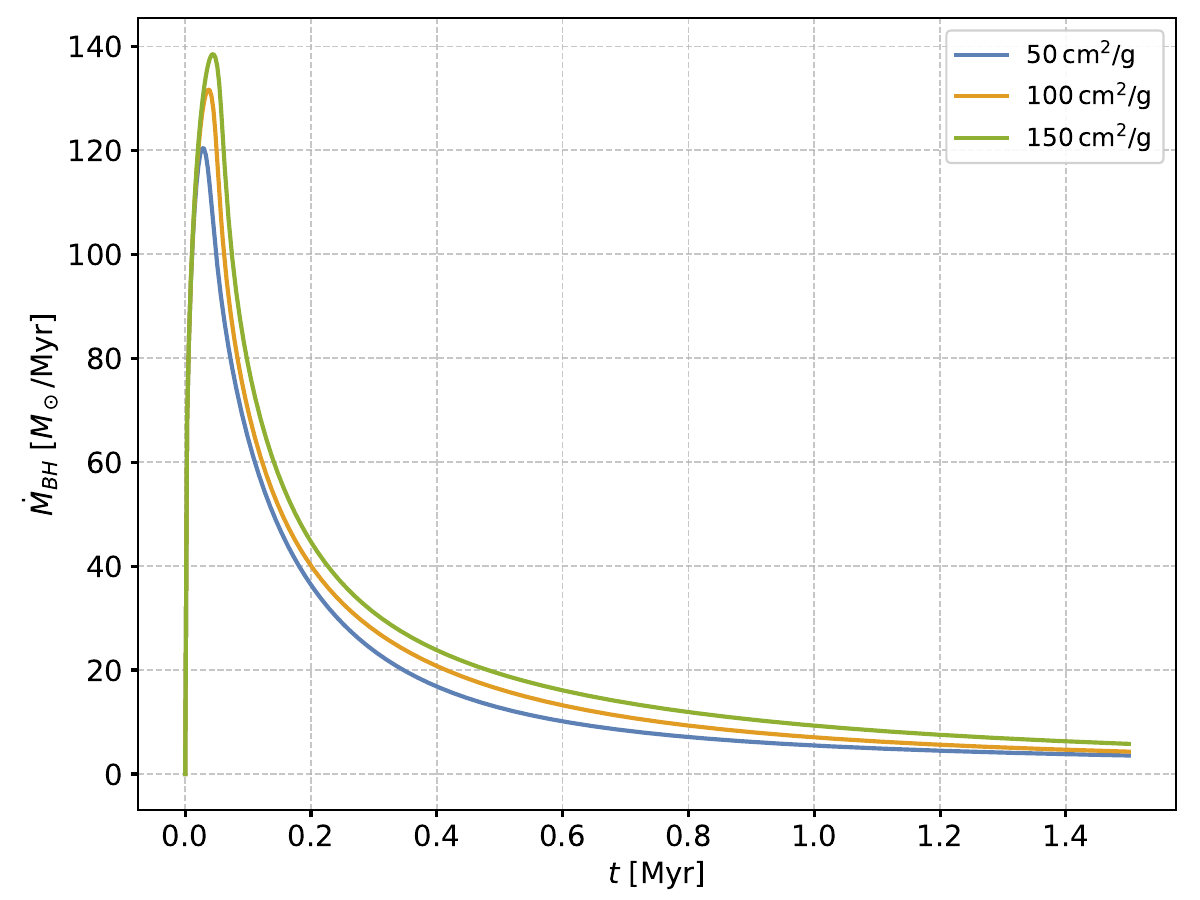}
    \end{subfigure} 
    \begin{subfigure}[b]{0.4\textwidth}
        \includegraphics[width=1\textwidth]{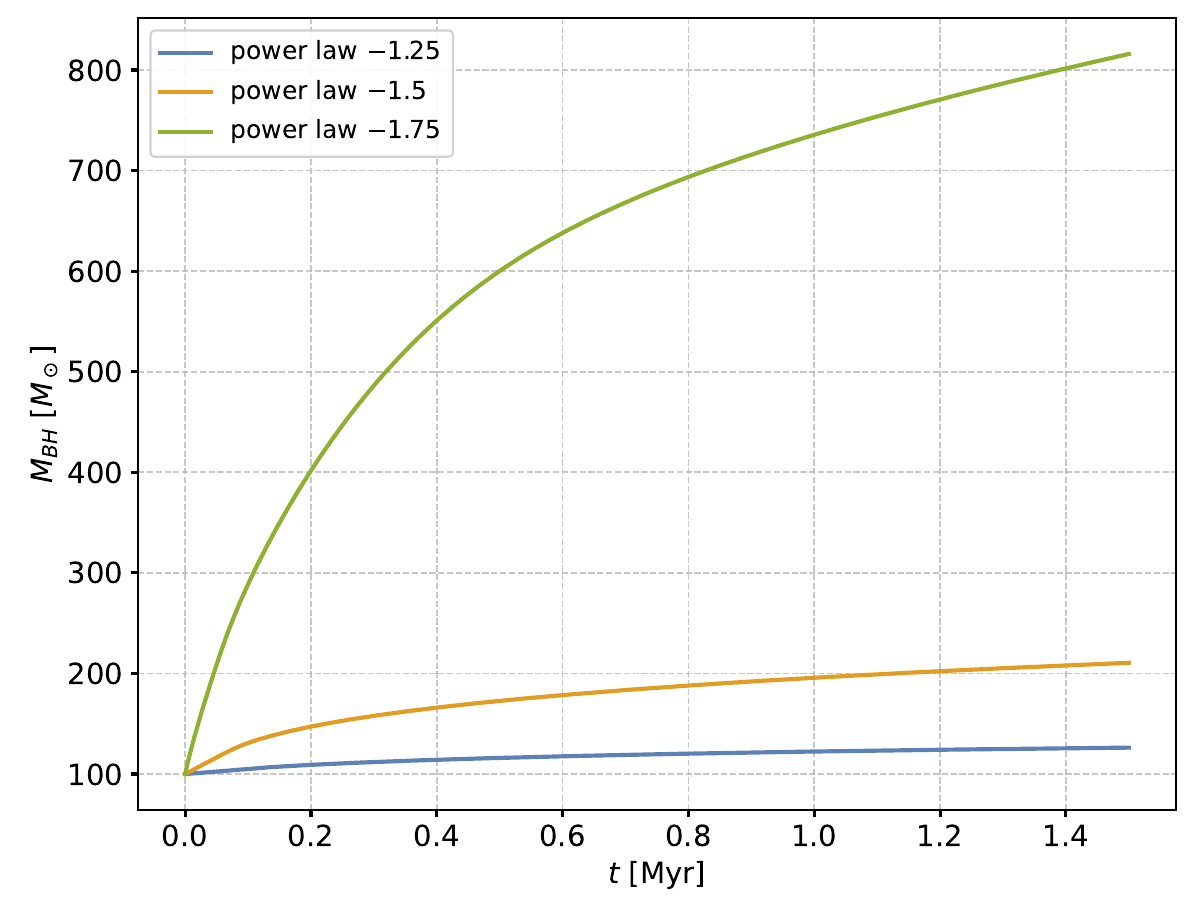}
    \end{subfigure}%
    \begin{subfigure}[b]{0.4\textwidth}
        \includegraphics[width=1\textwidth]{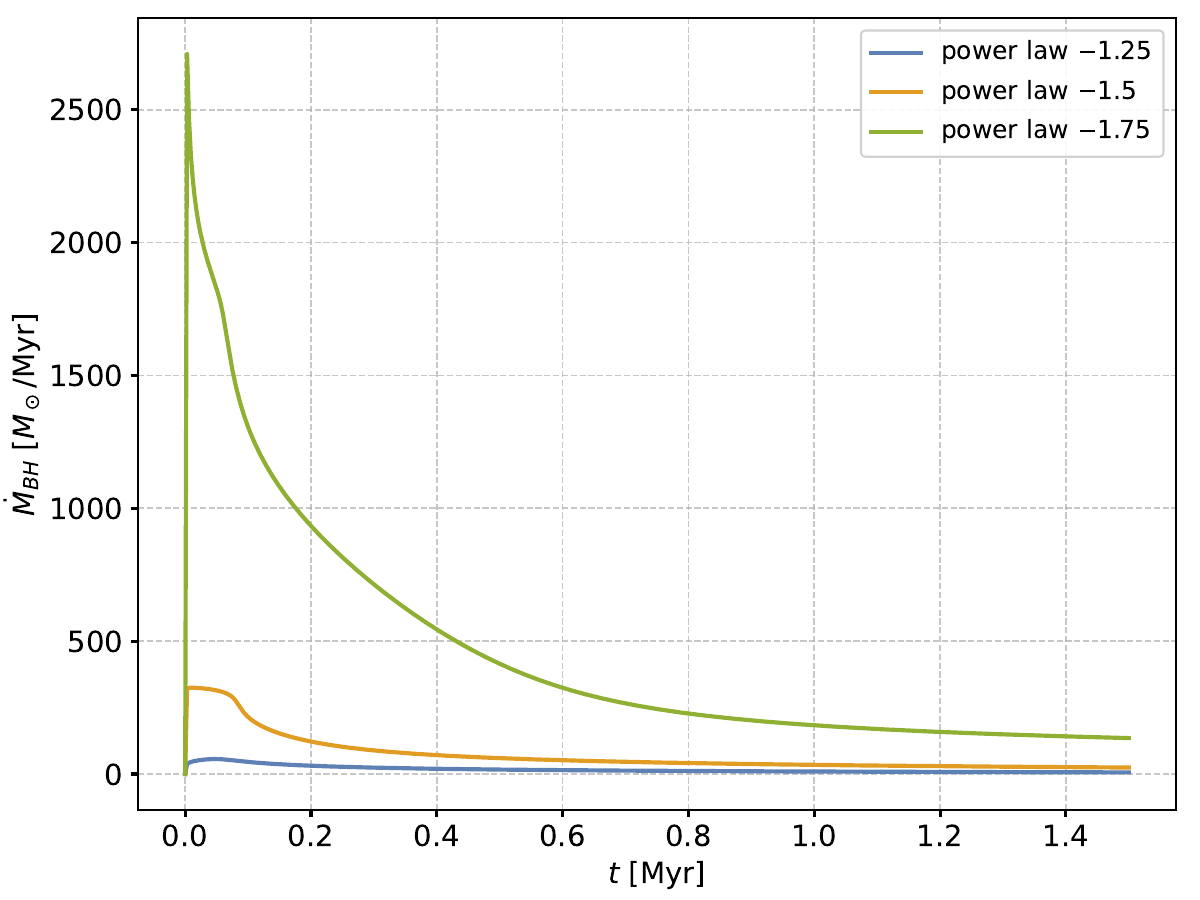}
    \end{subfigure}
    \caption{Time evolution of the black hole mass (left panel) and accretion rate (right panel) under various parameter configurations. 
    Parameters are identical to those in Figure~\ref{fig:physical_parameters_v_linear}.
    Rows (top to bottom) correspond to variations in initial black hole mass, scattering cross section, and density power-law index, respectively. 
    Line colors distinguish different parameter values within each row. In the top-left panel, the mass increment $\Delta M_{\text{BH}}$ is plotted to facilitate comparison between different initial masses.}
    \label{fig:BH_accretion}
\end{figure*}

Figure~\ref{fig:physical_parameters_v_linear} shows the evolution of the density, radial velocity, and velocity dispersion profiles, while Figure~\ref{fig:BH_accretion} presents the corresponding black hole mass growth and accretion histories. 
As shown in Figure~\ref{fig:physical_parameters_v_linear}, the density evolution is particularly sensitive to the initial BH mass and the initial density slope. 
A larger initial black hole mass leads to a steeper inner density profile, because its deeper gravitational potential more strongly shapes the central region.
Likewise, a steeper initial density profile leads to a higher central concentration of dark matter.
The radial velocity is governed mainly by the gravitational field and is therefore most strongly affected by the initial BH mass and the density slope.
By contrast, the self-interaction cross section influences the flow primarily through thermal conduction. 
This reflects the competition between gravity and conductive heat transport: gravity drives matter inward, whereas thermal conduction redistributes energy outward, promotes thermal relaxation, and tends to flatten the inner SIDM profile.
In addition, a larger initial BH mass and a steeper initial density profile both lead to a more rapid increase in the central velocity dispersion. 
The dependence on the self-interaction cross section is regime dependent.
In the SMFP regime, a smaller $\sigma_m$ increases the collisional mean free path and therefore enhances local thermal conduction, leading to faster heat redistribution and core flattening.
This trend should not be extrapolated to arbitrarily small cross sections: as $\sigma_m$ decreases further, the system eventually enters the LMFP regime, where heat transport is limited by the scattering rate rather than by the mean free path. 
Consequently, the effective gravothermal conductivity decreases with decreasing $\sigma_m$ in the LMFP regime.

Figure~\ref{fig:BH_accretion} shows how the BH accretion rate and mass growth are influenced by these three parameters.
A larger initial BH mass leads to more rapid accretion, because the deeper gravitational potential enhances both the central density and the inflow velocity. 
Similarly, a steeper initial density slope promotes faster accretion by supplying more dark matter to the inner region. 
The dependence on the scattering cross section is also clear: in the SMFP regime, a larger $\sigma_m$ reduces the efficiency of thermal conduction, so outward heat transport is less effective at counteracting gravity. As a result, the inner halo profile flattens more slowly, and the black hole can sustain a higher accretion rate.

\section{\label{sec:discuss} Discussion}
In this study, we model self-interacting dark matter (SIDM) as a self-gravitating fluid with thermal conduction and investigate its accretion onto a central black hole. 
Assuming spherical symmetry and Newtonian gravity, we solve the governing hydrodynamic equations with an operator-splitting method. 
The hyperbolic part is treated using the Monotonized Central (MC) limiter for spatial reconstruction and a Roe Riemann solver for the numerical fluxes, together with explicit Euler time integration. 
The thermal conduction term is evolved with an implicit Euler scheme to avoid the severe numerical stiffness associated with diffusive transport.
Benchmark comparisons demonstrate that the numerical scheme accurately reproduces the expected gravothermal evolution of SIDM halos.

Using this framework, we perform simulations for SIDM halos with initial NFW and SIS density profiles. 
Our results show that the black hole accretion rate generally exhibits an initial transient peak, which arises from the relaxation of the nonequilibrium initial condition.
After this early adjustment phase, the accretion rate gradually declines as the halo evolves under the combined effects of central inflow, conductive heat transport, and the depletion of the inner mass supply.
In the NFW case, thermal conduction efficiently heats the initially colder inner region and reduces the central density, which leads to a relatively modest accretion rate.
By contrast, in the SIS case, the initially isothermal halo has a much weaker thermal gradient, so conduction is less important at early times and the black hole can accrete much more rapidly, growing from $100\,\mathrm{M_{\odot}}$ to $10^4\,\mathrm{M_{\odot}}$. 
This result suggests that SIDM accretion in SIS-like environments may provide an efficient channel for the formation of massive black hole seeds.
We also investigate the dependence of the accretion process on the initial black hole mass, the self-interaction cross section, and the slope of the inner density profile. 
These results show that SIDM accretion is sensitive to both the initial halo structure and the efficiency of conductive heat transport.

This work establishes a fluid-dynamical framework for studying black hole accretion in SIDM halos beyond the quasi-static treatments commonly adopted in the literature. By directly solving the time-dependent hydrodynamic equations with thermal conduction, our approach makes it possible to follow the coupled evolution of halo structure, heat transport, and black hole growth in a self-consistent manner. In this sense, the present study serves as a first step toward a more complete dynamical description of black hole growth in SIDM environments. Our results show that the accretion history is sensitive not only to the initial halo profile, but also to the efficiency of conductive heat transport and to the initial black hole mass, indicating that SIDM microphysics may play an important role in determining the early growth of black hole seeds.

In future work, we plan to extend the present model by including additional physical ingredients that are expected to be important in realistic astrophysical settings. 
In particular, baryonic cooling, radiative feedback, and angular momentum transport may substantially alter the accretion flow and its long-term evolution. 
We also plan to explore a broader range of halo parameters and cosmological environments, and ultimately to generalize the present one-dimensional treatment to multidimensional simulations. 
Such developments should help clarify the conditions under which SIDM can significantly enhance black hole growth in the early Universe.

\begin{acknowledgments}
T. Chen is supported by the National Natural 
Science Foundation of China Grant No. 12547155. 
B. Hu is supported in part by the National Natural Science Foundation of China Grant No.~12541301 and No.~12333001.
R.-G. Cai is supported by the National Natural Science Foundation of China Grant No.~12588101.
L. Gao is supported by the National Natural Science Foundation of China Grant No.~11988101.
\end{acknowledgments}

\appendix

\section{The hyperbolic solver}
\label{app:hyperbolic}

\subsection{The MC slope limiter}
In computational fluid dynamics (CFD), slope limiters are used to reconstruct interfacial values from cell-averaged quantities~\cite{leveque2002finite}.
Given the cell-averaged state vector $\mathbf{U}_i$, the left and right reconstructed states at the interface $i+1/2$, denoted by $\mathbf{U}_{i+1/2}^{\text{L}}$ and $\mathbf{U}_{i+1/2}^{\text{R}}$, can be obtained through linear reconstruction as follows:
\begin{subequations}
    \begin{align}
        \mathbf{U}_{i+1/2}^{\text{L}} &= \mathbf{U}_{i} + \tilde{\bm{\sigma}}_i \frac{\Delta \til r_i}{2},\\
        \mathbf{U}_{i+1/2}^{\text{R}} &= \mathbf{U}_{i+1} - \tilde{\bm{\sigma}}_{i+1} \frac{\Delta \til r_{i+1}}{2},
    \end{align}
\end{subequations}
where $\Delta \til r_i$ denotes the width of the $i$-th cell and $\tilde{\bm{\sigma}}_i$ is the numerical slope vector. 
The selection of $\tilde{\bm{\sigma}}_i$ is critical in the FVM, since it determines the balance between higher-order spatial accuracy and numerical stability by suppressing spurious oscillations near sharp gradients.

A widely used choice is the Monotonized Central (MC) limiter introduced by van Leer~\cite{van1977towards}. In this case, the slope $\tilde{\bm{\sigma}}_i$ is given by
\begin{equation}
    \begin{aligned}
    \tilde{\bm{\sigma}}_i =\text{minmod}\Bigg[    &\left(\frac{\mathbf{U}_{i+1} - \mathbf{U}_{i-1}}{2\Delta \til r_i}\right),
    2\left(\frac{\mathbf{U}_{i} - \mathbf{U}_{i-1}}{\Delta \til r_i}\right),\\
    &2\left(\frac{\mathbf{U}_{i+1} - \mathbf{U}_{i}}{\Delta \til r_i}\right)
    \Bigg],
    \end{aligned}
\end{equation}
where the $\text{minmod}$ function for three arguments is defined as:
\begin{equation}
    \text{minmod}(a,b,c) = 
    \begin{cases}
        \min(a,b,c), & a>0,b>0,c>0, \\
        \max(a,b,c), & a<0,b<0,c<0, \\
        0, & \text{others}.
    \end{cases}
\end{equation}
Applying the MC limiter yields robust and accurate interfacial states, $\mathbf{U}_{i+1/2}^{\text{L}}$ and $\mathbf{U}_{i+1/2}^{\text{R}}$, which serve as the left and right input states for the Riemann problem at each cell interface.

\subsection{The Roe Riemann solver}
In computational fluid dynamics, a Riemann problem is an initial-value problem for a hyperbolic system of conservation laws with piecewise-constant initial data. 
When the computational domain is discretized into cells, a local Riemann problem is defined at each interface $i+1/2$ by the reconstructed left and right states, $\mathbf{U}_{i+1/2}^{\text{L}}$ and $\mathbf{U}_{i+1/2}^{\text{R}}$. 
Solving these local Riemann problems determines the numerical flux at each interface, which is then used to update the cell-averaged quantities to the next time step. 
This framework was pioneered by Godunov~\cite{godunov1959finite}.

Solving the exact Riemann problem at every interface is computationally expensive because of the intrinsic nonlinearity of the governing equations. 
To improve efficiency, a number of approximate Riemann solvers have been developed, including the Roe~\cite{roe1981approximate}, HLL~\cite{Amiram:2006zjz}, and HLLC~\cite{1994ShWav...4...25T} solvers.
In this work, we adopt the Roe approximate Riemann solver~\cite{roe1981approximate}. The central idea of Roe's method is to replace the nonlinear Jacobian matrix by a locally linearized counterpart, $\hat{\mathbf{A}}$, constructed from specifically designed Roe-averaged variables. This linearization ensures that the solver remains efficient while satisfying the Rankine-Hugoniot conditions~\cite{roe1981approximate}. The specific formulas used in our implementation are detailed below.

The convective flux vector $\mathbf{F}_{\text{c}}$ can be expressed as a function of the state vector $\mathbf{U}$:
\begin{equation}
    \mathbf{F}_{\text{c}} = 
    \begin{bmatrix}
        U_2\\
        \frac{2(U_2^2 + U_1 U_3)}{3U_1}\\
        \frac{5 U_1 U_2 U_3 - U_2^3}{3 U_1^2}
    \end{bmatrix},\,\,
    \mathbf{U} \equiv
    \begin{bmatrix}
        U_1 \\
        U_2 \\
        U_3
    \end{bmatrix}
    =
    \begin{bmatrix}
        \til \rho \\
        \til \rho \til u \\
        \til \rho \til E
    \end{bmatrix} .
\end{equation}
The corresponding Jacobian matrix $\mathbf{A} = \partial \mathbf{F}_c/\partial \mathbf{U}$ is
\begin{equation}
    \begin{aligned}
        \mathbf{A} &= 
        \begin{bmatrix}
            0 & 1 & 0\\
            -\frac{2U_2^2}{3U_1^2} & \frac{4U_2}{3U_1} & \frac23\\
            \frac{2U_2^3 - 5U_1U_2U_3}{3U_1^3} & \frac{5U_1U_3-3U_2^2}{3U_1^2} & \frac{5U_2}{3U_1}
        \end{bmatrix}\\
        &=
        \begin{bmatrix}
            0 & 1 & 0\\
            -\frac{2}{3}\til u^2 & \frac{4}{3}\til u & \frac23\\
            \frac13\til u^3 - \til H \til u & -\frac23\til u^2 + \til H & \frac{5}{3}\til u
        \end{bmatrix},
    \end{aligned}
\end{equation}
where $\til H = \til E + \til v^2$ denotes the total enthalpy. 
To linearize the Riemann problem, we introduce the Roe parameter vector $\mathbf{Q} = [\sqrt{\til \rho}, \sqrt{\til \rho}\til u, \sqrt{\til \rho}\til H]^\text{T}$. The Roe average is then defined by the arithmetic mean of the left and right parameter vectors,
\begin{equation}
    \mathbf{\hat{Q}} = \frac12(\mathbf{Q}^{\text{L}} + \mathbf{Q}^{\text{R}}).
\end{equation}
From this definition, the Roe-averaged density, velocity, and enthalpy are obtained as 
\begin{subequations}
    \begin{align}
        \hat{\til \rho} &= \frac14 \left( \sqrt{\til \rho^{\text{L}}} + \sqrt{\til \rho^{\text{R}}}\right)^2,\\
        \hat{\til u} & = \frac{\til u^{\text{L}}\sqrt{\til \rho^{\text{L}}} + \til u^{\text{R}}\sqrt{\til \rho^{\text{R}}}}{\sqrt{\til \rho^{\text{L}}} + \sqrt{\til \rho^{\text{R}}}},\\
        \hat{\til H} & = \frac{\til H^{\text{L}}\sqrt{\til \rho^{\text{L}}} + \til H^{\text{R}}\sqrt{\til \rho^{\text{R}}}}{\sqrt{\til \rho^{\text{L}}} + \sqrt{\til \rho^{\text{R}}}}.
    \end{align}
\end{subequations}
The Roe-averaged Jacobian matrix $\hat{\mathbf{A}}$ is then obtained by evaluating $\mathbf{A}$ at the Roe-averaged state:
\begin{equation}
    \hat{\mathbf{A}} = 
    \begin{bmatrix}
        0 & 1 & 0\\
        -\frac{2}{3}\hat{\til u}^2 & \frac{4}{3}\hat{\til u} & \frac23\\
        \frac13\hat{\til u}^3 - \hat{\til H} \hat{\til u} & -\frac23\hat{\til u}^2 + \hat{\til H} & \frac{5}{3}\hat{\til u}
    \end{bmatrix}.
\end{equation} 
This construction satisfies the Roe property
\begin{equation}
    \mathbf{F}^{\text{R}} - \mathbf{F}^{\text{L}} = \hat{\mathbf{A}} \cdot (\mathbf{U}^{\text{R}} - \mathbf{U}^{\text{L}}),
\end{equation}
where the superscripts $\text{L}$ and $\text{R}$ denote the left and right states, respectively. 
The resulting local linearized Jacobian matrix $\hat{\mathbf{A}}$ enables a characteristic decomposition of the flow at each cell interface.

The eigenvalues of $\hat{\mathbf{A}}$ are
\begin{subequations}
    \begin{align}
        \lambda_1 &= \hat{\til u}, \\
        \lambda_2 &= \hat{\til u} - \sqrt{\frac53}\hat{\til v}, \\
        \lambda_3 &= \hat{\til u} + \sqrt{\frac53}\hat{\til v},
    \end{align}
\end{subequations}
and the corresponding eigenvectors are
\begin{subequations}
    \begin{align}
        \mathbf{K}_1 &= [1, \hat{\til u}, \frac{\hat{\til u}^2}{2}]^\text{T}, \\
        \mathbf{K}_2 &= [1, \hat{\til u} - \sqrt{\frac53}\hat{\til v}, \frac12\hat{\til u}^2 - \sqrt{\frac53}\hat{\til u}\hat{\til v} + \frac52\hat{\til v}^2 ]^\text{T}, \\
        \mathbf{K}_3 &= [1, \hat{\til u} + \sqrt{\frac53}\hat{\til v}, \frac12\hat{\til u}^2 + \sqrt{\frac53}\hat{\til u}\hat{\til v} + \frac52\hat{\til v}^2]^\text{T},
    \end{align}
\end{subequations}
where $\hat{\til v} = \sqrt{2(\hat{\til H}-\hat{\til u}^2/2)/5}$.

These eigenvectors define the characteristic directions in state space, while the eigenvalues represent the corresponding wave speeds. At each cell interface, the jump in the state vector,
$\Delta\mathbf{U} = \mathbf{U}^{\text{R}} - \mathbf{U}^{\text{L}}$ can be decomposed along the characteristic directions as
\begin{equation}
    \Delta\mathbf{U} = \alpha_1 \mathbf{K}_1 + \alpha_2 \mathbf{K}_2 + \alpha_3 \mathbf{K}_3.
\end{equation}
By defining $\phi = \hat{\tilde{u}}^2\Delta U_1 - 2\hat{\tilde{u}}\Delta U_2 + 2\Delta U_3$, these coefficients are compactly expressed as
\begin{equation}
    \begin{aligned}
        \alpha_1 &= \Delta U_1 - \frac{\phi}{5\hat{\tilde{v}}^2}, \\
        \alpha_{2,3} &= \frac{\phi \pm \sqrt{15}\hat{\tilde{v}}(\hat{\tilde{u}}\Delta U_1 - \Delta U_2)}{10\hat{\tilde{v}}^2}.
    \end{aligned}
\end{equation}

The numerical flux is obtained by adding left-going wave contributions to the left-state flux,
\begin{equation}
    \mathbf{F} = \mathbf{F}(\mathbf{U}^{\text{L}}) + \sum_{\lambda_j\leq 0}\lambda_j \alpha_j \mathbf{K}_j,
\end{equation}
or by subtracting right-going wave contributions from the right-state flux,
\begin{equation}
    \mathbf{F} = \mathbf{F}(\mathbf{U}^{\text{R}}) - \sum_{\lambda_j\geq 0}\lambda_j \alpha_j \mathbf{K}_j.
\end{equation}
Combining these two forms yields the standard symmetric Roe flux:
\begin{equation}
    \mathbf{F} = \frac12\left(\mathbf{F}(\mathbf{U}^{\text{L}}) + \mathbf{F}(\mathbf{U}^{\text{R}})\right) - \frac12 \sum_{j=1}^3 \lvert\lambda_j\rvert \alpha_j \mathbf{K}_j.
\end{equation}

\section{Implicit Euler method}  
\label{app:implicit}
In this appendix, we describe the numerical treatment of the diffusive flux $\mathbf{F}_{\text{diff}}$. In our formulation, $\mathbf{F}_{\text{diff}}$ contains only a nonzero energy component associated with thermal conduction driven by the temperature gradient. 
Consequently, it affects only the energy component of the state vector $\mathbf{U}$, while the density and momentum components remain unchanged. 
Based on this property, we adopt a frozen-coefficient treatment during the conduction step, in which quantities such as the density $\til \rho$ and radial velocity $\til u$ are held fixed. 
This decoupling reduces the thermal update to an independent scalar diffusion problem.

The thermal conduction equation is written as
\begin{equation}\label{eq:conduction}
    \til \rho \frac{\partial}{\partial \til t} \left(\frac32 \til v^2 \right) =  \frac{1}{\til r^2} \frac{\partial}{\partial \til r} \left( \til r^2 \til \kappa \frac{\partial \til v^2}{\partial \til r} \right).
\end{equation}
Applying the finite volume method gives 
\begin{equation}
    \begin{aligned}
        \frac{3}{2} \til \rho_i\frac{(\til v^2)_i^{n+1} - (\til v^2)_i^{n}}{\Delta \til t} \til V_i =  
        \til A_{i+\frac12} \til\kappa_{i+\frac12}(\nabla \til v^2)^{n+1}_{i+\frac12} \\
        - \til A_{i-\frac12} \til\kappa_{i-\frac12}(\nabla \til v^2)^{n+1}_{i-\frac12},
    \end{aligned}
\end{equation}
where the numerical gradient is approximated by
\begin{equation}
    (\nabla \til v^2)^{n+1}_{i+\frac12} = 
    \frac{(\til v^2)_{i+1}^{n+1} - (\til v^2)_i^{n+1}}{\til r_{i+1}-\til r_{i}}.
\end{equation}
The discretized equation can be rewritten in tridiagonal form
\begin{equation}
    a_i (\til v^2)_{i-1}^{n+1} + b_i (\til v^2)_{i}^{n+1} + c_i (\til v^2)_{i+1}^{n+1} = d_i,
\end{equation}
where the coefficients are
\begin{subequations}
    \begin{align}
        a_i &= - \frac{2}{3} \frac{\Delta \til t}{\til V_i} \cdot \frac{\til A_{i-1/2}\til\kappa_{i-1/2}}{\til r_i - \til r_{i-1}} \\
        c_i &= - \frac{2}{3} \frac{\Delta \til t}{\til V_i} \cdot \frac{\til A_{i+1/2}\til\kappa_{i+1/2}}{\til r_{i+1} - \til r_{i}} \\
        b_i &= \til \rho_i - a_i - c_i \\
        d_i &= \til \rho_i (\til v^2)_i^n
    \end{align}
\end{subequations}
The resulting tridiagonal linear system is solved efficiently with the Thomas algorithm to obtain the updated velocity dispersion squared $(\til v^2)^{n+1}$.

\newpage


\bibliography{citeLib}

@article{Nishikawa:2019lsc,
    author = "Nishikawa, Hiroya and Boddy, Kimberly K. and Kaplinghat, Manoj",
    title = "{Accelerated core collapse in tidally stripped self-interacting dark matter halos}",
    eprint = "1901.00499",
    archivePrefix = "arXiv",
    primaryClass = "astro-ph.GA",
    doi = "10.1103/PhysRevD.101.063009",
    journal = "Phys. Rev. D",
    volume = "101",
    number = "6",
    pages = "063009",
    year = "2020"
}

@ARTICLE{van1977towards,
       author = {{van Leer}, Bram},
        title = "{Towards the Ultimate Conservative Difference Scheme. IV. A New Approach to Numerical Convection}",
      journal = {Journal of Computational Physics},
         year = 1977,
        month = mar,
       volume = {23},
        pages = {276},
          doi = {10.1016/0021-9991(77)90095-X},
       adsurl = {https://ui.adsabs.harvard.edu/abs/1977JCoPh..23..276V}
}

@article{godunov1959finite,
  title={Finite difference methods for the computation of discontinuous solutions of the equations of fluid dynamics},
  author={Godunov, SK0171},
  journal={Mat. Sb.},
  volume={47},
  pages={271--306},
  year={1959}
}

@ARTICLE{roe1981approximate,
       author = {{Roe}, P.~L.},
        title = "{Approximate Riemann Solvers, Parameter Vectors, and Difference Schemes}",
      journal = {Journal of Computational Physics},
         year = 1981,
        month = oct,
       volume = {43},
       number = {2},
        pages = {357-372},
          doi = {10.1016/0021-9991(81)90128-5},
       adsurl = {https://ui.adsabs.harvard.edu/abs/1981JCoPh..43..357R}
}

@book{toro2013riemann,
  title={Riemann solvers and numerical methods for fluid dynamics: a practical introduction},
  author={Toro, Eleuterio F},
  year={2013},
  publisher={Springer Science \& Business Media}
}

@book{leveque2002finite,
  title={Finite volume methods for hyperbolic problems},
  author={LeVeque, Randall J},
  volume={31},
  year={2002},
  publisher={Cambridge university press}
}

@book{blazek2015computational,
  title={Computational fluid dynamics: principles and applications},
  author={Blazek, Jiri},
  year={2015},
  publisher={Butterworth-Heinemann}
}

@article{Amiram:2006zjz,
    author = {Harten, Amiram and Lax, Peter D. and van Leer, Bram},
    title = "{On Upstream Differencing and Godunov-Type Schemes for Hyperbolic Conservation Laws}",
    doi = "10.1137/1025002",
    journal = "SIAM Rev.",
    volume = "25",
    number = "1",
    pages = "35--61",
    year = "1983"
}

@ARTICLE{1994ShWav...4...25T,
       author = {{Toro}, E.~F. and {Spruce}, M. and {Speares}, W.},
        title = "{Restoration of the contact surface in the HLL-Riemann solver}",
      journal = {Shock Waves},
         year = 1994,
        month = jul,
       volume = {4},
       number = {1},
        pages = {25-34},
          doi = {10.1007/BF01414629},
       adsurl = {https://ui.adsabs.harvard.edu/abs/1994ShWav...4...25T}
}

@article{Arcadi:2017kky,
    author = "Arcadi, Giorgio and Dutra, Ma{\'\i}ra and Ghosh, Pradipta and Lindner, Manfred and Mambrini, Yann and Pierre, Mathias and Profumo, Stefano and Queiroz, Farinaldo S.",
    title = "{The waning of the WIMP? A review of models, searches, and constraints}",
    eprint = "1703.07364",
    archivePrefix = "arXiv",
    primaryClass = "hep-ph",
    doi = "10.1140/epjc/s10052-018-5662-y",
    journal = "Eur. Phys. J. C",
    volume = "78",
    number = "3",
    pages = "203",
    year = "2018"
}

@article{XENON:2024hup,
    author = "Aprile, E. and others",
    collaboration = "XENON",
    title = "{First Search for Light Dark Matter in the Neutrino Fog with XENONnT}",
    eprint = "2409.17868",
    archivePrefix = "arXiv",
    primaryClass = "hep-ex",
    doi = "10.1103/PhysRevLett.134.111802",
    journal = "Phys. Rev. Lett.",
    volume = "134",
    number = "11",
    pages = "111802",
    year = "2025"
}

@article{LZ:2024zvo,
    author = "Aalbers, J. and others",
    collaboration = "LZ",
    title = "{Dark Matter Search Results from 4.2{\,}{\,}Tonne-Years of Exposure of the LUX-ZEPLIN (LZ) Experiment}",
    eprint = "2410.17036",
    archivePrefix = "arXiv",
    primaryClass = "hep-ex",
    reportNumber = "FERMILAB-PUB-24-0796-V",
    doi = "10.1103/4dyc-z8zf",
    journal = "Phys. Rev. Lett.",
    volume = "135",
    number = "1",
    pages = "011802",
    year = "2025"
}

@article{Spergel:1999mh,
    author = "Spergel, David N. and Steinhardt, Paul J.",
    title = "{Observational evidence for selfinteracting cold dark matter}",
    eprint = "astro-ph/9909386",
    archivePrefix = "arXiv",
    doi = "10.1103/PhysRevLett.84.3760",
    journal = "Phys. Rev. Lett.",
    volume = "84",
    pages = "3760--3763",
    year = "2000"
}

@article{Vogelsberger:2012ku,
    author = "Vogelsberger, Mark and Zavala, Jesus and Loeb, Abraham",
    title = "{Subhaloes in Self-Interacting Galactic Dark Matter Haloes}",
    eprint = "1201.5892",
    archivePrefix = "arXiv",
    primaryClass = "astro-ph.CO",
    doi = "10.1111/j.1365-2966.2012.21182.x",
    journal = "Mon. Not. Roy. Astron. Soc.",
    volume = "423",
    pages = "3740",
    year = "2012"
}

@article{Tulin:2017ara,
    author = "Tulin, Sean and Yu, Hai-Bo",
    title = "{Dark Matter Self-interactions and Small Scale Structure}",
    eprint = "1705.02358",
    archivePrefix = "arXiv",
    primaryClass = "hep-ph",
    doi = "10.1016/j.physrep.2017.11.004",
    journal = "Phys. Rept.",
    volume = "730",
    pages = "1--57",
    year = "2018"
}

@ARTICLE{2010AdAst2010E...5D,
       author = {{de Blok}, W.~J.~G.},
        title = "{The Core-Cusp Problem}",
      journal = {Advances in Astronomy},
         year = 2010,
        month = jan,
       volume = {2010},
          eid = {789293},
        pages = {789293},
          doi = {10.1155/2010/789293},
archivePrefix = {arXiv},
       eprint = {0910.3538},
 primaryClass = {astro-ph.CO},
       adsurl = {https://ui.adsabs.harvard.edu/abs/2010AdAst2010E...5D}
}

@article{Boylan-Kolchin:2011qkt,
    author = "Boylan-Kolchin, Michael and Bullock, James S. and Kaplinghat, Manoj",
    title = "{Too big to fail? The puzzling darkness of massive Milky Way subhaloes}",
    eprint = "1103.0007",
    archivePrefix = "arXiv",
    primaryClass = "astro-ph.CO",
    doi = "10.1111/j.1745-3933.2011.01074.x",
    journal = "Mon. Not. Roy. Astron. Soc.",
    volume = "415",
    pages = "L40",
    year = "2011"
}

@article{Zavala:2012us,
    author = "Zavala, Jesus and Vogelsberger, Mark and Walker, Matthew G.",
    title = "{Constraining Self-Interacting Dark Matter with the Milky Way's dwarf spheroidals}",
    eprint = "1211.6426",
    archivePrefix = "arXiv",
    primaryClass = "astro-ph.CO",
    doi = "10.1093/mnrasl/sls053",
    journal = "Mon. Not. Roy. Astron. Soc.",
    volume = "431",
    pages = "L20--L24",
    year = "2013"
}

@article{Oman:2015xda,
    author = "Oman, Kyle A. and others",
    title = "{The unexpected diversity of dwarf galaxy rotation curves}",
    eprint = "1504.01437",
    archivePrefix = "arXiv",
    primaryClass = "astro-ph.GA",
    doi = "10.1093/mnras/stv1504",
    journal = "Mon. Not. Roy. Astron. Soc.",
    volume = "452",
    number = "4",
    pages = "3650--3665",
    year = "2015"
}

@article{Kamada:2016euw,
    author = "Kamada, Ayuki and Kaplinghat, Manoj and Pace, Andrew B. and Yu, Hai-Bo",
    title = "{How the Self-Interacting Dark Matter Model Explains the Diverse Galactic Rotation Curves}",
    eprint = "1611.02716",
    archivePrefix = "arXiv",
    primaryClass = "astro-ph.GA",
    doi = "10.1103/PhysRevLett.119.111102",
    journal = "Phys. Rev. Lett.",
    volume = "119",
    number = "11",
    pages = "111102",
    year = "2017"
}

@article{Kaplinghat:2015aga,
    author = "Kaplinghat, Manoj and Tulin, Sean and Yu, Hai-Bo",
    title = "{Dark Matter Halos as Particle Colliders: Unified Solution to Small-Scale Structure Puzzles from Dwarfs to Clusters}",
    eprint = "1508.03339",
    archivePrefix = "arXiv",
    primaryClass = "astro-ph.CO",
    doi = "10.1103/PhysRevLett.116.041302",
    journal = "Phys. Rev. Lett.",
    volume = "116",
    number = "4",
    pages = "041302",
    year = "2016"
}

@article{Bullock:2017xww,
    author = "Bullock, James S. and Boylan-Kolchin, Michael",
    title = "{Small-Scale Challenges to the $\Lambda$CDM Paradigm}",
    eprint = "1707.04256",
    archivePrefix = "arXiv",
    primaryClass = "astro-ph.CO",
    doi = "10.1146/annurev-astro-091916-055313",
    journal = "Ann. Rev. Astron. Astrophys.",
    volume = "55",
    pages = "343--387",
    year = "2017"
}

@misc{Kong:2025sqx,
    author = "Kong, Demao and Nadler, Ethan O. and Yu, Hai-Bo",
    title = "{Strong Lensing Perturbers from the SIDM Concerto Suite}",
    eprint = "2510.01491",
    archivePrefix = "arXiv",
    primaryClass = "astro-ph.CO",
    month = "10",
    year = "2025"
}

@misc{Yu:2025tmp,
    author = "Yu, Hai-Bo",
    title = "{Three Birds with One Stone: Core-Collapsed SIDM Halos as the Common Origin of Dense Perturbers in Lenses, Streams, and Satellites}",
    eprint = "2510.11006",
    archivePrefix = "arXiv",
    primaryClass = "astro-ph.GA",
    month = "10",
    year = "2025"
}

@article{Matthee:2023utn,
    author = "Matthee, Jorryt and others",
    title = "{Little Red Dots: An Abundant Population of Faint Active Galactic Nuclei at z {\ensuremath{\sim}} 5 Revealed by the EIGER and FRESCO JWST Surveys}",
    eprint = "2306.05448",
    archivePrefix = "arXiv",
    primaryClass = "astro-ph.GA",
    doi = "10.3847/1538-4357/ad2345",
    journal = "Astrophys. J.",
    volume = "963",
    number = "2",
    pages = "129",
    year = "2024"
}

@misc{Feng:2025rzf,
    author = "Feng, Wei-Xiang and Yu, Hai-Bo and Zhong, Yi-Ming",
    title = "{Dark Bondi Accretion Aided by Baryons and the Origin of JWST Little Red Dots}",
    eprint = "2506.17641",
    archivePrefix = "arXiv",
    primaryClass = "astro-ph.GA",
    month = "6",
    year = "2025"
}

@article{Inayoshi:2025isg,
    author = "Inayoshi, Kohei",
    title = "{Little Red Dots as the Very First Activity of Black Hole Growth}",
    doi = "10.3847/2041-8213/adea66",
    journal = "Astrophys. J. Lett.",
    volume = "988",
    number = "1",
    pages = "L22",
    year = "2025"
}

@article{Shapiro:2023gpe,
    author = "Shapiro, Stuart L.",
    title = "{Spikes and accretion of unbound, collisionless matter around black holes}",
    eprint = "2310.13739",
    archivePrefix = "arXiv",
    primaryClass = "astro-ph.GA",
    doi = "10.1103/PhysRevD.108.083037",
    journal = "Phys. Rev. D",
    volume = "108",
    number = "8",
    pages = "083037",
    year = "2023"
}

@article{Ostriker:1999ee,
    author = "Ostriker, Jeremiah P.",
    title = "{Collisional dark matter and the origin of massive black holes}",
    eprint = "astro-ph/9912548",
    archivePrefix = "arXiv",
    reportNumber = "POPE-818",
    doi = "10.1103/PhysRevLett.84.5258",
    journal = "Phys. Rev. Lett.",
    volume = "84",
    pages = "5258--5260",
    year = "2000"
}

@article{Hennawi:2001be,
    author = "Hennawi, Joseph F. and Ostriker, Jeremiah P.",
    title = "{Observational constraints on the self interacting dark matter scenario and the growth of supermassive black holes}",
    eprint = "astro-ph/0108203",
    archivePrefix = "arXiv",
    doi = "10.1086/340226",
    journal = "Astrophys. J.",
    volume = "572",
    pages = "41",
    year = "2002"
}

@article{Hu:2005cd,
    author = "Hu, Jian and Shen, Yue and Lou, Yu-Qing and Zhang, Shuangnan",
    title = "{Forming supermassive black holes by accreting dark and baryon matter}",
    eprint = "astro-ph/0510222",
    archivePrefix = "arXiv",
    doi = "10.1111/j.1365-2966.2005.09712.x",
    journal = "Mon. Not. Roy. Astron. Soc.",
    volume = "365",
    pages = "345--351",
    year = "2006"
}

@article{VMSabarish:2025tya,
    author = {Sabarish, V. M. and Br{\"u}ggen, Marcus and Schmidt-Hoberg, Kai and Fischer, Moritz S.},
    title = "{Accretion of self-interacting dark matter onto supermassive black holes}",
    eprint = "2505.14779",
    archivePrefix = "arXiv",
    primaryClass = "astro-ph.CO",
    doi = "10.1051/0004-6361/202555586",
    journal = "Astron. Astrophys.",
    volume = "703",
    pages = "A142",
    year = "2025"
}

@article{Shapiro:2014oha,
    author = "Shapiro, Stuart L. and Paschalidis, Vasileios",
    title = "{Self-interacting dark matter cusps around massive black holes}",
    eprint = "1402.0005",
    archivePrefix = "arXiv",
    primaryClass = "astro-ph.CO",
    doi = "10.1103/PhysRevD.89.023506",
    journal = "Phys. Rev. D",
    volume = "89",
    number = "2",
    pages = "023506",
    year = "2014"
}

@article{Balberg:2002ue,
    author = "Balberg, Shmuel and Shapiro, Stuart L. and Inagaki, Shogo",
    title = "{Self-interacting dark matter halos and the gravothermal catastrophe}",
    eprint = "astro-ph/0110561",
    archivePrefix = "arXiv",
    doi = "10.1086/339038",
    journal = "Astrophys. J.",
    volume = "568",
    pages = "475--487",
    year = "2002"
}

@article{Navarro:1995iw,
    author = "Navarro, Julio F. and Frenk, Carlos S. and White, Simon D. M.",
    title = "{The Structure of cold dark matter halos}",
    eprint = "astro-ph/9508025",
    archivePrefix = "arXiv",
    doi = "10.1086/177173",
    journal = "Astrophys. J.",
    volume = "462",
    pages = "563--575",
    year = "1996"
}

@article{Lei:2025pky,
    author = "Lei, Lei and Wang, Yi-Ying and Li, Qiao and Dong, Jiang and Wang, Ze-Fan and Lin, Wei-Long and Shu, Yi-Ping and Cao, Xiao-Yue and Yang, Da-Neng and Fan, Yi-Zhong",
    title = "{A Dense Dark Matter Core of the Subhalo in the Strong Lensing System JVAS B1938+666}",
    eprint = "2509.07808",
    archivePrefix = "arXiv",
    primaryClass = "astro-ph.CO",
    doi = "10.3847/2041-8213/ae047c",
    journal = "Astrophys. J. Lett.",
    volume = "991",
    number = "1",
    pages = "L27",
    year = "2025"
}

@article{Read:2002wb,
    author = "Read, Justin I. and Gilmore, G.",
    title = "{Can supermassive black holes alter cold dark matter cusps through accretion?}",
    eprint = "astro-ph/0210658",
    archivePrefix = "arXiv",
    doi = "10.1046/j.1365-8711.2003.06232.x",
    journal = "Mon. Not. Roy. Astron. Soc.",
    volume = "339",
    pages = "949",
    year = "2003"
}

@article{Koda:2011yb,
    author = "Koda, Jun and Shapiro, Paul R.",
    title = "{Gravothermal collapse of isolated self-interacting dark matter haloes: N-body simulation versus the fluid model}",
    eprint = "1101.3097",
    archivePrefix = "arXiv",
    primaryClass = "astro-ph.CO",
    reportNumber = "TCC-001-11",
    doi = "10.1111/j.1365-2966.2011.18684.x",
    journal = "Mon. Not. Roy. Astron. Soc.",
    volume = "415",
    pages = "1125",
    year = "2011"
}

@article{Ahn:2004xt,
    author = "Ahn, Kyung-Jin and Shapiro, Paul R.",
    title = "{Formation and evolution of the self-interacting dark matter halos}",
    eprint = "astro-ph/0412169",
    archivePrefix = "arXiv",
    doi = "10.1111/j.1365-2966.2005.09492.x",
    journal = "Mon. Not. Roy. Astron. Soc.",
    volume = "363",
    pages = "1092--1124",
    year = "2005"
}

@article{Outmezguine:2022bhq,
    author = "Outmezguine, Nadav Joseph and Boddy, Kimberly K. and Gad-Nasr, Sophia and Kaplinghat, Manoj and Sagunski, Laura",
    title = "{Universal gravothermal evolution of isolated self-interacting dark matter halos for velocity-dependent cross-sections}",
    eprint = "2204.06568",
    archivePrefix = "arXiv",
    primaryClass = "astro-ph.GA",
    doi = "10.1093/mnras/stad1705",
    journal = "Mon. Not. Roy. Astron. Soc.",
    volume = "523",
    number = "3",
    pages = "4786--4800",
    year = "2023"
}

@article{Gad-Nasr:2023gvf,
    author = "Gad-Nasr, Sophia and Boddy, Kimberly K. and Kaplinghat, Manoj and Outmezguine, Nadav Joseph and Sagunski, Laura",
    title = "{On the late-time evolution of velocity-dependent self-interacting dark matter halos}",
    eprint = "2312.09296",
    archivePrefix = "arXiv",
    primaryClass = "astro-ph.GA",
    doi = "10.1088/1475-7516/2024/05/131",
    journal = "JCAP",
    volume = "05",
    number = "2024",
    pages = "131",
    year = "2024"
}

@article{Yang:2022hkm,
    author = "Yang, Daneng and Yu, Hai-Bo",
    title = "{Gravothermal evolution of dark matter halos with differential elastic scattering}",
    eprint = "2205.03392",
    archivePrefix = "arXiv",
    primaryClass = "astro-ph.CO",
    doi = "10.1088/1475-7516/2022/09/077",
    journal = "JCAP",
    volume = "09",
    number = "2022",
    pages = "077",
    year = "2022"
}

@article{Jiang:2025jtr,
    author = "Jiang, Fangzhou and Jia, Zixiang and Zheng, Haonan and Ho, Luis C. and Inayoshi, Kohei and Shen, Xuejian and Vogelsberger, Mark and Feng, Wei-Xiang",
    title = "{Formation of the Little Red Dots from the Core Collapse of Self-interacting Dark Matter Halos}",
    eprint = "2503.23710",
    archivePrefix = "arXiv",
    primaryClass = "astro-ph.GA",
    doi = "10.3847/2041-8213/ae247a",
    journal = "Astrophys. J. Lett.",
    volume = "996",
    number = "1",
    pages = "L19",
    year = "2026"
}

@article{Li:2025kpb,
    author = "Li, Shubo and others",
    title = "{The {\textquotedblleft}Little Dark Dot{\textquotedblright}: Evidence for Self-interacting Dark Matter in the Strong Lens SDSS J0946+1006?}",
    eprint = "2504.11800",
    archivePrefix = "arXiv",
    primaryClass = "astro-ph.GA",
    doi = "10.3847/1538-4357/ae1462",
    journal = "Astrophys. J.",
    volume = "994",
    number = "2",
    pages = "201",
    year = "2025"
}

@article{Feng:2020kxv,
    author = "Feng, Wei-Xiang and Yu, Hai-Bo and Zhong, Yi-Ming",
    title = "{Seeding Supermassive Black Holes with Self-interacting Dark Matter: A Unified Scenario with Baryons}",
    eprint = "2010.15132",
    archivePrefix = "arXiv",
    primaryClass = "astro-ph.CO",
    doi = "10.3847/2041-8213/ac04b0",
    journal = "Astrophys. J. Lett.",
    volume = "914",
    number = "2",
    pages = "L26",
    year = "2021"
}

@article{Hou:2025gmv,
    author = "Hou, Siyuan and Yang, Daneng and Li, Nan and Li, Guoliang",
    title = "{A universal analytic model for gravitational lensing by self-interacting dark matter halos}",
    eprint = "2502.14964",
    archivePrefix = "arXiv",
    primaryClass = "astro-ph.CO",
    doi = "10.1088/1475-7516/2025/08/048",
    journal = "JCAP",
    volume = "08",
    number = "2025",
    pages = "048",
    year = "2025"
}

@article{Hu:2023oiu,
    author = "Hu, Li and Cai, Rong-Gen and Wang, Shao-Jiang",
    title = "{Distinctive GWBs from eccentric inspiraling SMBH binaries with a DM spike}",
    eprint = "2312.14041",
    archivePrefix = "arXiv",
    primaryClass = "gr-qc",
    doi = "10.1088/1475-7516/2025/02/067",
    journal = "JCAP",
    volume = "02",
    number = "2025",
    pages = "067",
    year = "2025"
}

@misc{Gu:2026zzq,
    author = "Gu, Hua-Peng and Jiang, Fangzhou and Chen, Xian and Li, Ran",
    title = "{Non-Equilibrium Relativistic Core Collapse of Self-Interacting Dark Matter Halos -- Limits On Seed Black Hole Mass}",
    eprint = "2601.17117",
    archivePrefix = "arXiv",
    primaryClass = "astro-ph.CO",
    month = "1",
    year = "2026"
}

@article{Roberts:2024uyw,
    author = "Roberts, M. Grant and Kaplinghat, Manoj and Valli, Mauro and Yu, Hai-Bo",
    title = "{Gravothermal collapse and the diversity of galactic rotation curves}",
    eprint = "2407.15005",
    archivePrefix = "arXiv",
    primaryClass = "astro-ph.GA",
    doi = "10.1103/PhysRevD.111.103041",
    journal = "Phys. Rev. D",
    volume = "111",
    number = "10",
    pages = "103041",
    year = "2025"
}

@article{Zhong:2023yzk,
    author = "Zhong, Yi-Ming and Yang, Daneng and Yu, Hai-Bo",
    title = "{The impact of baryonic potentials on the gravothermal evolution of self-interacting dark matter haloes}",
    eprint = "2306.08028",
    archivePrefix = "arXiv",
    primaryClass = "astro-ph.CO",
    doi = "10.1093/mnras/stad2765",
    journal = "Mon. Not. Roy. Astron. Soc.",
    volume = "526",
    number = "1",
    pages = "758--770",
    year = "2023"
}

@article{Yang:2021kdf,
    author = "Yang, Daneng and Yu, Hai-Bo",
    title = "{Self-interacting dark matter and small-scale gravitational lenses in galaxy clusters}",
    eprint = "2102.02375",
    archivePrefix = "arXiv",
    primaryClass = "astro-ph.GA",
    doi = "10.1103/PhysRevD.104.103031",
    journal = "Phys. Rev. D",
    volume = "104",
    number = "10",
    pages = "103031",
    year = "2021"
}

@article{Kaplinghat:2019dhn,
    author = "Kaplinghat, Manoj and Ren, Tao and Yu, Hai-Bo",
    title = "{Dark Matter Cores and Cusps in Spiral Galaxies and their Explanations}",
    eprint = "1911.00544",
    archivePrefix = "arXiv",
    primaryClass = "astro-ph.GA",
    doi = "10.1088/1475-7516/2020/06/027",
    journal = "JCAP",
    volume = "06",
    number = "2020",
    pages = "027",
    year = "2020"
}

@misc{Jia:2026ocr,
    author = "Jia, Zixiang and Jiang, Fangzhou and Li, Shubo and Li, Ran and Wang, Jing and Zhu, Ling",
    title = "{An Enhanced Isothermal Jeans Approach to Constraining Dark Matter Self-Interactions from Galactic Kinematics}",
    eprint = "2601.17118",
    archivePrefix = "arXiv",
    primaryClass = "astro-ph.GA",
    month = "1",
    year = "2026"
}

@article{Jiang:2022aqw,
    author = "Jiang, Fangzhou and others",
    title = "{A semi-analytic study of self-interacting dark-matter haloes with baryons}",
    eprint = "2206.12425",
    archivePrefix = "arXiv",
    primaryClass = "astro-ph.CO",
    doi = "10.1093/mnras/stad705",
    journal = "Mon. Not. Roy. Astron. Soc.",
    volume = "521",
    number = "3",
    pages = "4630--4644",
    year = "2023"
}

@misc{Li:2026duw,
    author = "Li, Shubo and Fischer, Moritz S. and Jia, Zixiang and Jiang, Fangzhou and Li, Ran and Yu, Hai-Bo",
    title = "{Testing the isothermal Jeans model for self-interacting dark matter halos in the collapse phase}",
    eprint = "2603.01772",
    archivePrefix = "arXiv",
    primaryClass = "astro-ph.CO",
    month = "3",
    year = "2026"
}

@article{Yang:2025xsp,
    author = "Yang, Daneng and Fan, Yi-Zhong and Hou, Siyuan and Tsai, Yue-Lin Sming",
    title = "{Self-interacting dark matter with mass segregation: a unified explanation of dwarf cores and small-scale lenses}",
    eprint = "2506.14898",
    archivePrefix = "arXiv",
    primaryClass = "astro-ph.CO",
    doi = "10.1016/j.scib.2026.01.077",
    journal = "Sci. Bull.",
    volume = "71",
    pages = "1349--1356",
    year = "2026"
}

@article{Yang:2022mxl,
    author = "Yang, Daneng and Nadler, Ethan O. and Yu, Hai-Bo",
    title = "{Strong Dark Matter Self-interactions Diversify Halo Populations within and surrounding the Milky Way}",
    eprint = "2211.13768",
    archivePrefix = "arXiv",
    primaryClass = "astro-ph.GA",
    doi = "10.3847/1538-4357/acc73e",
    journal = "Astrophys. J.",
    volume = "949",
    number = "2",
    pages = "67",
    year = "2023"
}

@article{Li:2025nvu,
    author = "Li, Zhengrong and Inayoshi, Kohei and Chen, Kejian and Ichikawa, Kohei and Ho, Luis C.",
    title = "{Little Red Dots: Rapidly Growing Black Holes Reddened by Extended Dusty Flows}",
    doi = "10.3847/1538-4357/ada5fb",
    journal = "Astrophys. J.",
    volume = "980",
    number = "1",
    pages = "36",
    year = "2025"
}

@article{Yang:2023jwn,
    author = "Yang, Daneng and Nadler, Ethan O. and Yu, Hai-Bo and Zhong, Yi-Ming",
    title = "{A parametric model for self-interacting dark matter halos}",
    eprint = "2305.16176",
    archivePrefix = "arXiv",
    primaryClass = "astro-ph.CO",
    doi = "10.1088/1475-7516/2024/02/032",
    journal = "JCAP",
    volume = "02",
    number = "2024",
    pages = "032",
    year = "2024"
}

@article{Feng:2021rst,
    author = "Feng, Wei-Xiang and Yu, Hai-Bo and Zhong, Yi-Ming",
    title = "{Dynamical instability of collapsed dark matter halos}",
    eprint = "2108.11967",
    archivePrefix = "arXiv",
    primaryClass = "astro-ph.CO",
    doi = "10.1088/1475-7516/2022/05/036",
    journal = "JCAP",
    volume = "05",
    number = "2022",
    pages = "036",
    year = "2022"
}

@article{Gilman:2022ida,
    author = "Gilman, Daniel and Zhong, Yi-Ming and Bovy, Jo",
    title = "{Constraining resonant dark matter self-interactions with strong gravitational lenses}",
    eprint = "2207.13111",
    archivePrefix = "arXiv",
    primaryClass = "astro-ph.CO",
    doi = "10.1103/PhysRevD.107.103008",
    journal = "Phys. Rev. D",
    volume = "107",
    number = "10",
    pages = "103008",
    year = "2023"
}

\end{document}